\documentclass[aps,prd,superscriptaddress,nofootinbib,onecolumn,preprintnumbers,amsmath,amsfonts,amssymb]{revtex4} 

\usepackage{graphicx}
\usepackage{dcolumn}
\usepackage{bm}

\usepackage{hyperref}
\usepackage[utf8]{inputenc}
\usepackage[dvipsnames]{xcolor}
\usepackage[normalem]{ulem}
\usepackage{xspace}
\usepackage{fontawesome} 
\usepackage{multirow}

\usepackage{textgreek}

\usepackage{dsfont}

\usepackage[dvipsnames]{xcolor}
\usepackage[utf8]{inputenc}
\hypersetup{
    colorlinks=true,
    citecolor = green,
    linkcolor=red,
    filecolor=magenta,      
    urlcolor=blue,
}
\hypersetup{allcolors=[rgb]{0.0 0.0 0.9},linkcolor=[rgb]{0.75 0.05 0.05},urlcolor=[rgb]{0.6 0.0 0.6}, }

\usepackage{url}

\newcommand{\e}  {\oplus}
\newcommand{\s}  {\odot}

\begin{document}

\preprint{BARI-TH/788-26}

\title{Matter Flavor Conversion Mediated by Pseudo-Sterile States\\
as the Possible Origin of Neutrino Oscillation Anomalies}


\author{		Sabya Sachi Chatterjee}
\email{sabya.chatterjee@kit.edu}
\affiliation{		Institut f\"{u}r Astroteilchenphysik, Karlsruher Institut f\"{u}r
Technologie (KIT), Hermann-von-Helmholtz-Platz 1, 76344
Eggenstein-Leopoldshafen, Germany}

\author{		Antonio Palazzo}
\email{palazzo@ba.infn.it}
\affiliation{ 	Dipartimento Interateneo di Fisica ``Michelangelo Merlin,'' Via Amendola 173, 70126 Bari, Italy}
\affiliation{ 	Istituto Nazionale di Fisica Nucleare, Sezione di Bari, Via Orabona 4, 70126 Bari, Italy}


\begin{abstract}


Neutrino oscillation experiments present anomalous results across a vast range of baselines and energies.
Here we show that a 3+1 scenario in which sterile neutrinos feel 
a novel matter potential $V_s$ proportional to background density of ordinary or (asymmetric)
dark matter is able to explain several 
anomalies. In the limit of low-energies $E\lesssim$ 1 TeV
the model gives rise to effective NSI-like couplings among active flavors arranged
in a distinctive non-generic rank-1 Hamiltonian. Remarkably,
unlike the ordinary NSI, the induced NSI-like couplings, being generated via mixing with a gauge-singlet state,
are immune to the stringent constraints from $SU(2)$ gauge invariance involving the charged leptons.
The NSI-like couplings eliminate the tension between the two LBL experiments 
NOvA and T2K provided that the potential is negative
and the two sterile mixing angles $\theta_{14}$ and $\theta_{24}$ are non-zero.
Solar neutrino data further constrain the strength of the potential and
 impose the bound  $|U_{e4}|^2 \simeq \sin^2\theta_{14}\lesssim 0.04$, while
LBL and low-energy atmospheric data establish the hierarchical pattern  $|U_{\mu4}|^2 \ll |U_{e4}|^2$ and
 $|U_{\tau4}|^2 \ll |U_{e4}|^2$.
 A further indication in favor of a non-zero negative potential comes from the anomalous excess of $\nu_e$-like events 
observed in  Super-Kamiokande multi-GeV atmospheric neutrinos,
which, in the new scenario can be explained by a modification of the  3-flavor resonance 
at energies of a few GeV.
In the high-energy regime ($E\gtrsim $ 1 TeV)
 the new framework  reveals its genuine 4-flavor nature and can produce a resonant behavior
 at $E \simeq$ 10 TeV as hinted at by IceCube which, for the preferred negative values of the potential,  
 occurs in the neutrino channel. The resonance presents unconventional 
 features, reflecting the complex structure of the amplification phenomenon.
Specifically, we identify an irreducible three-level dynamics generating a new resonance in the  $(\nu_e, \nu_\mu)$ sector
intertwined with two conventional resonances in the $(\nu_e, \nu_s$) and $(\nu_\mu, \nu_s)$ sub-systems.
The novel amplification mechanism
manifests with the emergence of 
effective mixing angles in matter ($\theta_{12}^m$ or $\theta_{13}^m$) involving active neutrinos
unrelated to their vacuum counterparts and can be operative even in the absence 
of conventional resonances.
In order to have maximal phenomenological relevance the scenario requires
values of $f = V_s/|V_{NC}| \sim -20 $, $\Delta m^2_{41} \sim 60 $ eV$^2$ (both roughly uncertain by a factor of two),
$|U_{e4}|^2\simeq \sin^2\theta_{14} \simeq 0.01-0.03$ and $|U_{\mu4}|^2 \simeq \sin^2\theta_{24}\simeq 10^{-4}-10^{-3}$.
 Such a very small size of $|U_{\mu4}|^2$ eliminates the tension between IceCube and the other (all negative) $\nu_\mu$ disappearance searches.
 The third mixing angle can have a sizable impact even for tiny values
as small as $|U_{\tau4}|^2 \simeq \sin^2\theta_{34}\sim3 \times 10^{-4}$.
Concerning the nature (Dirac vs Majorana) of the pseudo-sterile neutrino, the null results  from $0\nu2\beta$ decay 
strongly favor the Dirac case. However, this conclusion can be evaded in 3+2 and 3+3 frameworks if
an underlying symmetry forces the effective Majorana mass close to zero.
The model can play a substantial role in the Reactor anomaly and can partially explain the Gallium one.
More crucially, it can be directly 
probed by KATRIN, which is very sensitive to the electron-sterile neutrino admixture in the region of high $\Delta m^2_{41}$.

\end{abstract}

\pacs{13.15.+g, 14.60.Pq}

\maketitle

\section{Introduction} 

The question of the anomalies in the neutrino sector is becoming 
increasingly intricate. Together with the longstanding results of LSND~\cite{LSND:2001aii},
the Reactor~\cite{Mention:2011rk}, Gallium~\cite{GALLEX:1994rym,GALLEX:1997lja,Kaether:2010ag,Abdurashitov:1996dp,SAGE:2009eeu,Barinov:2021asz,Barinov:2022wfh}  
and Neutrino-4~\cite{NEUTRINO-4:2018huq} anomalies and the MiniBooNE excess~\cite{MiniBooNE:2020pnu},%
\footnote{The anomalous excess of electron neutrinos recorded by MiniBooNE has been recently 
challenged by the negative results of a dedicated search of MicroBooNE~\cite{MicroBooNE:2024tym}.}
new anomalous results  are gradually emerging and consolidating. This is the case of IceCube~\cite{IceCubeCollaboration:2024nle,IceCube:2024pky},
 which hints at a resonant-like behavior
in atmospheric muon neutrinos for energies of $\sim$ 10 TeV. In addition,
an apparently unrelated tension  
between the two long-baseline (LBL) experiments NOvA~\cite{NOVA_talk_nu2024} and T2K~\cite{T2K_talk_nu2024} has recently 
appeared~\cite{Denton:2020uda,Chatterjee:2020kkm,Chatterjee:2024kbn}. Finally, here we evidence 
that there is a possibly anomalous excess of electron-like events in the atmospheric neutrino data 
collected by Super-Kamiokande~\cite{Super-Kamiokande:2023ahc},
which in our opinion deserves attention. According to these latest developments,
the anomalous results involve a vast range of energies and distances. 
Hence, in our view, a more complex picture is emerging, in which  the ``traditional'' 
 short-baseline (SBL) anomalies represent only one piece of a broader mosaic. 
In addition, the new discrepancies involving LBL, low-energy and high-energy atmospheric neutrinos, 
strongly suggest a key role of matter effects in the anomalies.

The so-called 3+1 scenario, entailing one sterile neutrino in addition to the three active species, 
has been thoroughly explored as a possible solution of the anomalies with prominent attention
to those recorded in SBL experiments  (for a recent review see~\cite{Dasgupta:2021ies,Acero:2022wqg}). 
However, it has been gradually
recognized that this simple scheme is not able to address simultaneously all the SBL anomalies and it can only explain 
a limited subset of them. Adding further light sterile degrees of freedom (the so-called 3$+N$ schemes) does not help. 
 Perhaps, the most important drawback of these scenarios is their inability to explain simultaneously
 the positive signal of electron neutrino appearance and the joint disappearance searches of electron neutrinos (with positive outcome)
 and muon neutrinos (with negative outcome)~\cite{Gariazzo:2015rra,Dentler:2018sju,Diaz:2019fwt}. 
 Additionally, a tension between the IceCube high-energy results and the other muon neutrino disappearance searches
 is consolidating and has been recently quantified to lie around the 3$\sigma$  level~\cite{Villarreal:2025ged}.
 Also, very recently the 3+1 scheme has been challenged by a dedicated analysis of MicroBooNE~\cite{MicroBooNE:2025nll}. Finally, in the background,  it remains the question of the incompatibility of such ``vanilla'' sterile neutrinos
 with cosmology (see for example~\cite{Berryman:2019nvr,Hagstotz:2020ukm,Archidiacono:2022ich}).
In an effort to overcome these drawbacks, it has been gradually accepted that sterile neutrinos, if 
still deemed relevant for the anomalies, should possess additional exotic properties.
This is not implausible and in retrospect it seems quite natural given that we are scrutinizing  
completely unknown fundamental particles. Examples of ideas going in this direction are given by the  recent hypothesis of
the non-unitary 3+1 scheme~\cite{Minakata:2025azk}, the invisible sterile neutrino 
decay~\cite{Diaz:2019fwt,Moulai:2019gpi,Hardin:2022muu} or decoherence~\cite{Hardin:2022muu} as a means to alleviate
the appearance-disappearance tension, the decay in visible products~\cite{Palomares-Ruiz:2005zbh,Bai:2015ztj,Dentler:2019dhz,deGouvea:2019qre,Hostert:2024etd}
to explain the LSND and MiniBooNE excesses,
and the so-called secret interactions~\cite{Hannestad:2013ana,Dasgupta:2013zpn,Archidiacono:2014nda,Saviano:2014esa,Mirizzi:2014ama,Chu:2015ipa,Cherry:2016jol,Archidiacono:2016kkh,Chu:2018gxk,Farzan:2019yvo,Cline:2019seo,Archidiacono:2020yey},
put forward to reconcile terrestrial neutrino data with cosmology.

Here we consider a 3+1 scenario which entails  a new matter potential felt by the sterile neutrinos,
already investigated in various contexts related to neutrino oscillations~\cite{Belotsky:2001fb,Nelson:2007yq,Engelhardt:2010dx,Pospelov:2011ha, Karagiorgi:2012kw,Kopp:2014fha,Berryman:2018jxt,Liao:2018mbg,Esmaili:2018qzu,Denton:2018dqq,Alves:2022vgn,Sponsler:2024iej},
but without reaching a concrete conclusion on its role in the anomalies.%
\footnote{Interestingly, such a scenario has been recently considered in~\cite{Brdar:2025azm} as a possible explanation
of the tension between KM3NeT and IceCube concerning the high-energy 220 PeV event recorded in KM3NeT.}
We will refer to such a kind of neutrinos as {\em pseudo-sterile neutrinos} to emphasize that these
states, albeit being singlets under the  Standard Model gauge group, can have other kinds of interactions in addition
to the gravitational one.%
\footnote{Pseudo comes from Greek $\psi\varepsilon\upsilon\delta\acute{\eta}\varsigma$ (pseud\={e}s) meaning ``false''.}
We will show that the pseudo-sterile states can concretely help to explain several anomalies.
The key point of this scenario, unrecognized in the previous literature, is that these exotic states can
act as mediators of new unconventional matter flavor conversion phenomena among the active flavors, 
which under appropriate conditions can be even resonantly amplified. 
We will explore the rich phenomenology implied by this scenario having
in mind two goals. First, we will ensure that the new scenario respects the constraints coming from the 
entire neutrino oscillation phenomenology. Second, we will show how it can provide a satisfying explanation
of several anomalies, without being affected by the internal tensions plaguing the ordinary 3+1 scheme.
 
 We will demonstrate that in the limit of relatively low energies ($E\lesssim$ 1 TeV)
 the sterile state $\nu_s$ can be integrated out from the dynamics,
 and the model behaves as an effective 3-flavor NSI-like scheme
among active flavors. The NSI-like couplings are arranged 
in a distinctive non-generic rank-1 Hamiltonian. Remarkably,
unlike ordinary NSI, the induced NSI-like couplings, being generated via mixing with a gauge-singlet state,
are immune to the stringent constraints from $SU(2)$ gauge invariance involving charged leptons.
At the phenomenological level, the NSI-like couplings allow us to explain the tension between the two LBL experiments 
NOvA and T2K provided that the new potential is negative%
\footnote{Here we tacitly refer to neutrinos. For antineutrinos the potential has the opposite sign.}
and the two sterile mixing angles $\theta_{14}$ and $\theta_{24}$ are non-zero.
With a dedicated analytical and numerical study we will show that solar neutrino data 
further constrain the strength of the new potential and impose the bound  $|U_{e4}|^2 \simeq \sin^2\theta_{14}\lesssim 0.04$. 
In addition,  we will see that LBL and low-energy atmospheric data establish the hierarchical pattern
 $|U_{\mu4}|^2 \ll |U_{e4}|^2$ and $|U_{\tau4}|^2 \ll |U_{e4}|^2$.
This information will guide us to identify the benchmark parameters of the model which seem able to help
in resolving other experimental anomalies. 
In particular, remaining in the low-energy regime, we will show that in addition to the LBL discrepancy,
the model can explain an anomalous excess of $\nu_e$-like events recorded in 
multi-GeV atmospheric data in Super-Kamiokande. This is possible
because the NSI-like couplings lead to a substantial modification of the standard 
3-flavor resonant behavior occurring in the atmospheric neutrino flavor conversion at energies of a few GeV.

At high energies $E\gtrsim $ 1 TeV, the fourth state $\nu_s$ cannot be integrated out%
\footnote{In particular cases, the fourth state $\nu_s$ can be integrated out also at high energies $E\gtrsim $ 1 TeV, 
giving rise to a novel effective two-level resonance (see Section~\ref{Sec:Dark}).}
and the new framework  reveals its genuine 4-flavor nature producing a resonant behavior
 at $E \simeq$ 10 TeV as hinted at by IceCube which, for the preferred negative values of the potential,  
 occurs in the neutrino channel.
 The possibility of  the realization of a resonance in high-energy neutrinos is not new and indeed it is expected 
in the ordinary 3+1 scheme. However, we will demonstrate that in the new
scenario the resonant amplification mechanism becomes much more complex and
presents unconventional features. More precisely, the model predicts 
new resonances in the active $(\nu_e,\nu_\mu,\nu_\tau)$ sector, which
are mediated by the fourth pseudo-sterile state $\nu_s$ through amplified NSI-like effects.
We enucleate an irreducible three-level dynamics 
in which a new resonance in the active sector, for example the $(\nu_e, \nu_\mu)$ sub-system, 
 is intertwined with two  conventional resonances in the $(\nu_e, \nu_s$) and $(\nu_\mu, \nu_s)$ sub-systems, 
 giving rise to a triple-avoided crossing phenomenon.
This  three-level behavior manifests  
with the emergence of large matter mixing angles and mass-squared differences 
in the active sector unrelated to their (irrelevant) vacuum counterparts.
While three-level systems are widely studied
in other fields of physics (photonics, quantum optics and quantum information among the others),
they are basically unexplored in the neutrino field. This circumstance has deep roots in the
structure of the 3-flavor Hamiltonian in which the two mass-squared differences (solar and atmospheric) 
are hierarchically separated and the potential has a particularly simple form,
 leading to two distinct resonances~\cite{Kuo:1987zx}. 
Therefore, the pseudo-sterile neutrino 
scenario, in addition to its prominent purpose of providing a possible interpretation of
the experimental anomalies, offers the opportunity to study and hopefully observe 
the rich dynamics of  multi-level systems in neutrino oscillations.

In order to work and be of maximal phenomenological interest, the scenario under investigation requires
values of $f = V_s/|V_{NC}| \simeq -20 $, $\Delta m^2_{41} \simeq 60 $ eV$^2$ (both roughly uncertain by a factor of two),
$|U_{e4}|^2\simeq \sin^2\theta_{14} \simeq 0.01-0.03$ and $|U_{\mu4}|^2 \simeq \sin^2\theta_{24}\simeq 10^{-4}-10^{-3}$.
Notably, such small values of $\sin^2\theta_{24}$ eliminate the tension 
of IceCube with the stringent bounds deriving from the negative searches of $\nu_\mu$ disappearance performed
by MINOS/MINOS+~\cite{MINOS:2017cae} and NOvA~\cite{NOvA:2024imi}. 
Beyond this, the model predicts several additional experimental signatures. 
Most importantly, for the non-negligible value of $|U_{e4}|^2 \simeq \sin^2\theta_{14}\sim 0.01-0.03$, it should
show up with a distinctive kink in the $\beta$-decay energy spectrum measured by KATRIN,
which is very sensitive to the electron neutrino mixing at the high values of $\Delta m^2_{41}$ considered here.
Concerning the nature (Dirac vs Majorana) of the pseudo-sterile neutrino, the null results  from $0\nu2\beta$ decay 
strongly favor the Dirac case. However, this conclusion can be evaded in 3+2 and 3+3 frameworks if
an underlying symmetry forces the effective Majorana mass close to zero.
 
The paper is structured as follows. In Sec.~\ref{Sec:Model} we introduce the 3+1 scheme with the new matter potential.
In Sec.~\ref{Sec:LBL} we demonstrate how this framework at low energies can be approximated
for small values of the new mixing angles and
elucidate the formal similarity of the ordinary NSI Hamiltonian involving the active neutrinos. 
In Sec.~\ref{Sec:Tension} we show that the tension between NOvA and T2K can be resolved
within the new framework. In Sec.~\ref{Sec:Solar} we perform an analytical and numerical study of the new model in
the context of solar neutrino flavor conversions showing their role in constraining its parameters.
We consider the two cases in which the new matter potential is proportional to the 
electron (or proton) number density and to neutron number density. We show that only in the first case,
a degenerate solution is present in the upper (dubbed as ``dark'') octant of the solar mixing angle $\theta_{12}$. 
In Sec.~\ref{Sec:LBL_ATM}, guided by NOvA and T2K findings, the solar neutrino results and 
the existing bounds/hints on NSI from low energy atmospheric neutrino experiments, 
we identify the benchmark parameters of the model of phenomenological interest.
In Sec.~\ref{Sec:Res_6GeV} we describe the modifications induced by the new matter potential
on the standard 3-flavor resonance occurring in atmospheric neutrinos at energies of a few GeV, 
showing that they can explain the excess of $\nu_e$-like events observed in Super-Kamiokande. 
Also, we assess the status of the neutrino mass ordering (NMO) determination
in the new framework. In Sec.~\ref{Sec:IceCube} we show that for the benchmark parameters of the model 
an unconventional resonant behavior is expected around 10 TeV in IceCube, 
qualitatively different from the one realized in the ordinary  3+1 scheme. We clarify the framework
behind the conventional resonances  and explain the basic features of the new resonances  aided
by a numerical study of the behavior of the eigenvalues and mixing angles (eigenvectors) in matter.
We point out the irreducible three-level origin of the new resonant phenomenon.
In Sec.~\ref{Sec:Dark} we demonstrate that for the solar mixing angle $\theta_{12}$ in the dark octant,
a novel $(\nu_e, \nu_\mu)$ resonance occurs at a high energies which is regulated by 
an effective two-level Hamiltonian, providing a full analytical treatment of this regime. 
In Sec.~\ref{Sec:ADM} we briefly discuss the possibility that the new
potential may be related to the interaction of the pseudo-sterile neutrinos with a background of  asymmetric dark matter.
In Sec.~\ref{Sec:Discussion} we discuss the possible origin of the new matter potential, the nature
(Dirac vs Majorana) of the pseudo-sterile neutrinos and 
the phenomenological implications of the proposed scenario. 
Finally, we present our conclusions in Sec.~\ref{Sec:Conclusions}.

\section{The Model}
\label{Sec:Model}

In the presence of a sterile neutrino, the mixing between the flavor ($\nu_e,\nu_\mu,\nu_\tau, \nu_s$)
and the mass eigenstates ($\nu_1,\nu_2,\nu_3,\nu_4$) is parametrized by a $4\times4$ 
unitary matrix%
\footnote{We recall that in the 3-flavor framework one introduces three mass eigenstates $\nu_i$ with masses
$m_i\, (i = 1,2,3)$, three mixing angles $\theta_{12},\theta_{13}, \theta_{23}$, and one CP-phase $\delta_{\mathrm {CP}}$.
In this work we will indicate $\delta_{\mathrm {CP}}\equiv \delta_{13}$ both in the 3-flavor framework and in the 3+1 scheme. 
The neutrino mass ordering (NMO) is said to be
 normal (inverted) if $m_3>m_{1,2}$  ($m_3<m_{1,2}$). We will abbreviate normal (inverted) ordering as NO (IO).}
\begin{equation}
\label{eq:U}
U =   \tilde R_{34}  R_{24} \tilde R_{14} R_{23} \tilde R_{13} R_{12}\,, 
\end{equation} 
where, $R_{ij}$ ($\tilde R_{ij}$) represents a real (complex) $4\times4$ rotation 
of a mixing angle $\theta_{ij}$ which contains the $2\times2$ submatrix 
\begin{eqnarray}
\label{eq:R_ij_2dim}
     R^{2\times2}_{ij} =
    \begin{pmatrix}
         c_{ij} &  s_{ij}  \\
         - s_{ij}  &  c_{ij}
    \end{pmatrix} ,
\,\,\,\,\,\,\,   
     \tilde R^{2\times2}_{ij} =
    \begin{pmatrix}
         c_{ij} &  \tilde s_{ij}  \\
         - \tilde s_{ij}^*  &  c_{ij}
    \end{pmatrix}
\,,    
\end{eqnarray}
in the  $(i,j)$ sub-block, having defined
\begin{eqnarray}
 c_{ij} \equiv \cos \theta_{ij}, \qquad s_{ij} \equiv \sin \theta_{ij}, \qquad  \tilde s_{ij} \equiv s_{ij} e^{-i\delta_{ij}}.
\end{eqnarray}
Some useful properties of the matrix in Eq.~(\ref{eq:U}) can be evidenced as follows: i) The 
3-flavor PMNS matrix is recovered if one sets $\theta_{14} = \theta_{24} = \theta_{34} =0$.
ii) For small values of the mixing angles $\theta_{13}$, $\theta_{14}$, and $\theta_{24}$, 
we have $|U_{e3}|^2 \simeq s^2_{13}$, $|U_{e4}|^2 = s^2_{14}$, 
$|U_{\mu4}|^2 \simeq s^2_{24}$, and $|U_{\tau4}|^2 \simeq s^2_{34}$, 
with a simple connection between the matrix elements and the mixing angles. 
iii) The leftmost positioning of the matrix $\tilde R_{34}$ guarantees that the $\nu_{\mu} \to \nu_{e}$ probability 
in vacuum is independent of $\theta_{34}$ and of the associated CP phase $\delta_{34}$
(see~\cite{Klop:2014ima}).

The presence of a new sterile neutrino potential modifies the Hamiltonian of neutrino propagation in matter, which in the flavor basis becomes
\begin{flalign}
H \,=\,\, &U\, K\, U^{\dagger} + V\,,
\label{Eq:H4nu}
\end{flalign}
%
where, $K={\mathrm{diag}}\left(0, k_{21}, k_{31}, k_{41}\right)$ denotes the diagonal matrix containing the wave numbers
 $k_{ij} = (m_i^2 - m_j^2)/2E$, while $V$ represents the diagonal matrix encoding the matter effects defined as
\begin{flalign}
V = {\mathrm{diag}}\left(V_{CC}, 0, 0, V_S - V_{NC} \right)\,,
\end{flalign}
where
\begin{flalign}
\label{eq:V_CC}
V_{CC} = \sqrt{2}G_F N_e\,,
\end{flalign}
is the standard charged current (CC) potential, $N_e$ being the electron number density of the background matter, and
\begin{flalign}
\label{eq:V_NC}
V_{NC} = -\frac{1}{2}\sqrt{2}G_F N_n\,,
\end{flalign}
is the standard neutral current (NC)  potential, $N_n$ being the neutron number density.
The additional term $V_S$ is the new sterile potential which, for definiteness,  we measure in units
of the standard NC potential as
\begin{equation}
\label{eq:V_S}
V_S = - f  \times V_{NC}
\end{equation}
by introducing the rescaling parameter $f$. With this definition, the new potential has the same
sign of $f$, since $V_{NC}$ is negative.
For the sake of convenience, we introduce the auxiliary parameter
\begin{equation}
\bar f =  r(1 + f) 
\end{equation}
where $r =|V_{NC}|/|V_{CC}|$, which in the Earth is constant and approximately $r = 0.5$,
while in Sun it depends on the radial position $x = R/R_{sun}$,
where $R_{sun}$ is the Sun radius. 
Using this notation the matrix representing the matter potential simply reads
\begin{flalign}
\label{eq:V2}
V = V_{CC}\times {\mathrm{diag}}\left(1, 0, 0, \bar f \right)\,.
\end{flalign}
We see that $\bar f $ allows us to gauge the potential felt by sterile neutrinos in units of the charged-current potential $V_{CC}$.
We note that $\bar f = r$ in the standard model case (being $f=0$), while for the big values of $f$,
which will be  phenomenologically relevant, the approximation $\bar f \simeq rf$
holds. When considering the evolution of antineutrinos, in addition to the replacement $U\to U^*$,
one has to change the sign of the potentials $V_{CC}, V_{NC}$ and $V_{S}$, which is equivalent to change
the sign of $V_{CC}$ in Eq.~(\ref{eq:V2}).

Here, an important remark is in order. Equation~(\ref{eq:V_S}) assumes that the new matter potential 
is proportional to the neutron number density. 
While we will adopt this choice as a benchmark throughout our work, it should be stressed that 
the nature of the interaction of the sterile neutrinos is unknown and it can involve electrons, 
protons and neutrons. For the propagation in the Earth the particular choice of  the fermion is irrelevant since the 
number densities of the three particles are approximately equal $N_e = N_p \simeq N_n$,
being the Earth an iso-scalar medium. Differently, for solar neutrinos we still have  $N_e = N_p$ to ensure
neutrality, while $N_n$  is suppressed by a factor which depends on the radial position $x = R/R_{sun}$. 
In particular, the neutron number density close to the center of the sun, which is relevant 
for adiabatic flavor conversions, is approximately a factor of two smaller than $N_e$.
Therefore, in the treatment of solar neutrinos this aspect must be taken into account if one considers
a matter potential different from our benchmark choice. In case of  interaction with fermions
different from neutrons (electrons or protons), the new potential should be 
rescaled consistently with the adopted choice (see Sec.~\ref{Sec:Solar}  for further clarifications).

\section{Small-mixing NSI-like approximation in LBL and atmospheric neutrinos}
\label{Sec:LBL}

Let us now consider the impact of the new potential for the phenomenology of LBL
experiments such as NOvA and T2K. The treatment is valid also for atmospheric
neutrinos for energies which are much lower than the resonance energy, which as we
will see in Sec.~\ref{Sec:IceCube}, lies around 10 TeV. Before treating the matter effects, 
we consider the kinematical ones by giving a look to the appearance probability in vacuum.
As first pointed out~\cite{Klop:2014ima}, the $\nu_{\mu} \to \nu_{e}$ 
probability can be approximated as the sum of three terms,
\begin{eqnarray}
\label{eq:Pme_4nu_3_terms}
P^{4\nu}_{\mu e}  \simeq  P^{\rm{ATM}} + P^{\rm {INT}}_{\rm I}+   P^{\rm {INT}}_{\rm II}\,,
\end{eqnarray}
where,
\begin{flalign}
\label{eq:Pme_atm}
& P^{\rm {ATM}} \,\, \simeq\,  4 s_{23}^2 s^2_{13}  \sin^2{\Delta},\\
 \label{eq:Pme_int_1}
& P^{\rm {INT}}_{\rm I} \,\,  \simeq\,    8 s_{13} s_{12} c_{12} s_{23} c_{23} (\alpha \Delta)\sin \Delta \cos({\Delta + \delta_{13}}),\\
 \label{eq:Pme_int_2}
& P^{\rm {INT}}_{\rm II} \,\,  \simeq\,   4 s_{14} s_{24} s_{13} s_{23} \sin\Delta \sin (\Delta + \delta_{13} - \delta_{14})\,,
\end{flalign}
where $\Delta \equiv  \Delta m^2_{31}L/4E$ is the atmospheric oscillating factor, 
$L$  and $E$ being the neutrino baseline and energy, respectively. 
We have also defined the ratio of the solar over the atmospheric mass-squared splitting 
$\alpha \equiv \Delta m^2_{21}/ \Delta m^2_{31} \simeq 0.03$. The term $P^{\rm ATM}$ 
represents the leading order standard three neutrino term which is related to the oscillations driven by 
the atmospheric mass-squared splitting. The other two terms, namely 
$P^{\rm {INT}}_{\rm I}$ and $P^{\rm {INT}}_{\rm II}$ represent two interference terms and each one of them is generated by the interference of two independent frequencies. In particular, the term $P^{\rm INT}_{\rm I}$ arises from the interference between the solar and atmospheric frequencies, whereas the term $P^{\rm INT}_{\rm II}$, a genuine 4-flavor effect, arises from the interference between the atmospheric and the new frequency introduced by the fourth mass eigenstate. Note that these terms survive the averaging process of the fast oscillations induced by the large
eV$^2$-scale value of $\Delta m^2_{41}$.
For the small values of the mixing angles that we will find phenomenologically relevant in the next sections 
($s^2_{14} \simeq 0.02$ and $s^2_{24} \simeq 10^{-3}$), 
the amplitude of the new interference term is $2.5 \times 10^{-3}$, almost one order of magnitude smaller than that of the
standard interference effect which lies around $10^{-2}$. Therefore, the new interference term has negligible 
impact in LBL experiments and in atmospheric neutrinos as we have explicitly checked.
An analogous derivation can be done for the disappearance probability of muon neutrinos $P_{\mu\mu}$ 
(see~\cite{Parveen:2023ixk}), which shows that the corrections to the 3--flavor case are proportional
to the product of $s_{14} s_{24} s_{13}$ and thus below $10^{-3}$.
Hence, the kinematical/vacuum effects induced by the sterile neutrinos are negligible, and 
we must be concerned only with the perturbations induced in the oscillations probabilities
by the matter effects. 

Let's consider the matter effects explicitly. As shown in~\cite{Klop:2014ima} (see the appendix), 
in order to simplify the treatment of matter effects in LBL setups, it is useful to introduce the new basis
\begin{equation}
\bar \nu = \bar U{^\dag} \nu\,,
\label{eq:newbasis}
\end{equation}
where, $\bar U = \tilde R_{34} R_{24} \tilde R_{14}$,\,
is the part of the mixing matrix defined in Eq.~(\ref{eq:U}) that contains only the rotations involving the fourth neutrino mass eigenstate. 
The mixing matrix $U$ is split as follows  
$U = \bar U U_{3\nu}$,\,
where $U_{3\nu}$ is the $4\times4$ matrix which contains the standard 3-flavor mixing
matrix in the (1,2,3) sub-block. In the new basis, the Hamiltonian takes the form
\begin{equation} \label{eq:Hbar}
   \bar H = \bar H^{\rm {vac}} + \bar H^{\rm {mat}} = U_{3\nu} K U_{3\nu}^\dagger
    + {\bar U}^\dagger V {\bar U} \,,
\end{equation}
where the first term describes the oscillations in vacuum, and the second one represents a non-standard dynamical term.
Since the wavenumber $k_{41}$ is much bigger than $k_{21}$ and $|k_{31}|$,
we can reduce the dynamics to that of an effective 3-flavor system. In fact, 
from Eq.~(\ref{eq:Hbar}) one can see that the (4,4) entry of $\bar H$ is much larger than 
all the other entries and the fourth eigenvalue of  $\bar H$ is much larger than
the other three ones. As a consequence, the eigenstate $\bar \nu_s$ evolves independently of the others.
Extracting from  $\bar H$ the submatrix with indices  $(1,2,3)$, one obtains the effective $3\times 3$ 
Hamiltonian
\begin{equation} \label{eq:Hbar_3nu}
   \bar H_{3\nu} = \bar H_{3\nu}^{\rm {vac}} + \bar H_{3\nu}^{\rm {mat}},\,
 \end{equation}
which governs the evolution of the rotated eigenstates $(\bar \nu_{e}, \bar \nu_{\mu}, \bar \nu_{\tau})$.
It should be noted that labeling the rotated states with the flavor index makes sense, since
for very small mixing angles they are very close to the original flavor eigenstates.
In order to calculate
the oscillation probabilities, it is useful to introduce the evolution operator,
which, in the new basis, takes the expression
\begin{equation}
    \bar {S} \equiv e^{-i \bar H L} \approx
    \begin{pmatrix}
   	e^{-i \bar H_{3\nu} L} & \boldsymbol{0} \\
	\boldsymbol{0} & e^{-i k_{41} L}
    \end{pmatrix}\,,
\end{equation}
and is related to the evolution operator in the flavor basis via the
unitary transformation
\begin{equation}
     S = \bar{U} \bar{S} \bar{U}^\dagger \,.
\end{equation}
Considering the block-diagonal form of $\bar{S}$ and the identities $\bar U_{e2} = \bar U_{e3} = \bar U_{\mu 3} = 0$, 
one has for the relevant transition amplitudes (see~\cite{Klop:2014ima,Agarwalla:2016xxa})
\begin{eqnarray}
S_{e\mu} &= &\bar U_{e1} \bar {S}_{ee}  \bar {U}_{\mu 1}^*   + \bar U_{e1}  \bar {S}_{e\mu} \bar {U}_{\mu 2}^*
+ \bar U_{e4} \bar S_{ss} \bar U_{\mu4}^*\,,\\
S_{\mu\mu} &=& \bar U_{\mu1} \bar {S}_{ee}  \bar {U}_{\mu 1}^* + \bar U_{\mu1} \bar {S}_{e\mu}  \bar {U}_{\mu 2}^*
+   \bar U_{\mu2} \bar {S}_{\mu e}  \bar {U}_{\mu 1}^* + \bar U_{\mu2} \bar {S}_{\mu\mu}  \bar {U}_{\mu 2}^*
+ \bar U_{\mu4} \bar S_{ss} \bar U_{\mu4}^*\,.
\end{eqnarray}
Since  $\bar S_{ss} =  e^{-i k_{41} L}$ oscillates very fast, 
the associated terms are averaged out by the finite energy resolution of the detector.
Considering the explicit expressions of the elements of the matrix $\bar U$, 
and expanding in the small values of the mixing angles, one has 
 for the oscillation probabilities the approximate expressions 
\begin{eqnarray}
\label{eq:pme_exp_mix}
    P_{\mu e}^{4\nu} \equiv |S_{e\mu}|^2 &\simeq& (1 - s_{14}^2 -s_{24}^2) \bar P_{\mu e}^{3\nu} -2 s_{14} s_{24} {\rm Re}  (e^{-i \delta_{14}} \bar S_{ee} \bar S_{e \mu}^*)\,,\\
    P_{\mu \mu}^{4\nu} \equiv |S_{\mu\mu}|^2 &\simeq& (1 - 2 s_{24}^2) \bar P_{\mu \mu}^{3\nu} -2 s_{14} s_{24} {\rm Re} 
     (e^{-i \delta_{14}} \bar S_{\mu\mu} \bar S_{e \mu}^* + e^{i \delta_{14}} \bar S_{\mu\mu} \bar S_{\mu e}^*)\,,
      \nonumber
  \end{eqnarray}
where $\bar P_{\mu e}^{3\nu} \equiv  |\bar S_{e \mu}|^2$ and $\bar P_{\mu \mu}^{3\nu} \equiv  |\bar S_{\mu\mu}|^2$. 
Therefore, the  4-flavor problem is reduced to a more familiar 3-flavor one, being the probabilities
expressed in terms of the amplitudes $\bar S_{ee}$, $\bar S_{e\mu}$, $\bar S_{\mu e}$  and  $\bar S_{\mu\mu}$
of the 3-flavor system governed by the effective Hamiltonian $\bar H_{3\nu}$ in Eq.~(\ref{eq:Hbar_3nu}). 
More specifically, we see that the 4-flavor probabilities are the sum of two contributions,
the first one being the 3-flavor probability (with rotated Hamiltonian $\bar H_{3\nu}$) with a prefactor close to unity, and a term deriving
from the interference of the sterile frequency with the atmospheric and solar ones. As first noted
in~\cite{Klop:2014ima} such an interference, maybe counterintuitively,  survives the process of averaging the oscillations.
 It is easy to check (see~\cite{Klop:2014ima}) that in the vacuum limit one obtains the probability expression 
given in Eqs.~(\ref{eq:Pme_atm}-\ref{eq:Pme_int_2}), and in particular the 4-flavor interference
 term in Eq.~(\ref{eq:Pme_int_2}). If the active-sterile mixing angles are very small like in our case, in which 
we will consider $s^2_{14}\simeq0.02$ and $s^2_{24}\simeq 10^{-3}$, and also taking into account that%
\footnote{In normal conditions  $|\bar S_{e\mu}| \propto s_{13}$, but in the presence of the 1-3 resonance it can be bigger.
We have checked that even in this case, for the small values of the sterile mixing angles considered in the present work, 
the relations in the Eqs.~(\ref{eq:pme_exp_analogy}) provide a good estimate of the oscillation probabilities.}
 $|\bar S_{e\mu}| \ll 1$, 
one can neglect the interference terms and arrive at the approximate expressions  
\begin{eqnarray}
\label{eq:pme_exp_analogy}
    P_{\mu e}^{4\nu} &\simeq& \bar P_{\mu e}^{3\nu}\, \\
    P_{\mu \mu}^{4\nu} &\simeq&  \bar P_{\mu \mu}^{3\nu}\,.
      \nonumber
  \end{eqnarray}
Therefore, the 4-flavor probabilities with the original Hamiltonian in Eq.~(\ref{Eq:H4nu}) can be approximated by the 3-flavor probabilities
with the effective Hamiltonian $\bar H_{3\nu}$. While these expressions cannot be used for precision calculations 
(in fact we use full 4-flavor evolution in our codes), they can be very insightful to grasp the behavior
of the 4-flavor probabilities in the limit of very small mixing angles.
In fact, using a result found in~\cite{Klop:2014ima} and making the appropriate modifications needed
to incorporate the new potential, the matter part of the effective Hamiltonian in Eq.~(\ref{eq:Hbar_3nu}), 
for small values of the mixing angles, can be approximated by  
\begin{eqnarray} \label{eq:Hdyn_2}
\arraycolsep=3pt
\medmuskip = 1mu
    \bar H^{\rm{mat}}_{3\nu}  \approx  
   V_{\rm{CC}} \!   \begin{bmatrix}
	1 + \bar f_\e s^2_{14}  & \bar f_\e  \tilde s_{14} s_{24}  & \bar f_\e \tilde s_{14} \tilde s_{34}^*	\\
	\dagger &\bar f_\e s_{24}^2 & \bar f_\e s_{24} \tilde s_{34}^* 
	\\
	\dagger & \dagger
	&\bar f_\e s_{34}^2
    \end{bmatrix}, \,
\end{eqnarray}
where we have denoted $\bar f_\e =  r_\e(1 + f) \simeq 0.5 (1+f)$. The Hamiltonian in Eq.~(\ref{eq:Hdyn_2})
returns the standard 3-flavor matter effect when one sets $\theta_{14} = \theta_{24} = \theta_{34} = 0$. 
On the other hand for the case $f=0$ it reproduces the 3+1 scheme.
Now, one can notice that this matter Hamiltonian, apart from the term equal to unity in the $(1,1)$ entry, 
represents a perturbation matrix which is formally equivalent to the one which regulates the usual
non-standard interactions (NSI), and can be written as
\begin{eqnarray} \label{eq:Hdyn_3_LBL}
\arraycolsep=3pt
\medmuskip = 1mu
    \bar H^{\rm{NSI}}_{3\nu}  \approx  
   V_{\rm{CC}} \!   \begin{bmatrix}
	\varepsilon_{ee}^\e  &   \varepsilon_{e\mu}^\e  & \varepsilon_{e\tau}^\e
	\\
	\dagger & \varepsilon_{\mu\mu}^\e & \varepsilon_{\mu\tau}^\e	\\
	\dagger & \dagger
	& \varepsilon_{\tau\tau}^\e    \end{bmatrix} \,.
\end{eqnarray}
It is important to observe that these couplings do not describe interactions which involve directly the active neutrinos,
since the interaction intervenes only mediated by their mixing with the fourth mass eigenstate and by the new
potential. Remarkably, unlike ordinary NSI, the induced NSI-like couplings, being generated via mixing with a 
gauge-singlet state, are immune to the stringent constraints from $SU(2)$ gauge invariance involving charged leptons.

 Also, we note that in the NSI-like Hamiltonian in Eqs.~(\ref{eq:Hdyn_3_LBL})
there are not nine independent parameters as in a generic hermitian matrix but only six,%
\footnote{In neutrino oscillations a diagonal matrix proportional to the identity can be always added/subtracted, 
so the number of independent parameters decreases by one unit.}
since there are three mixing 
angles $\theta_{14}, \theta_{24}, \theta_{34}$,
two CP-phases $\delta_{14}$ and $\delta_{34}$, and the rescaling parameter $\bar f_\e$.%
\footnote{This parameter counting can be understood also by observing that the NSI-like Hamiltonian can be 
constructed with the real scalar $\bar f $ and the outer product of the complex vector $u^T = (\tilde s_{14}, s_{24}, 
\tilde s_{34})$ being $H^{\rm{NSI}}_{3\nu}   = \bar f u u^\dag$.
This also implies that  the matrix is rank-deficient being of rank 1, having only one linearly independent 
column (or row).} 
It should be also stressed that $\bar f_\e$ is an overall multiplicative factor that,
apart from its sign, can be reabsorbed in the couplings by recasting the  NSI-like hamiltonian
in the form
\begin{eqnarray} \label{eq:Hdyn_2_v2}
    \bar H^{\rm{NSI}}_{3\nu}  \approx  
\arraycolsep=3pt
\medmuskip = 1mu  
 (\operatorname{sgn}{\bar f_\e})
 V_{\rm{CC}} \!   \begin{bmatrix}
	|\varepsilon_{ee}^\e|  &   \sqrt{|\varepsilon_{ee}^\e||\varepsilon_{\mu\mu}^\e|}e^{i\phi_{e\mu}}	 & \sqrt{|\varepsilon_{ee}^\e||\varepsilon_{\tau\tau}^\e}|e^{i\phi_{e\tau}}\\
	\dagger & |\varepsilon_{\mu\mu}^\e| &  \sqrt{|\varepsilon_{\mu\mu}^\e|\varepsilon_{\tau\tau}^\e|}e^{i\phi_{\mu\tau}}	\\
	\dagger & \dagger
	& |\varepsilon_{\tau\tau}^\e|
    \end{bmatrix} \,,
\end{eqnarray}
which depends on the three independent diagonal amplitudes $|\varepsilon^\e_{\alpha\alpha}|$ ($\alpha = e,\mu,\tau$) 
and three CP-phases 
 $\phi_{e\mu} \equiv - \delta_{14}$,  $\phi_{\mu\tau} \equiv \delta_{34}$,
 the third one being a linear combination of the first two  ones
 $\phi_{e\tau} =  \phi_{e\mu} + \phi_{\mu\tau} = \delta_{34} - \delta_{14}$. 
 This observation allows us to derive three important consequences. 
 First, the Hamiltonian depends on five effective parameters instead of six.
 Second, there is a complete degeneracy among the rescaling parameter $\bar f_\e$ and
 the products of the mixing angles $s_{i4}s_{j4}$ with $(i =1,2,3)$ appearing in the
 $(i,j)$ position of the matrix. In order to break such a degeneracy, one will
 need other kind of systems, different from LBL and low-energy neutrinos, like
 for example, SBL setups, solar neutrinos and the resonant behavior like that occurring in IceCube
 high-energy neutrinos, which present a different analytical dependency on the model
 parameters. Third, despite this complete degeneracy, the system
 (in our case LBL and atmospheric oscillations) is sensitive to the sign of $\bar f_\e$.
 We will see that this last property will be crucial for our findings concerning NOvA and T2K.  
 
 We underline that the sterile NSI-like Hamiltonian in Eqs.~(\ref{eq:Hdyn_2}) and (\ref{eq:Hdyn_2_v2})
  has a very distinctive pattern and constitutes one of the main signatures
  of the model under consideration, which could be the target of future LBL experiments
  such as DUNE, Hyper-Kamiokande and ESS$\nu$SB.
In the absence of the new potential of sterile neutrinos one has $\bar f_\e = r_\e \simeq 0.5$
and all the couplings in Eq.~(\ref{eq:Hdyn_2}) are very small, being of the second
order in powers of the $s_{ij}$.  In this limit, one recovers the ordinary 3+1 scenario in which
matter effects have almost no relevance.%
\footnote{The role of the ordinary 3+1 scheme as a possible solution of the NOvA and T2K tension
has been considered in~\cite{Chatterjee:2020yak}, where it has been shown that it is not able to explain the discrepancy
assuming for the mixing angles $\theta_{14}$ and $\theta_{24}$ the values preferred by the 
global fits performed in the 3+1 scheme. For the choice of the mixing angles which will be relevant for
the present work, we have checked that the 3+1 fit to NOvA and T2K is almost indistinguishable from that obtained in the 3-flavor scheme.}
However, things are expected to qualitatively 
change when the rescaling factor $\bar f_\e$  is allowed to assume large $O(10)$ values,
in which case some of the NSI-like couplings will have a non-negligible impact.

\section{Resolution of the tension between NO\lowercase{v}A and T2K}
\label{Sec:Tension}

As we underlined in~\cite{Chatterjee:2020kkm}, the two long-baseline accelerator experiments
NOvA and T2K represent the ideal setup where
to seek non-standard matter effects in neutrino propagation due to their different and complementary configurations. 
In particular, the two experiments work at different peak energies  (2 GeV for NOvA and 0.6 GeV for T2K) 
because of  the different baselines (810 km for NOvA and 295 km for T2K). As a consequence, in NOvA  
matter effects are approximately three/four times larger with respect to T2K. In this situation, one can use T2K
as a  quasi-vacuum experiment which is capable of measuring the CP-phase $\delta_{13}$
almost independently of the presence of NSI. On the other hand, NOvA can exploit this external information 
to acquire sensitivity to the NSI couplings. The interplay between NOvA and T2K presents similarities with
the synergy of solar neutrinos with the KamLAND experiment. In fact, the irrefutable 
proof of the existence  of matter effects in solar neutrino flavor conversion could be obtained 
only by exploiting the external information provided by KamLAND, which
fixes the oscillation parameters (in particular $\Delta m^2_{21}$) independently of matter effects~\cite{Fogli:2003vj}.

NOvA and T2K since 2020 display a persisting discrepancy in the $\nu_e$ 
appearance channel, which shows up
in three ways when the data are interpreted in the standard 3-flavor scheme.%
\footnote{An alternative way to interpret the observed discrepancy, which is conceptually more correct when adopting a 
non-standard scenario, is to consider it as an internal tension of the theoretical 3-flavor framework, which is not flexible
enough to provide a satisfying fit of the experimental data.}
 First, in the normal neutrino mass 
ordering the estimates of the CP-phase $\delta_{13}$ provided by the two experiments are in disagreement, 
being in tension with a statistical significance of $\sim2\sigma$, as recently reevaluated  in~\cite{Chatterjee:2024kbn}.  
While T2K~\cite{T2K_talk_nu2024}  prefers values of $\delta_{13}\simeq 1.5\pi$, 
NOvA~\cite{NOVA_talk_nu2024} points towards $\delta_{13}\simeq \pi$. Second, while each of the
two experiments taken separately prefers the NO, their combination prefer the IO. Third, 
their joint preference for IO is in contrast with all the remaining world
neutrino oscillation data, which prefer NO~\cite{Capozzi:2025ovi,Esteban:2026phq,Capozzi:2025wyn,Esteban:2024eli,deSalas:2020pgw}. 
In~\cite{Chatterjee:2020kkm} (see also~\cite{Denton:2020uda}) we pointed out that all these three discrepancies 
could be solved by hypothesizing the existence of NSI. More specifically,  we found a $\sim 2\sigma$ level
 preference for non-zero complex neutral-current NSI of the flavor changing type involving the
 $e-\mu$ or the $e-\tau$ sectors with couplings $|\varepsilon^\e_{e\mu}| \sim |\varepsilon^\e_{e\tau}|\sim 0.1$.

In the present work, as a preliminary step, we generalize the NSI  analysis of NOvA and T2K
by taking into account two NSI couplings simultaneously. More specifically, we consider the 
off-diagonal complex coupling $\varepsilon^\e_{e\mu}$ (already considered alone in our previous work) 
and the diagonal coupling $\varepsilon^\e_{ee}$ (not considered in our previous work). 
We have done a similar analysis (whose results will be not shown for brevity) 
by replacing  $\varepsilon^\e_{e\mu}$  with $\varepsilon^\e_{e\tau}$. Thanks to the formal analogy expounded in the previous
section, the results of the NSI analysis will allow us to gain insight when interpreting the findings
concerning the 3+1 scenario with the novel matter potential. In addition, this  analysis will allow us to
compare the hint of non-zero NSI coming from NOvA and T2K with the bounds/hints deriving from low
energy atmospheric neutrino experiments. For the analysis, we have made use of the same datasets
for  NOvA~\cite{NOVA_talk_nu2024} and T2K~\cite{T2K_talk_nu2024} experiments used in~\cite{Chatterjee:2024kbn}.
We have fully accounted for both the disappearance and appearance channels for each experiment. 
We use the GLoBES software package~\cite{Huber:2004ka,Huber:2007ji} 
along with an additional public tool~\cite{Kopp_NSI} designed to implement non-standard interactions. 
We marginalized over $\theta_{13}$ using a 2.4\% 1 sigma prior, 
with a central value of $\sin^2\theta_{13}= 0.0223$, as determined by the global analysis~\cite{Capozzi:2025wyn}.
The solar parameters $\Delta m^2_{21}$ and $\sin^2\theta_{12}$ are taken at their 
best fit values found in the updated global fit~\cite{Capozzi:2025ovi}.%
\footnote{When considering the so called Dark-LMA solution~\cite{Miranda:2004nb}, in NOvA and in T2K we fix 
$\sin^2\theta_{12}$ at the octant symmetric value with respect to the value obtained for the ordinary LMA solution.}
\begin{figure*}[t!]
\vspace*{0.1cm}
\hspace*{-0.1cm}
\includegraphics[height=8.80cm,width=8.80cm]{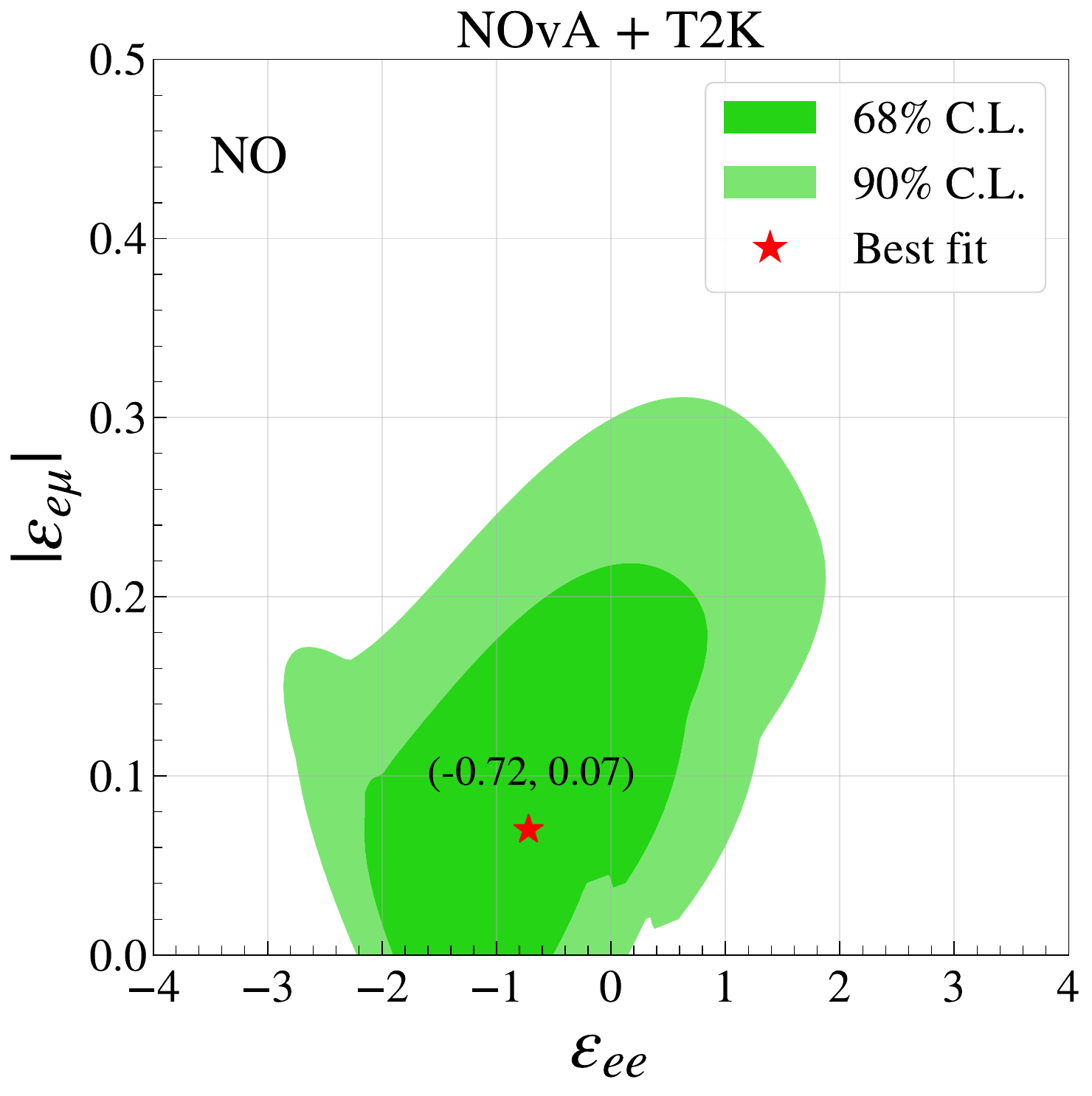}
\includegraphics[height=8.80cm,width=8.80cm]{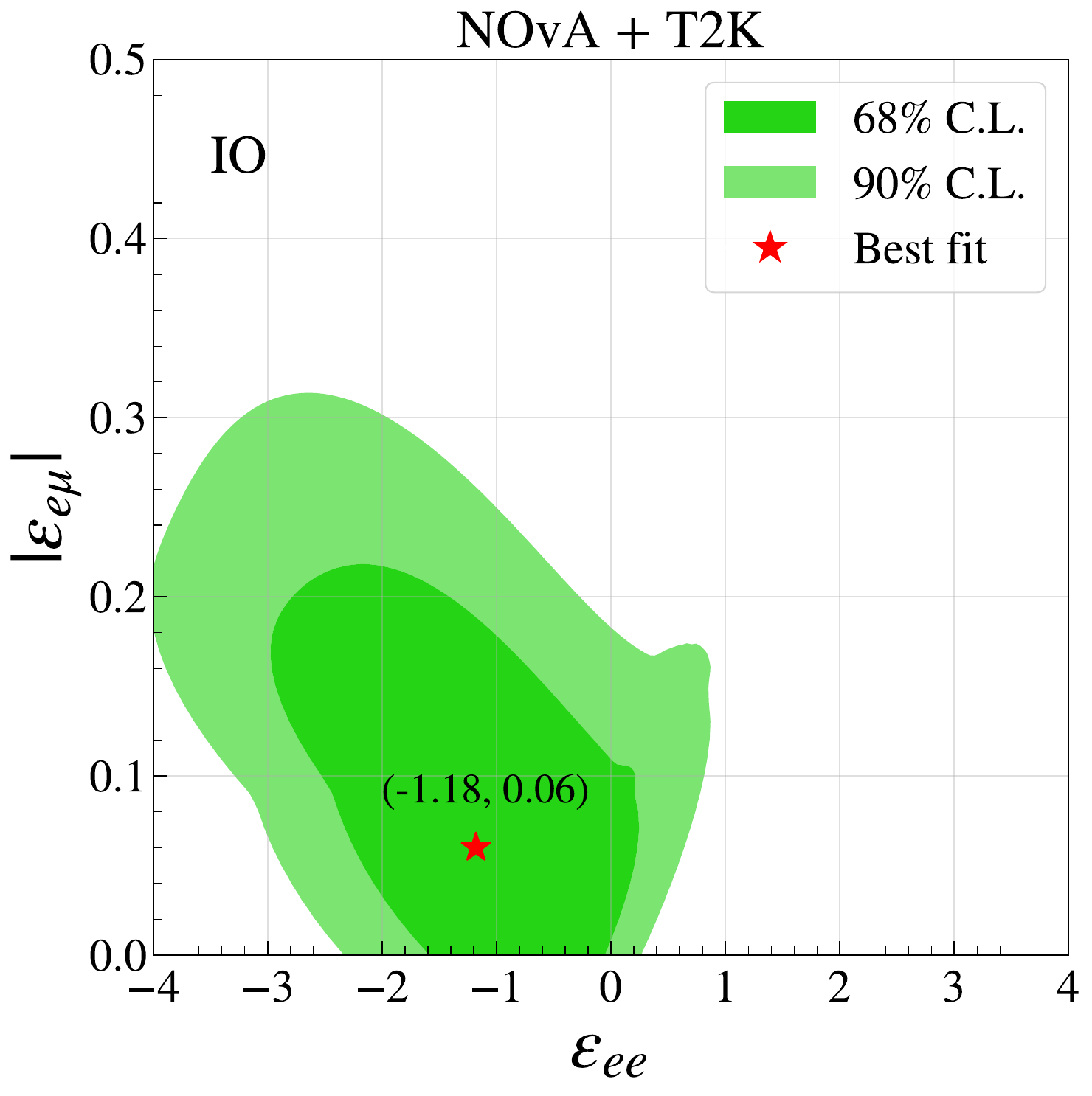}
\vspace*{-0.0cm}
\caption{Allowed regions for the two NSI couplings $\varepsilon_{ee}$ and $|\varepsilon_{e\mu}|$
determined by the combined analysis of NOvA and T2K.  The left (right)  panel refers to NO (IO).
 The contours correspond to the 68$\%$ and 90$\%$ C.L. for 2 d.o.f..}
\label{fig_eps_LBL}
\end{figure*} 

Figure~\ref{fig_eps_LBL} displays the results of this analysis for NO (IO) in the left (right) panel,
where the CP phase $\phi_{e\mu}$ is marginalized. In addition, we marginalize the
standard parameters $\theta_{13}$, $\delta_{13}$, $\theta_{23}$ and $\Delta m^2_{31}$.
We see that in both orderings there is a preference for non-zero NSI couplings, which is more 
pronounced for the case of NO. Specifically, in NO the case of no NSI is disfavored with 
$\Delta \chi^2\simeq 4.1$, while in IO, it is disfavored at $\Delta \chi^2 \simeq 2.6$.
In both plots we observe a pronounced correlation between $\varepsilon^\e_{ee}$ and
$|\varepsilon^\e_{e\mu}|$. We also observe that the plots corresponding to NO and IO 
are symmetrical with respect to the replacement $\varepsilon^\e_{ee} \to -2 -\varepsilon^\e_{ee}$.
Also we have checked that for such a transformation the marginalization  process provides
the replacement  $\phi_{e\mu} \to 2\pi - \phi_{e\mu}$ i.e.  $\varepsilon^\e_{e\mu} \to  \varepsilon^{\e *}_{e\mu}$.
This behavior is expected due to a well known degeneracy property of the neutrino flavor evolution
(see~\cite{Coloma:2016gei}). We notice that while the preferred values of the off-diagonal coupling lie
around $0.05$, the diagonal coupling $|\varepsilon^\e_{ee}|$ can assume large $O(1)$ negative
values. To this regard, one should note that in both NMO,
the case $\varepsilon^\e_{ee} = -1$ (for $\varepsilon^\e_{e\mu} = 0$), corresponds to the complete absence 
of matter effects (standard and non-standard), i.e. to propagation in vacuum. 
Graphically, this would correspond to coinciding biprobability plots for
the two cases of NO and IO for both experiments NOvA and T2K, which, as it is well known, get separated
due to (standard) matter effects. Hence, it is very interesting to note that the two LBL experiments
jointly prefer the vacuum case, and that this weird circumstance can be seen as a further declination
of their mutual tension. 

\begin{figure*}[t!]
\vspace*{0.1cm}
\hspace*{-0.1cm}
\includegraphics[height=8.80cm,width=8.80cm]{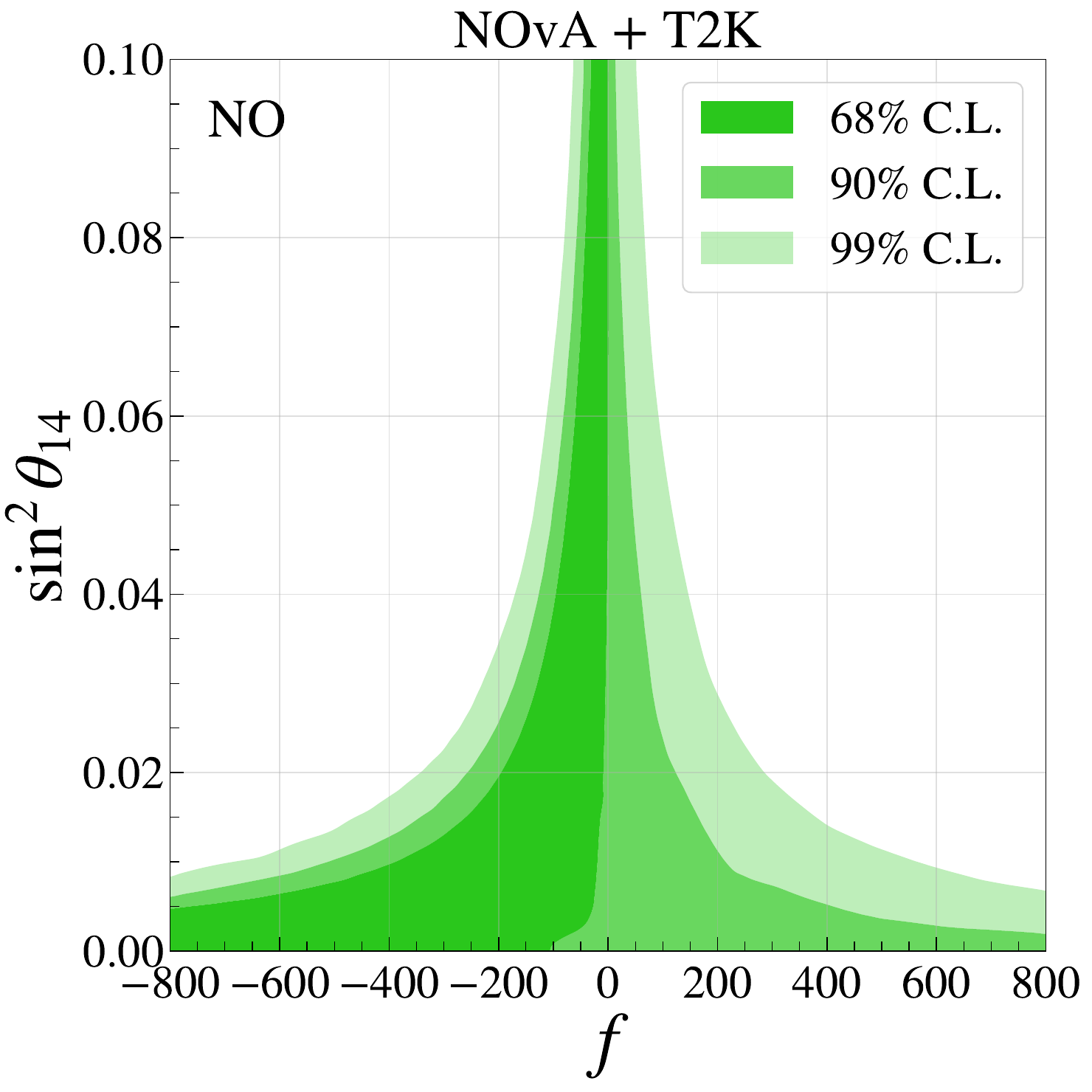}
\hspace*{0.0cm}
\includegraphics[height=8.80cm,width=8.80cm]{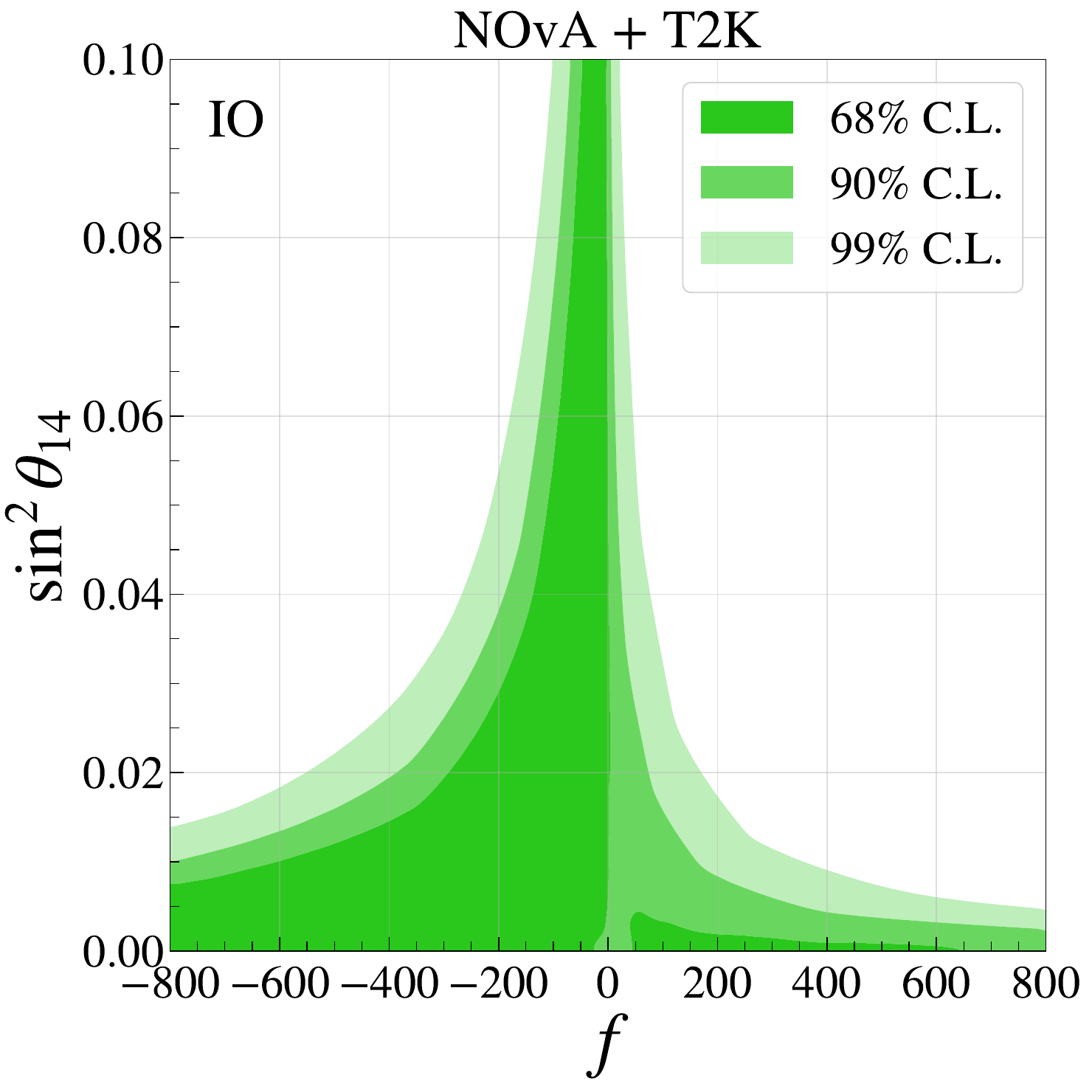}
\vspace*{-0.1cm}
\caption{
The left (right) panel represents the allowed regions for NO (IO)
at the 68\%, 90\%  and 99\% C.L. for 2 d.o.f. by the combination of NOvA and T2K  in the plane spanned by the two 
parameters $[f, s^2_{14}]$. We have marginalized over the CP phase $\delta_{14}$ and imposed
the prior $|f s^2_{24}| < 0.1$  at $90\%$ C.L. to take into account the bound on $|\varepsilon_{\mu\mu}|$
deriving from atmospheric neutrino experiments.}
\label{fig_LBL_f_s14}
\end{figure*}

Let's now consider the 3+1 scenario with the novel potential. Figure~\ref{fig_LBL_f_s14}  presents the 
constraints in the plane spanned by the two parameters $f$ and $s^2_{14}$ obtained by the combined analysis 
of NOvA and T2K, having marginalized away the new CP phase $\delta_{14}$. 
The left (right) panel refers to NO (IO).
The allowed regions are drawn at the 68\%, 90\% and 99\% C.L. in 2 d.o.f..
Concerning the 3-flavor parameters, we have fixed  
$\Delta m^2_{21} = 7.48\times 
10^{-5}$\,eV$^2$, $\sin^2 \theta_{12}= 0.308$, $\sin^2 \theta_{13}= 0.0223$, and marginalized over $\theta_{23}$ and $\delta_{13}$.
In these plots we have assumed $s^2_{34} = 0$, 
 thus automatically setting to zero the NSI-like couplings involving the $\tau$ neutrino 
 ($\varepsilon^\e_{e\tau} = \varepsilon^\e_{\mu\tau} = \varepsilon^\e_{\tau\tau}=0$).
In addition, we have imposed the prior $|\bar f_\e s^2_{24}| < 0.05$ (equivalent to $|f s^2_{24}| < 0.1$)
 at $90\%$ C.L. justified by
the stringent bound on the $\varepsilon_{\mu\mu}$-like copling deriving from atmospheric
neutrino experiments (see below). Under these working assumptions the 3+1 model
with the new potential becomes easily comparable with the NSI scheme analyzed  above, which
can be used to gain insight in the numerical results. 

From these two plots we learn that in 
NO (IO) the 3-flavor scheme (corresponding to $f=0$ and $s^2_{14}=0$) is disfavored
at more than the 90\% C.L (68\% C.L). Specifically we find $\Delta \chi^2 \simeq 4.3$ in NO
and $\Delta \chi^2 \simeq 2.5$ in IO. 
The allowed regions exhibit an inverted funnel-like structure,
which can be explained thanks to the NSI analogy, by observing that
NOvA and T2K are mostly sensitive to the coupling  $\varepsilon_{ee}^\e = \bar f_\e s^2_{14}$,
and therefore they can limit only the product of the two parameters $f$ and $s^2_{14}$.
The asymmetry of the allowed regions, which indicates a preference for negative 
values of $f$, is related to the fact that the combination of NOvA and T2K prefers
a negative value of the  $\varepsilon_{ee}^\e$. Also the preference for a negative 
real part of the complex NSI coupling $\varepsilon_{e\mu}^\e = \bar f_\e \tilde s_{14}s_{24}$
goes in the same direction. In summary, our combined analysis of NOvA and T2K
shows that the LBL data provide an indication in favor of the 3+1 scheme with the novel potential
over the 3-flavor scheme with a preference for negative values of the potential. In addition, the data tell us that the
indication is stronger in the NO case with respect to the IO.

\begin{figure*}[t!]
\vspace*{-0.4cm}
\hspace*{-0.1cm}
\includegraphics[height=8.80cm,width=8.80cm]{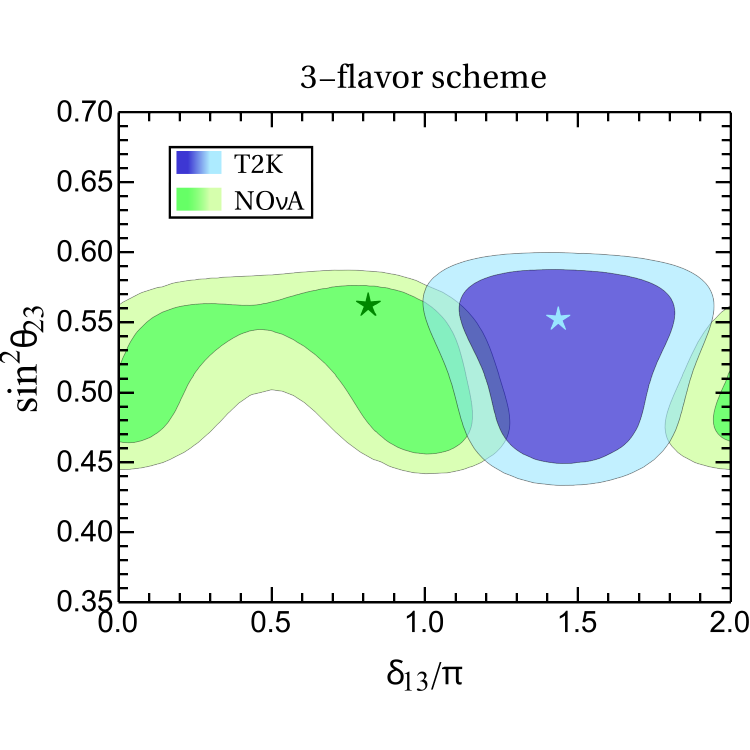}
\includegraphics[height=8.80cm,width=8.80cm]{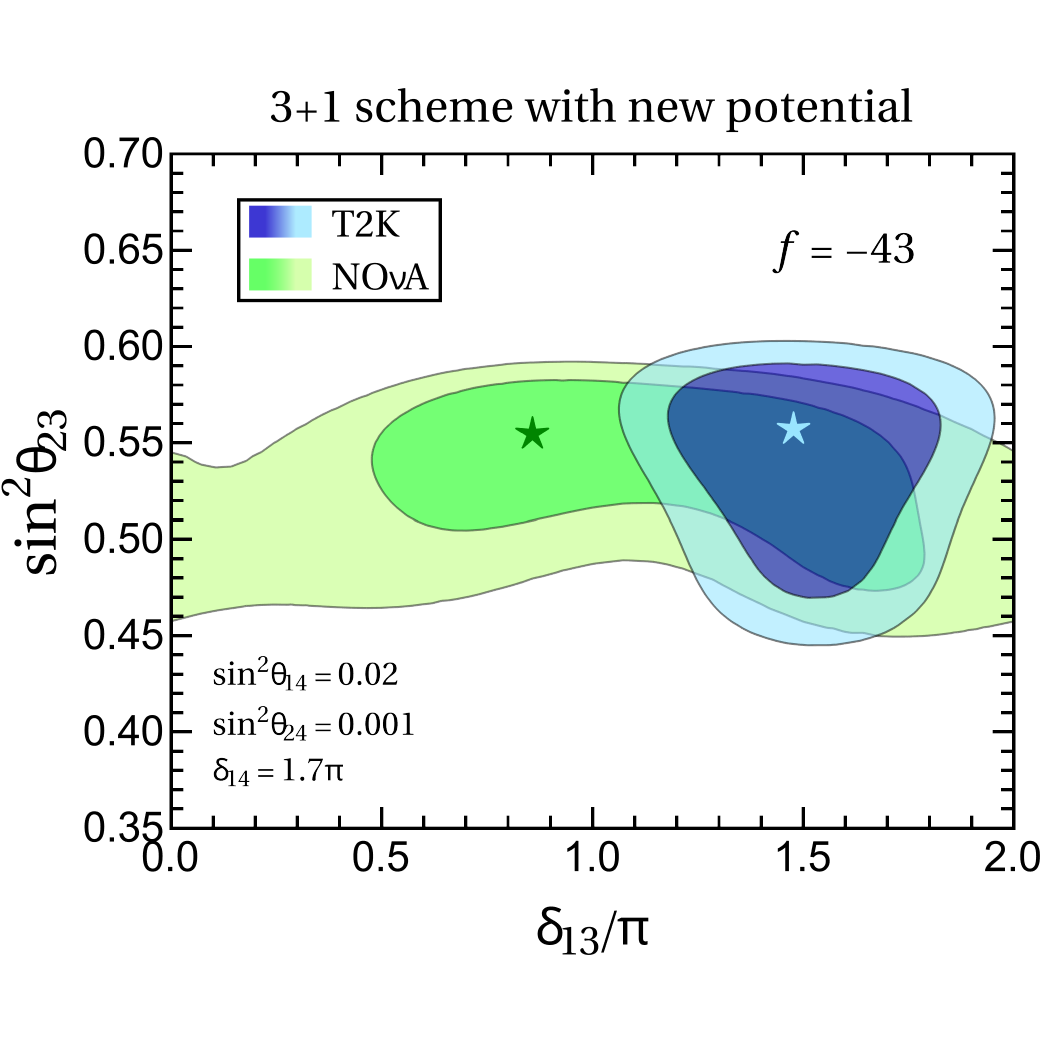}
\vspace*{-0.6cm}
\caption{Allowed regions by NOvA and T2K for normal ordering. The left  panel 
represents the 3-flavor framework, while the right panel refers to the 3+1 scheme with the new potential.
The contours correspond to the 68$\%$ and 90$\%$ C.L. for 2 d.o.f.}
\label{fig_tension_NO}
\end{figure*}

 In order to illustrate that such an improved fit is related to the capability of the model of resolving the tension 
 between the two experiments, we present in Figure~\ref{fig_tension_NO} the regions allowed for the
 two parameters  $\delta_{13}$ and $\sin^2\theta_{23}$ for the case of NO. The left plot represents the
standard 3-favor fit, where a clear tension emerges in the determination of the CP phase $\delta_{13}$.
 The right plot refers to the 3+1 scheme with enhanced potential. In this case we have fixed $s^2_{14} = 0.02$, 
 all the other parameters being constrained exactly as in Fig.~\ref{fig_LBL_f_s14}.
 For this value of $s^2_{14}$, we find the best fit values $\bar f_\e =-21.0$ (or equivalently $f = -43.0$)
 and $\delta_{14} = 1.7\pi$.
In the new 3+1 framework there is a high level of overlap 
of the allowed regions for values of $\delta_{13}$ 
close to $1.5\pi$, showing that the new model is very effective in reducing
the tension between the two experiments. 
We can observe that the T2K regions are basically unchanged 
as expected because of the low sensitivity to matter effects. Differently,
the NOvA regions appreciably change because of its pronounced sensitivity
to matter effects. These findings are in agreement with our analytical discussion and 
evidence the vacuum-matter synergy of the two experiments.

It is interesting and useful to compare the indication in favor of non-zero NSI 
coming from NOvA and T2K with the bounds/hints provided by the low-energy
atmospheric neutrino experiments. For this purpose, we list such bounds
in  Table~\ref{Tab:values_eps} also providing their combination performed
assuming gaussian (asymmetric where appropriate) uncertainties, which may serve
as a benchmark. From this table we can observe that only 
IceCube provides an interval for the diagonal coupling  $\varepsilon^\e_{ee} - \varepsilon^\e_{\mu\mu}$,
which interestingly weakly favors non-zero negative values like NOvA in T2K.
In IceCube the case $\varepsilon^\e_{ee} = -1$ (no matter effects) is disfavored
at $\Delta \chi^2 \simeq 7.0$, demonstrating an interesting sensitivity to this coupling.
Concerning the off-diagonal parameters $\varepsilon^\e_{e\mu}$  and $\varepsilon^\e_{e\tau}$,
we note that the atmospheric experiments provide a sensitivity for couplings of magnitude
$\sim 0.05$, and in some cases there are feeble hints in favor of a non-zero value.
On the third off-diagonal coupling  $\varepsilon^\e_{\mu\tau}$ the same experiments
provide the very stringent bound $\sim 3.3\times 10^{-3}$. 
Finally, for the diagonal coupling  $|\varepsilon^\e_{\mu\mu}-\varepsilon^\e_{\tau\tau}|$ 
there is an upper bound around 0.02. Such a bound justifies the prior 
$|f s^2_{24}| < 0.1$ used in the analysis of NOvA and T2K, 
since in the limit $\theta_{34}=0$ we have considered, one has  
$\varepsilon^\e_{\tau\tau} = \bar f_\e s^2_{34} = 0 $, and the equality $\varepsilon^\e_{\mu\mu}-\varepsilon^\e_{\tau\tau} = \varepsilon^\e_{\mu\mu}$
holds.
 
\section{Constraints from solar neutrino data}
\label{Sec:Solar}

Solar neutrinos have played a pivotal role in establishing the existence of  matter effects
as predicted by Wolfenstein~\cite{Wolfenstein:1977ue}. This is due to the high
electron number density in the central
region of the Sun, where neutrinos are produced in nuclear fusion reactions.  
In fact, solar neutrino data, with the solid anchor of the quasi-vacuum measurements performed with KamLAND,
where matter effects are much smaller,
provided the first irrefutable evidence for the coherent forward scattering of electron neutrinos mediated by charged currents~\cite{Fogli:2003vj}.
Therefore, one naturally expects that solar neutrinos should play an important role in constraining the scenario we are studying. 
For this reason we provide an analytical and numerical study of the solar neutrino conversion in the model under investigation. 

\begin{table}[t!]
\centering
\resizebox{.99\textwidth}{!}{\begin{minipage}{\textwidth}
\caption{\label{Tab:values_eps} 
Best estimates with 1 $\sigma$ error of the listed NSI couplings. The intervals/bounds refer to the 90\% C.L. for 1 d.o.f.}
\begin{ruledtabular}
\begin{tabular}{lccccc}
Experiment & $\varepsilon^\e_{ee}-\varepsilon^\e_{\mu\mu}$    &$\varepsilon^\e_{\tau\tau} -\varepsilon^\e_{\mu\mu}$ & 
  $|\varepsilon^\e_{e\mu}|$ &   $|\varepsilon^\e_{e\tau}|$  & $|\varepsilon^\e_{\mu\tau}|$     \\
\hline
$ \mathrm{ANTARES}$~\cite{ANTARES:2021crm} & $ - $ & $-0.032^{+0.015}_{-0.025}$ $\cup$ $0.032^{+0.035}_{-0.015}$ & -- &  --  & $<0.0045$  \\
\hline
$ \mathrm{IceCube}$~\cite{IceCubeCollaboration:2021euf} &  $[-2.26, -1.27] \cup [-0.74, 0.32]$ & $0.002\pm{0.025}$ &  $0.072^{+0.038}_{-0.069}$  & $0.060^{+0.08}{}$ & $< 0.0232$   \\
\hline
$ \mathrm{KM3NeT/ORCA}$~\cite{KM3NeT:2024pte} &   --  & $0.00\pm0.01$ &  $0.03^{+0.02}_{-0.03}$  & $0.04^{+0.02}_{-0.04}$ & $< 0.0055$  \\\hline
$ \mathrm{SuperKamiokande}$~\cite{Super-Kamiokande:2011dam} &  --  & $[-0.049, 0.049]$ &  --  &  --  & $<  0.011$  \\
\hline
$ \mathrm{Combination}$ &  $[-2.26, -1.27] \cup [-0.74, 0.32]$ & $[-0.018, 0.019]$ &  $0.038^{+0.018}_{-0.028}$  & $0.041^{+0.019}_{-0.040}$  & $<0.0033$\\
\end{tabular}
\end{ruledtabular}
\end{minipage}}
\end{table}

The treatment of the matter effects relevant for solar neutrinos 
in the ordinary 3+1 scheme has been performed in~\cite{Palazzo:2011rj}
(see also~\cite{Giunti:2009xz}), to which we refer the 
reader for more details (see the Appendixes). Here, we report the basic results with the
 appropriate modifications necessary to include the new matter potential.
 It is convenient to introduce the new basis
\begin{equation}
\bar \nu = A^\dag\nu
\,,
\label{eq:nu_bar}
\end{equation}
where we have defined the matrix $A$ as 
\begin{equation}
A =  \tilde R_{34}  R_{24} \tilde R_{14} R_{23} \tilde R_{13} \equiv U R_{12}^T \,.
\end{equation}
In this new basis the Hamiltonian assumes the form
\begin{equation}
\label{eq:Hrot2}
  \bar H = {\bar H}^{\mathrm{vac}} + \bar H^{\mathrm{mat}} = R_{12} K R_{12}^T + A^\dag  V A\,, 
\end{equation}
and, in the hierarchical limit $\Delta m^2_{21} \ll \Delta m^2_{31}  \ll \Delta m^2_{41}$,
the dynamics can be reduced to an effective 2-flavor one with Hamiltonian%
\begin{equation}
\label{eq:H_sol}
    \bar H_{2\nu} = \bar H^{\mathrm{vac}}_{2\nu}  + \bar H^{\mathrm{mat}}_{2\nu}  \,,    
\end{equation}
where the matter term, using the expressions of the elements of the matrix $A$, 
can be written in the form
\begin{equation}
\label{eq:H_dyn_rot}
    \bar H^{\mathrm{mat}}_{2\nu} = V_{CC}(x)
    \begin{pmatrix}
        \gamma^2 +  \bar f_\s \, |\alpha|^2 \  & \bar f_\s\, \alpha \beta\,  \\
        \bar f_\s\, (\alpha\beta)^*\,         & \bar f_\s \, |\beta|^2
    \end{pmatrix}
\,,    
\end{equation}
where the parameter $\bar f_\s = r_\s(x) (1+f)$ 
depends on  the radial position $x = R/R_{sun}$ ($R_{sun}$ being the Sun radius),
through $r_\s(x)$, which assumes the maximal value $r^{max}_\s \simeq 0.25$ at the center  of the Sun. 
The constant parameters ($\alpha, \beta, \gamma$) are given by%
\footnote{Note that in~\cite{Palazzo:2011rj} a different parametrization of the
mixing matrix was used and the parameters ($\alpha, \beta, \gamma$) do not depend 
on the mixing angle $\theta_{23}$. Here consistently we adopt the
parametrization introduced in Eq.~(\ref{eq:U}). Similar results were achieved in~\cite{Capozzi:2017auw}
considering a different scenario where sterile neutrinos are supposed to interact with
dark matter particles accreted by the Sun and confined in its central region.}
\begin{align}
\alpha
& = c_{13}c_{24}c_{34} \tilde s_{14} - \tilde s_{13}\left(c_{34}s_{23}s_{24} + c_{23}\tilde s_{34}\right),\, \\
\beta
& = c_{23}c_{34}s_{24}-s_{23}\tilde s_{34}^*\,,\\
\gamma
& = c_{14}c_{13}\,.
\end{align}
The matrix in Eq.~(\ref{eq:H_dyn_rot}) can be written, modulo an irrelevant diagonal factor, as a NSI-like Hamiltonian
\begin{eqnarray} \label{eq:Hdyn_3}
\arraycolsep=3pt
\medmuskip = 1mu
    \bar H^{\rm{mat}}_{2\nu}  \approx  
   V_{\rm{CC}} \!   \begin{bmatrix}
	\gamma^2 +\varepsilon_{ee}^\s  &   \varepsilon_{ex}^\s 
	\\
	\dagger & 0 
    \end{bmatrix} \,,
\end{eqnarray}
with
\begin{align} 
\label{eq:eps_ee}
\varepsilon_{ee}^\s \
& =  \bar f_\s (|\alpha|^2- |\beta|^2)\,, \\
\label{eq:eps_ex}
\varepsilon_{ex}^\s \
& =  \bar f_\s  \alpha \beta\,.
\end{align}
We see that in the case $f = 0$ one has $\bar f_\s = r_\s$ and 
the ordinary 3+1 scheme is recovered (see the appendix A in~\cite{Palazzo:2011rj}).
At the second order in the small values of $s^2_{i4}\, (i=1,2,3)$  we have the approximate expressions
\begin{align} 
\label{eq:eps_ee}
\varepsilon_{ee}^\s \
&  \simeq \bar f_\s s^2_{14}\,, \\
\label{eq:eps_ex}
\varepsilon_{ex}^\s \
& \simeq \bar f_\s  \tilde s_{14} (c_{23} s_{24}- s_{23}\tilde s_{34}^*)\,.
\end{align}
In this limit the expressions of the NSI-like couplings are similar to those found in the
treatment of LBL experiments in Eq.~(\ref{eq:Hdyn_2}), with $\varepsilon^\s_{ex}$ 
being a linear combination of $\varepsilon^\s_{e\mu}$ and  $\varepsilon^\s_{e\tau}$.
Note, however, that the size of the couplings is different because 
$\bar f_\s \ne \bar f_\e$. In particular in the central region of the Sun, which is relevant for
solar neutrinos,  we have $r_\s \simeq 0.25 \simeq 0.5 r_\e$ and $\bar f_\s \simeq 0.5\bar f_\e$.
Therefore, the amplitude of the solar NSI-like couplings is approximately 
one half of that acting in the Earth. The matrix $\bar H_{2\nu}$ in Eq.~(\ref{eq:H_sol}) can be exactly diagonalized by
 a $2\times2$ rotation $R_{12}^m$
which defines the mixing angle in matter $\theta_{12}^{m}$ and the full 4-flavor Hamiltonian 
in the flavor basis in Eq.~(\ref{Eq:H4nu}) is diagonalized by the matrix $U^m = AR_{12}^m$.
For adiabatic propagation of solar neutrinos the transition probabilities only depend upon their production and detection points.
Neglecting Earth-induced matter effects one has the general expressions for the conversion probabilities
\begin{equation}
\label{eq:Pee_adia}
  P_{e \alpha} = \sum_{i = 1}^4 |U_{\alpha i}|^2 |U^{m}_{ei}|^2 \,\,\,\,\,\,\, (\alpha = e,\mu,\tau, s)\,,
\end{equation}
where the $U_{ei}^m$'s denote the mixing matrix elements involving
the electron neutrino  in the production point. 
One can find  that the mixing angle $\theta_{12}^m$ is given by
\begin{align}
\cos2\theta_{12}^m
& = \frac{\cos2\theta_{12}- v(\gamma^2 + \varepsilon_{ee}^\s)}{\sqrt{|\sin2\theta_{12}+ 2v \varepsilon_{ex}^\s |^2
+ [\cos2\theta_{12}- v(\gamma^2 + \varepsilon_{ee}^\s)] ^2}}\,,
\end{align}
where $v = V_{CC}/k_{21} = 2 V_{CC}E/\Delta m^2_{21}$.
The 4-flavor survival probability can be expressed as
\begin{align}
\label{eq:Pee_4nu_13_14}
P_{ee} & =  c^4_{14}c^4_{13} \bar P_{ee}^{2\nu}  + c^4_{14} s^4_{13} + s^4_{14}\,, 
\end{align}
where 
\begin{align}
\label{eq:Pee_2nu_rot}
\bar P_{ee}^{2\nu} & =  \frac{1}{2}\left(1+\cos2\theta_{12}\cos2\theta_{12}^{m}\right)\, 
\end{align}
is the well known expression of the 2-flavor survival probability in the adiabatic regime~\cite{Kuo:1989qe}.
The expression of $P_{es}$ is less straightforward. In the limit of small mixing angles it depends
on second order combinations $s_{i4} s_{j4}$ $(i,j =1,2,3)$ and presents a 
dependency on the CP-phases $\delta_{14}$ and $\delta_{34}$.
In the analytical treatment above we have assumed that the propagation of neutrinos is adiabatic.
We have checked that this approximation is valid exploiting the analytical criterion described in~\cite{Palazzo:2011rj}.%
 \footnote{Violation from adiabaticity can occur if along the neutrino trajectory inside the Sun there is a point
 $x = R/R_\s$ where the difference between the two energy eigenstates in matter approaches zero. As shown 
in~\cite{Palazzo:2011rj}, this occurs if the two conditions: i) $v(x) (\gamma^2 + \varepsilon_{ee}^\s) = \cos 2\theta_{12}$
 and  ii) $|2v(x) \varepsilon_{ex}^\s + \sin 2 \theta_{12}| = 0$ hold simultaneously. The second condition can be verified 
 only if $\varepsilon_{ex}^\s$ is real and negative (i.e. $f<0$), and if $|\varepsilon_{ex}^\s|$ is very large. In the limit case $\theta_{34}=0$ 
 we will consider in the analysis, one has $|\varepsilon_{ex}^\s \simeq f_\s c_{23} \tilde s_{14} s_{24}|$. In addition, justified 
 by the  atmospheric neutrino data we will consider only small values of the product $f_\s s^2_{24}$ and this ensures 
 that the off-diagonal coupling $|\varepsilon_{ex}^\s|$ is always small enough to guaranty adiabaticity in the allowed regions.} 
 The Earth-regeneration effects
 are expected to modify the picture above and have been included as described in~\cite{Palazzo:2011rj}.

\begin{figure*}[t!]
\vspace*{-1.5cm}
\hspace*{0.1cm}
\includegraphics[height=10.80cm,width=8.80cm]{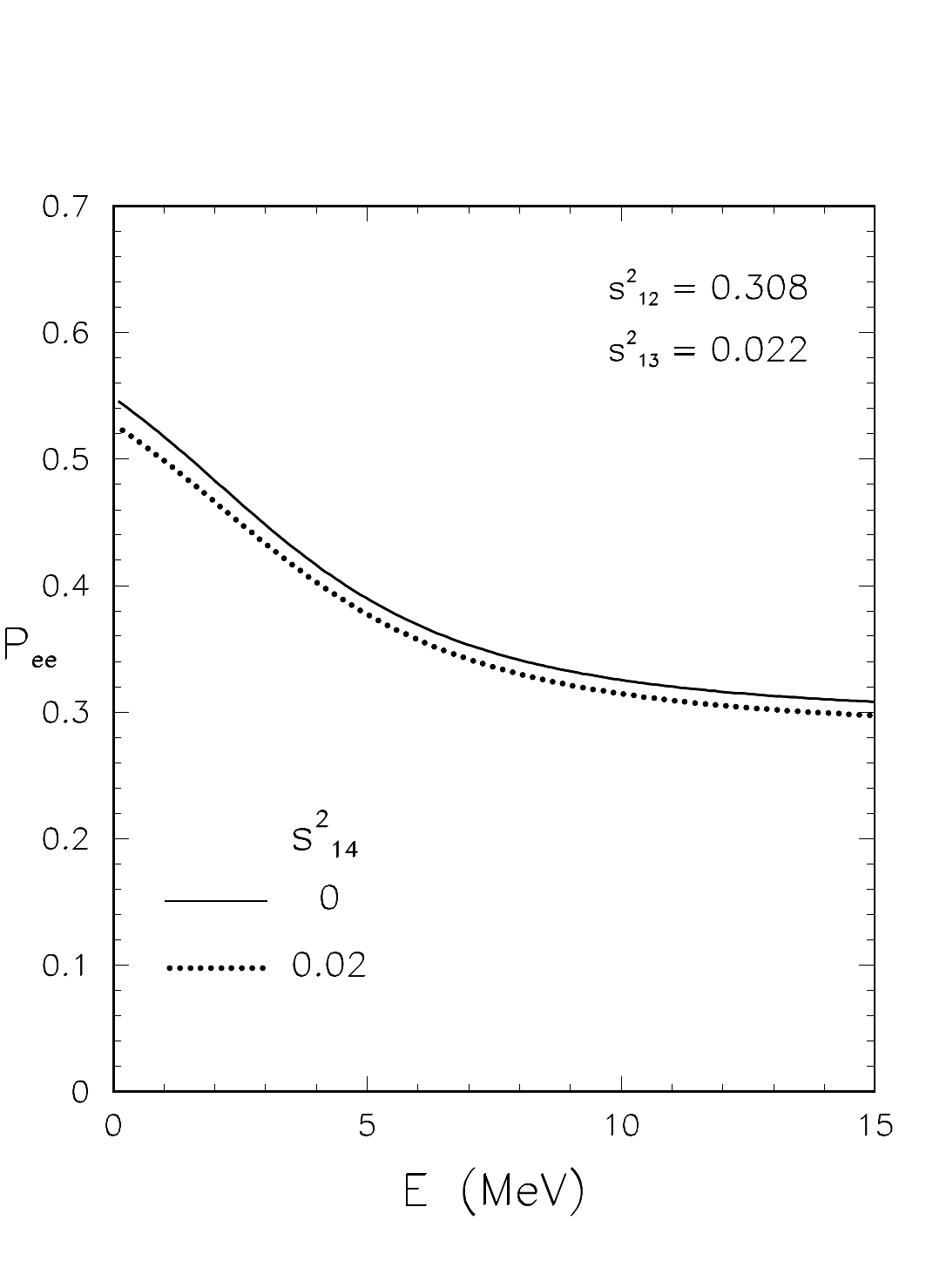}
\includegraphics[height=10.80cm,width=8.80cm]{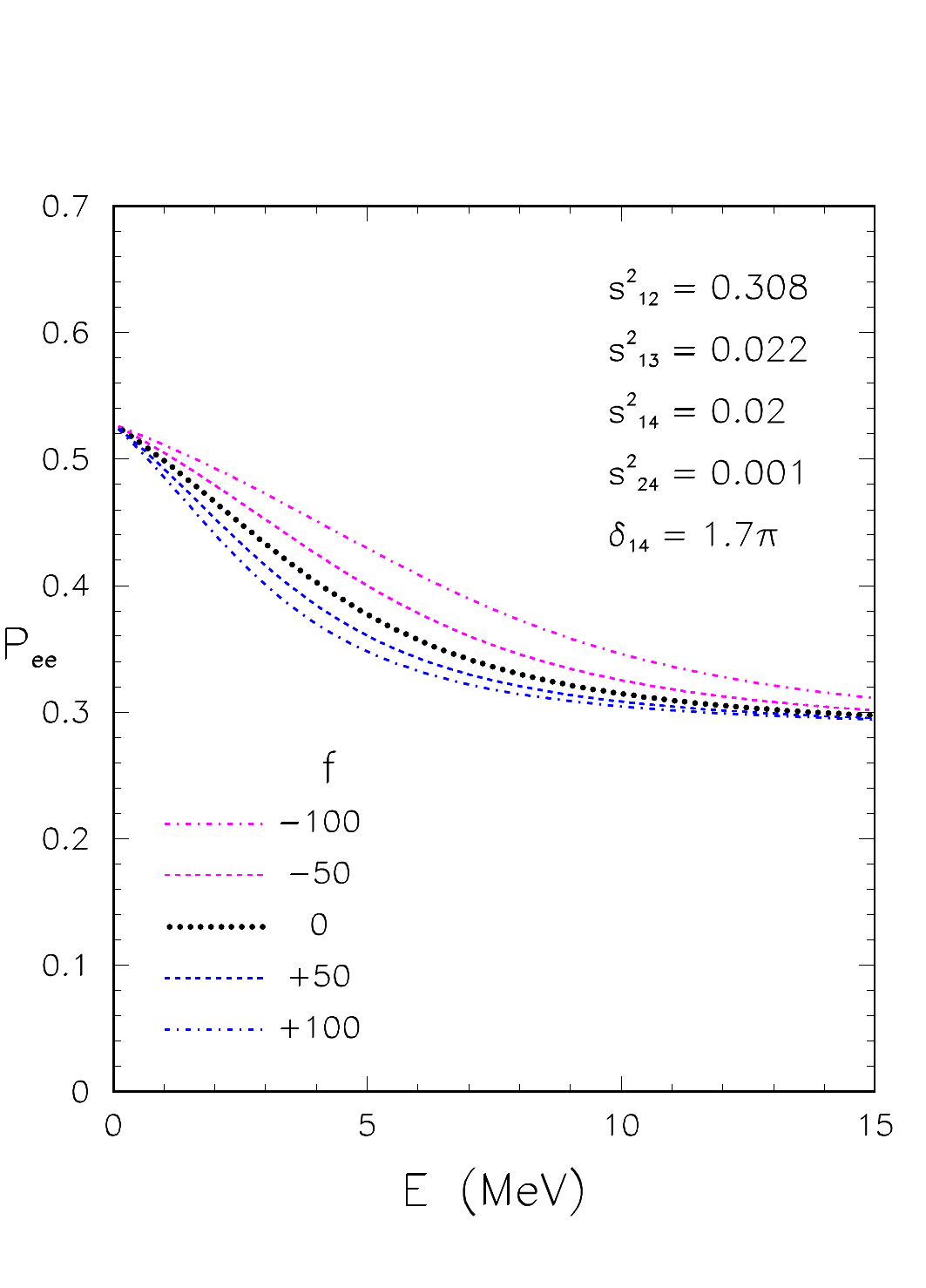}
\vspace*{-0.9cm}
\caption{Daytime survival probability of solar neutrinos averaged over the $^8$B neutrinos production region.
The left panel shows the impact of the non-zero value of $s^2_{14} = 0.02$ with respect to the 3-flavor case
in the ordinary 3+1 scheme. The right panel confronts the ordinary 3+1 scheme with the 3+1 scenario with the new
potential for four choices of the parameter $f$. In both panels we fix $\Delta m^2_{21} = 7.48\times 10^{-5}$\,eV$^2$
 as determined by KamLAND and JUNO.}
\label{fig_sol_prob}
\end{figure*}

In the left panel of Fig.~\ref{fig_sol_prob}, we show the probability for the 3-flavor and
the ordinary 3+1 framework. The solid curve represents the best fit 3-flavor scheme, 
while the dotted curve refers to the ordinary 3+1 scheme with $s^2_{14}= 0.02$.
In this scheme, the dynamical couplings in the Hamiltonian Eq.~(\ref{eq:H_dyn_rot}) 
are negligible, and the other sterile mixing angles are irrelevant. The only effect of the 
sterile neutrinos is the kinematical one, and consists in an energy independent suppression induced
by the non-zero value of $s^2_{14}$, 
which can be quantified as $c^4_{14} \simeq 1-2 s^2_{14}$.
It is clear that if one increases $s^2_{14}$ the suppression becomes too large to be tolerated by the data.
In fact, it is precisely this behavior that allows one to set an upper bound on the admixture of the electron neutrino
with the fourth mass eigenstate, as first deduced in~\cite{Palazzo:2011rj,Palazzo:2012yf}.
The existing analyses~\cite{Palazzo:2012yf,Goldhagen:2021kxe,Gonzalez-Garcia:2024hmf}
set a $90\%$ C.L. upper limit on $s^2_{14}$ lying in the interval $[0.02-0.04]$. The bound we
obtain in the 3+1 scheme (see below the case $f=0$) is in agreement with that 
obtained by one of us in the analysis performed in~\cite{Palazzo:2011rj,Palazzo:2012yf},
where a very similar data set was used.

In the right panel of Fig.~\ref{fig_sol_prob}, we show a few representative profiles of the electron
neutrino survival probability in the 3+1 scheme with the new potential. We assume
interaction with neutrons. All curves refers to the same value of the standard parameters 
$\Delta m^2_{21} = 7.48\times 10^{-5}$\,eV$^2$,
$\sin^2 \theta_{12}= 0.308$ and $\sin^2 \theta_{13}= 0.0223$.
The different curves represent the probability for different values of $f$ as listed
in the legend. The black profile corresponds to the case $f=0$, coinciding with
the ordinary 3+1 scheme. The colored profiles refer to four cases in which 
$f\ne 0$ and are chosen  to be relatively large in order to magnify the visibility of the effect.
The parameters related to the sterile sector are fixed
 at the values $s^2_{14} = 0.02$, $s^2_{24} = 10^{-3}$ and $\delta_{14} = 1.7\pi$.
We see that in the four cases in which $f\ne0$ 
there are energy dependent modifications of the
probability with respect to the normal 3+1 scheme. 
For positive (negative) values of $f$ the survival probability in the presence 
of the new potential is lower (higher) compared with the case $f = 0$ and this
occurs at all energies. Also, we note that the modifications of the profile of the 
probability are more (less) pronounced for negative (positive) values of $f$.

In order to better evaluate the impact of the new sterile matter potential scenario
in solar neutrino oscillations, we have performed a numerical analysis of solar neutrino
data including  the radiochemical experiments Homestake for Chlorine~\cite{Cleveland:1998nv}
and GALLEX+GNO and SAGE for Gallium~\cite{SAGE:2009eeu,Kaether:2010ag}, 
the three-phase data~\cite{SNO:2011hxd} from the Sudbury Neutrino Observatory (SNO), 
the Super-Kamiokande data from 1496-day phase I (energy and zenith spectrum)~\cite{Super-Kamiokande:2023jbt}
and 2970-day phase IV (day and night energy spectrum)~\cite{Super-Kamiokande:2023jbt}, and 
the low energy Borexino data~\cite{Bellini:2011rx,BOREXINO:2014pcl,Borexino:2017rsf}.
We adopt the recent solar model denoted MB22m in~\cite{SSM2023}.
We have included the first JUNO results imposing a prior on the solar mixing angle using
the estimate determined in the official analysis of the JUNO Collaboration~\cite{JUNO:2025gmd}. 
To this purpose, we note
that JUNO, probing vacuum oscillations, cannot distinguish
between the two octants of the solar mixing angle $\theta_{12}$, and one obtains a mirror
(dubbed ``dark'') solution~\cite{Miranda:2004nb} with the replacement $\theta_{12} \to \pi/2 - \theta_{12}$.  
While this is irrelevant within the standard 3-flavor framework, where solar neutrinos fix 
$\theta_{12}$ in the first octant thanks to matter effects, it becomes relevant
in our new scenario in which solar neutrinos in general cannot fix the $\theta_{12}$ octant
because the model entails the possibility of large NSI-like couplings, in particular $\varepsilon_{ee}^\s = -2$. 
In this case, due to a well known degeneracy theorem the octant of $\theta_{12}$ cannot be
distinguished. We will see that this conclusion holds only in the case of a potential proportional
to the electron or proton number density. In the case of a potential proportional to the neutron
number density things are different  because the NSI couplings in the Sun and in the Earth
scale differently. In such a case the inclusion of Earth matter effects breaks the degeneracy
distinguishing the ordinary LMA solution from the dark-LMA solution, strongly disfavoring the
second one.
For completeness and illustrative purposes we consider both cases in which the potential is proportional to
the electron/proton and to neutron number density. In the first case we define the new potential
 as $V_S = f^\prime\, V_{CC}$ introducing the parameter $f^\prime$ used only in the
 present section. In the second case we define $V_S = -f\, V_{NC}$ as in the rest of the paper.
 
\begin{figure*}[t!]
\vspace*{-4.45cm}
\hspace*{-1.5cm}
\includegraphics[height=25.60cm,width=21.0cm]
{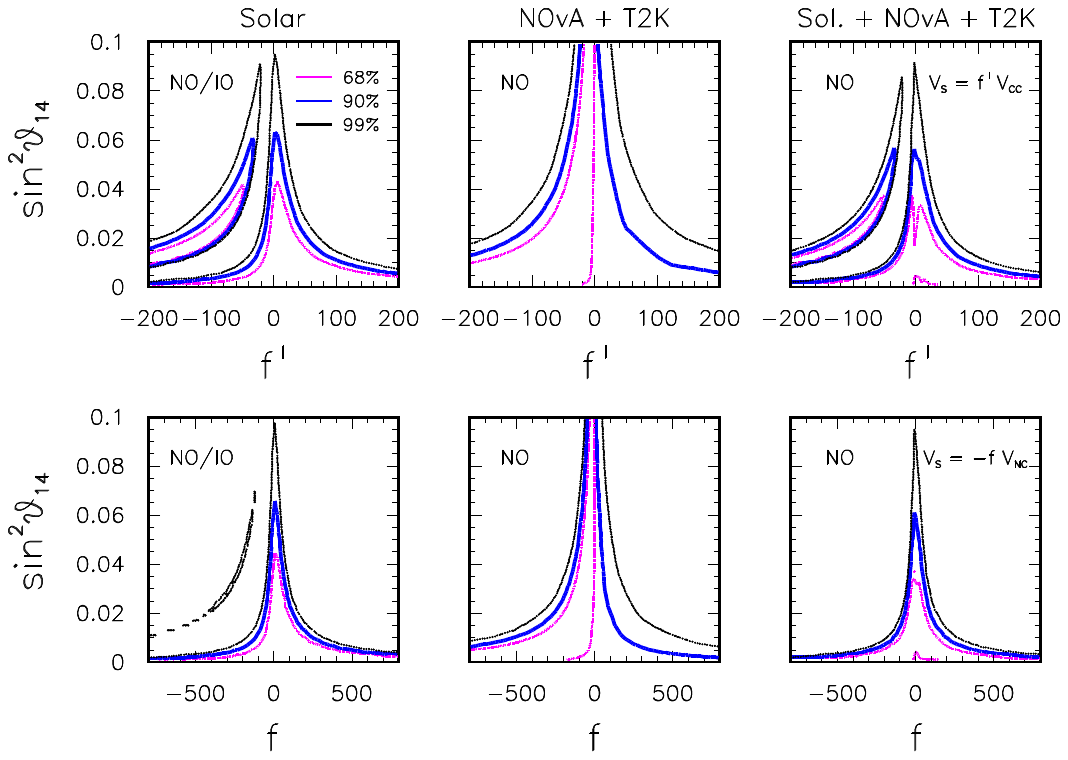}
\vspace*{-9.9cm}
\caption{Allowed regions of the two displayed parameters assuming interaction with electron/protons $V_s = f^\prime\, V_{CC}$ 
(upper panels) and with neutrons $V_s = - f\, V_{NC}$ (lower panels),  as
obtained from the solar neutrinos (left panels), LBL experiments NOvA and T2K (central panels) 
and their combination (right panels). The allowed regions in the left panels are 
independent of the NMO, while those represented in the central and right panels refer to NO.
The contours are obtained fixing $s^2_{34} = 0$, imposing the prior $f s^2_{24} < 0.1$,
and marginalizing over  $\delta_{14}$. Concerning the 3-flavor parameters we have fixed  
$\Delta m^2_{21} = 7.48\times 10^{-5}$\,eV$^2$, $\sin^2 \theta_{13}= 0.0223$,
and marginalized over $\theta_{12}$,  $\theta_{23}$ and $\delta_{13}$.}
\label{fig_6panel_horizontal_comb}
\end{figure*} 

Fig.~\ref{fig_6panel_horizontal_comb}  displays in the three upper (lower) panels 
the regions allowed  in the plane spanned by the two parameters $f^\prime$ ($f$) and $\theta_{14}$. 
The left panels represent the regions allowed by solar neutrinos which are identical for the 
two neutrino mass orderings (NO/IO). The central panels represent the regions allowed 
by the two LBL experiments NOvA and T2K, while the right panels show the combination
of solar with LBL experiments. The LBL experiments are sensitive to the NMO and 
we have adopted the NO choice for all the panels.
In order to determine the solar neutrino regions in the left panels, we have applied exactly the same procedure
adopted for NOvA and T2K by assuming $s^2_{34} = 0$
and imposing the prior $|f^\prime s^2_{24}| < 0.1$ at $90\%$ C.L., that as we stress again, is justified by
the stringent bound on the $\varepsilon_{\mu\mu}$-like coupling from atmospheric
neutrino experiments (see Sec.\ref{Sec:LBL_ATM}). Focusing on the left-upper panel
referring to a potential proportional to the electron/proton number density, we see that the allowed 
regions are composed by two separate branches. The right branch, approximately centered on $f^\prime=0$, 
corresponds to (marginalized values of)  $s^2_{12} < 0.5$ (lying in the ordinary LMA solution), while the left branch
is obtained for $s^2_{12} > 0.5$ (lying in the so-called dark-LMA solution~\cite{Miranda:2004nb}). 
The borders of the allowed regions follow approximately iso-contours of constant $\varepsilon_{ee}^\s =  f^\prime s^2_{14}$.
By inspection of the contours, we derive that this coupling is limited at the 90\% C.L. 
in the two intervals $[-0.3, 0.8]$ and $[-2.7, 1.7]$. These ranges are in agreement with those derived
in NSI analyses (see~\cite{Coloma:2023ixt}).   The dark branch is approximately centered 
on $\varepsilon^\s_{ee} =  f^\prime s^2_{14} \simeq -2$, as expected.
The $\chi^2$ minima of the two branches are basically degenerate, as expected on the basis of the so-called generalized
mass-ordering degeneracy~\cite{Coloma:2016gei}. 
The value $\varepsilon_{ee}^\s = -1$ is excluded at a very high confidence level (more than 10 standard 
deviations), reflecting the strong rejection of solar neutrino data of a mere vacuum-like flavor conversion,
in agreement with the early study~\cite{Fogli:2003vj}. We observe that
solar neutrinos provide an upper bound on $s^2_{14}$, which is allowed to reach the maximal 
value for $f^\prime \simeq0$ or along the iso-contour corresponding to  $\varepsilon_{ee}^\s \simeq -2$. 
It is now very interesting to investigate what is the result of the combined analysis of the solar neutrino sector
data with the two LBL experiments NOvA and T2K. 
The upper-central panel reports the 
regions allowed by NOvA and T2K for the NO case. Note that these regions are different with respect
to those obtained for a potential proportional to $V_{NC}$ (shown in the lower-central panel),
because in the Earth $V_{CC}$ and $V_{NC}$ used in the definition of the new potential differ by a factor of 
two (being $|V_{CC}| = 2 |V_{NC}|$). The contours for the two kinds of potential are
related by a simple rescaling of a factor of two. In any case the standard
3-flavor scheme corresponding to the origin of the axes ($f^\prime = 0,\,s^2_{14}=0$) is disfavored
at $\Delta \chi^2 \simeq 4.3$. 
In the right upper-panel, we see that in the combined fit there are still two degenerate branches
as expected. Also, we see that a slight preference for a non-zero value of $f^\prime$
persists around the 90\% C.L. with $\Delta \chi^2 \simeq 2.7$. 

Let's now consider the three lower panels in Fig.~\ref{fig_6panel_horizontal_comb},
corresponding to a new potential proportional to the neutron number density.
In the left-lower panel we see that only the ordinary LMA branch survives, 
while the dark-LMA branch is excluded almost at the 99\% C.L. The degeneracy is 
broken due to the fact that in the Sun the effective coupling with neutrons of the new potential 
 is different with respect to the Earth, being 
approximately two times smaller.
In particular, the dark branch approximately centered 
on $\varepsilon^\s_{ee} = \bar f_\s s^2_{14} \simeq -2$, corresponds to the larger coupling in the Earth
 $\varepsilon^\e_{ee} = \bar f_\e s^2_{14} \simeq -4$,
which leads to an enhanced regeneration effects in the Earth and a consequent higher day-night asymmetry
in tension with data at approximately $\Delta \chi^2 \simeq 4$. We have checked that JUNO constraints indirectly 
contribute to further disfavor the dark-LMA by $\Delta \chi^2\simeq 4$ by rendering the solar neutrino data 
less flexible in tolerating the different NSI-like couplings in the Sun and the Earth. The lower-central panel
reports the regions allowed by NOvA and T2K (already shown in the left panel of  Fig.~\ref{fig_LBL_f_s14}).
The lower-right panel displays the combination of LBL with solar neutrino data. We can observe that the
LBL data further disfavor the dark-LMA solution, which is excluded at more than 99\% C.L. This can
be understood, as already observed above, because the NSI-like couplings in the Earth probed by LBL neutrinos 
are a factor of two larger than those probed by solar neutrinos in the Sun. The 3-flavor case,
coinciding with the origin of the axes is disfavored at more than the 90\% C.L. with $\Delta \chi^2 \simeq 2.7$.

It is very interesting to observe that the size of the Earth matter effects in the LBL experiments
NOvA and T2K is similar to that experienced by solar neutrinos traversing the Earth.
This can be seen by looking at the dimensionless parameter  $v= 2 V_{CC}/k = 2 V_{CC} E/\Delta m^2$,
where $E$ is the neutrino energy ($E\simeq $ 10 MeV for solar and $E\simeq $ 1 GeV for LBL neutrinos) 
and $\Delta m^2$ is the relevant mass-squared difference ($\Delta m^2_{21}$ for solar and  $\Delta m^2_{31}$
for LBL neutrinos). Assuming for solar neutrinos propagating in the Earth its average density ($\rho_\mathrm{ave} = 5.5\, \mathrm{g/cm^3}$)
and for LBL neutrinos the density of the crust  ($\rho_\mathrm{crust} = 2.7\, \mathrm{g/cm^3}$), one has
$v_\mathrm{sol}/v_\mathrm{LBL} = 0.6\, (E/10\, \mathrm{MeV})(E/\mathrm{GeV})$.
Hence, it is not surprising to find that NOvA and T2K disfavor the LMA-dark solution
similarly to the day-night effects experienced by solar neutrinos in the case of a potential proportional to the neutron number density.
While Fig.~\ref{fig_6panel_horizontal_comb} refers to the case of NO,
we find similar results in IO, where the dark-LMA branch is still disfavored at more than
the 99\% C.L. in the combination of solar and LBL neutrinos. 
In conclusion, we observe that in the combination of solar with LBL data a slight
indication at $90\%$ C.L $(\Delta \chi^2 \simeq 2.7)$ persists
in favor of non-zero $f$  (or $f^\prime$).
\footnote{We find that the preferred value of $s^2_{14}$ is very sensitive
to the gallium cross section. By increasing or decreasing by 5\% such a cross section we obtain a
smaller/larger best fit. In literature there are several evaluations of the cross section
driven by the Gallium anomaly. Therefore, the exact best fit value of $s^2_{14}$ is quite uncertain.}
We can observe that by selecting a value of $s^2_{14}\simeq 0.01-0.03$ of some relevance for the short-baseline anomalies,  
the combination of solar with LBL data is compatible with values of $f$ in the range $[-60, 120]$
($f^\prime$ in the range $[-30, 60]$). These results will be of guide in the next section 
in identifying a benchmark set of parameters towards which to address our attention.

\section{Identification of the benchmark parameters}
\label{Sec:LBL_ATM}

 Now, on the basis of all the indication/bounds at our disposal and of the insight gained in 
 the structure of the model, we try to
 acquire further information on its parameters.
In the previous section we have seen that NOvA and T2K, together with IceCube indicate a 
preference for a non-zero negative value of  $\varepsilon^\e_{ee}$ and a non-zero 
 $|\varepsilon^\e_{e\mu}|$.
In addition, LBL, IceCube DeepCore and KM3NeT/ORCA tend to weakly
prefer a non-zero value of  the complex coupling $\varepsilon^\e_{e\mu}$.
 We now observe, that in our model with the extra matter potential, in order to
 construct a non-zero value of these two NSI-like couplings
 one needs a non-zero value of  all the three parameters  $f$, $s_{14}$ and $s_{24}$.
Then, by construction the other diagonal coupling  $\varepsilon^\e_{\mu\mu}$ 
is automatically non-zero as well. 
The preference of a negative $\varepsilon^\e_{ee}$ 
translates into a preference for $f<0$ with respect to $f>0$. 
Concerning the $\mu-\mu$ diagonal coupling, the limits are much stringent.
In fact, the atmospheric data provide for the combination ($\varepsilon^\e_{\mu\mu} -\varepsilon^\e_{\tau\tau}$), 
the 90\% range [-0.022, 0.019] (see Table~\ref{Tab:values_eps}).
Therefore, the  amplitude of the
NSI-like coupling $|\varepsilon^\e_{\mu\mu}|  = |\bar f_\e |s^2_{24}$ must be at least one order of magnitude smaller%
\footnote{In principle the bounds on the difference $\varepsilon^\e_{\mu\mu} -\varepsilon^\e_{\tau\tau}$ would allow
big values for the single couplings which cancel out in the linear combination. However, we will see 
that the IceCube resonance will require very small values of both mixing angles $s^2_{24}\sim s^2_{34}\ll 10^{-2}$.
Therefore, this possibility is excluded.}
 than that of $|\varepsilon^\e_{ee}| = |\bar f_\e| s^2_{14}$. This leads us to consider  $s^2_{24} \ll s^2_{14}$,
with  $s^2_{24}$ being one order of magnitude smaller than  $s^2_{14}$.
 Taking into account all the existing constraints we are led to the following benchmark values for the NSI-like couplings
entering the Hamiltonian, which are compatible at $1\sigma$ level with the existing combined bounds listed in the
last row of Table~\ref{Tab:values_eps} and with the solar neutrino constraints,  respecting at the same time
 the relations among the three parameters
($f, s_{14}, s_{24}$) imposed by our model
\begin{flalign} 
\label{eq:eps_equivalent_2}
 & \varepsilon_{ee}^\e \simeq  \bar f_\e   s^2_{14} = - 0.2\,,\\
\label{eq:eps_equivalent_1}
& \varepsilon_{e\mu}^\e \simeq  \bar f_\e  s_{14} s_{24} e^{-i\delta_{14}}= -0.045 \,, \\
 \label{eq:eps_equivalent_3}
& \varepsilon_{\mu\mu}^\e \simeq  \bar f_\e  s^2_{24} =  -0.01\,.
\end{flalign}
We have fixed the CP-phase $\delta_{14}=0$, which is allowed within the 68\% C.L. in the NOvA and T2K analysis.
\footnote{This choice will simplify the numerical simulations performed in the following sections
concerning the IceCube resonance at $\sim$ 10 TeV. We will underline the situations
in which the CP-phase $\delta_{14}$ can play an important role as in the case of the 1-3 resonance 
occurring in multi-GeV atmospheric neutrinos measured by Super-Kamiokande (see Sec.~\ref{Sec:Res_6GeV}).}
We stress once again that, as it is clear from the Eqs.~(\ref{eq:eps_equivalent_1}-\ref{eq:eps_equivalent_3}),
the LBL neutrinos and low-energy atmospheric data 
give information only on the product of $\bar f_\e$ with $s^2_{i4}$  or $s_{i4} s_{j4}$
with $(i,j) = 1,2$. Therefore, although they allow us to identify the hierarchical pattern
 $s^2_{14}\gg s^2_{24}$, in order to fix the values of the mixing angles, 
 one needs to know at least one of them or the value of  $f$.
 Since $f$, based on the data considered until now is completely unknown, we proceed by fixing 
 one of the two mixing angles, in particular $s^2_{14}$.
Quite interestingly, we will see that the IceCube resonance at $E \simeq 10$ TeV, in conjunction with the limits 
 determined by KATRIN, will provide us a preferred interval for $f$ compatible with that determined
 below.
 
The choice of fixing $\theta_{14}$ is motivated by the strong evidences of disappearance of electron neutrinos 
provided by the Reactor and Gallium anomalies as well as Neutrino-4, which naturally 
lead to assume  a non-negligible value of $s^2_{14}$.  However, in doing this choice 
we must take into account all the available upper bounds on 
this parameter, and in particular the strongest ones, which come from KATRIN and from solar neutrino data. 
Based on the numerical analysis of solar neutrinos presented in Sec.~\ref{Sec:Solar}
and on the constraints from KATRIN~\cite{KATRIN:2025lph},
we take $s^2_{14} = 0.02$ as a benchmark value, which is compatible at 90\% C.L. 
with $f$ in the range [-50, 50] according to the combined analysis of LBL and solar neutrino data. 
 With this choice Eqs.~(\ref{eq:eps_equivalent_2})
provides a benchmark value of $f = -20$ (corresponding to $\bar f_\e \simeq -10$). 
We choose a negative value of $f$ since NOvA and T2K show a clear preference
for such values. In the next section, we will see that also Super-Kamiokande atmospheric
neutrino data point towards a negative potential.
We also note that, once the benchmark values for  $s^2_{14}$ and $f$ have been
chosen, from Eqs.~(\ref{eq:eps_equivalent_2})-(\ref{eq:eps_equivalent_3}),
a benchmark value of the second sterile mixing angle is fixed to be $s^2_{24} \simeq 10^{-3}$
being the ratio $s^2_{14}/s^2_{24} \simeq 20$. 
We underline that, even after fixing $s^2_{14} = 0.02$, the uncertainty on the estimates of the NSI couplings
will translate into an uncertainty on $s^2_{24}$. 
The estimate of $\varepsilon^\e_{\mu\mu}$ in the range $[-0.018, 0.019]$ makes
it compatible with zero. So, the actual value of  $s^2_{24}$,
albeit necessarily non-zero by construction due to the structure of the model, may be
substantially smaller than the benchmark value  $s^2_{24} \simeq 10^{-3}$. 
It is interesting to note that the indicated very small values of $s^2_{24}$
eliminate altogether the tension even with the most stringent bounds deriving from the negative 
searches of muon neutrino disappearance performed by MINOS/MINOS+~\cite{MINOS:2017cae}
 and NOvA~\cite{NOvA:2024imi}, which put a 90\% upper limit $s^2_{24} \lesssim 0.01$
 for eV$^2$ mass-squared differences.%
 \footnote{We have explicitly checked that in MINOS and NOvA the muon neutrino survival probability $P_{\mu\mu}$ 
 in the presence of the new matter potential is basically indistinguishable from the 3-flavor case.
 This can be understood because the NSI-like couplings $\varepsilon^\e_{\mu\mu}$
and $\varepsilon^\e_{\mu\tau}$, which have more impact on such a channel,
are negligibly small in our scenario. Also, we have verified that the transition probability $P_{\mu s}$
of muon neutrinos into sterile neutrinos is very small being proportional to $s^2_{24} = 0.001$. Therefore,
the scenario is unobservable in the present dual-baseline searches performed 
using $\nu_\mu$ charged-current and neutral-currents as in MINOS/MINOS+ and NOvA.}
  
For the  $\varepsilon^\e_{e\tau}$ coupling, we can reason similarly to $\varepsilon^\e_{e\mu}$.
The indication for a non-zero $\varepsilon^\e_{e\tau}$ from NOvA and T2K leads
to consider a non-zero $s_{34}$. This automatically switches on the other two
couplings $\varepsilon^\e_{\mu\tau} = \bar f_\e \tilde s_{24}s_{34}^* $ and  $\varepsilon^\e_{\tau\tau} = \bar f_\e s^2_{34}$.
The limits on the combination $\varepsilon^\e_{\mu\mu} -\varepsilon^\e_{\tau\tau}$ lead
us to assume a non-zero very small value for $s^2_{34}$ similarly to $s^2_{24}$.
 The low-energy atmospheric data put a very stringent bound on this NSI coupling%
 \footnote{We point out that there exists an independent bound $|\varepsilon^\e_{\mu\tau}|\lesssim 4\times 10^{-3}$ 
 derived from IceCube~\cite{IceCube:2022ubv}, which interestingly indicates a $\sim 90\%$ C.L. preference for the
 non-zero coupling $|\varepsilon^\e_{\mu\tau}| \simeq 2.9\times 10^{-3}$, which would imply an estimate  
 of a non-zero $s^2_{34}$.
 However, this indication is derived
 by using a neutrino sample with very high energies, included those involved in the resonant-like behavior, 
 a region where our NSI-like treatment does not hold. For this reason we do not consider this indication.}
 $|\varepsilon^\e_{\mu\tau}| \lesssim 3.3\times 10^{-3}$. Hence, once the benchmark  value  $f=-20$ is fixed,
 the product of $s_{24}s_{34}$ is constrained to be smaller than $\lesssim 3 \times 10^{-4}$.
 One can have both mixing angles of similar size $s^2_{24} \simeq  s^2_{34} \simeq 10^{-4}$,
 or different size with the larger one being not bigger than $10^{-3}$ (and consequently
 the smaller one being not lower than $10^{-5}$)  in order
 to respect the upper bound on $|\varepsilon^\e_{\mu\mu}  -\varepsilon^\e_{\tau\tau}|$.
 In summary, we see that the current limits on the NSI lead us to identify the hierarchical pattern  
 $s^2_{14}\gg s^2_{24}$ and $s^2_{14}\gg s^2_{34}$, while the ratio $s^2_{34}/s^2_{24}$ lie
  in the interval $[0.1, 10]$. 
  As it will be clear from Sec.~\ref{Sec:IceCube}, one of the most important purposes
  of our work is to show that the new 3+1 model is able to produce in IceCube the 
  high-energy resonance in the $\mu-\mu$ channel.   
  Similarly to the standard 3+1 scheme the leading mixing angle regulating such a resonance 
  is $s^2_{24}$.
  For this reason, for simplicity, we consider the benchmark case $s^2_{34}=0$,
  commenting later on the impact of non-zero  $s^2_{34}$. 
  For clarity we summarize the benchmark values of the model parameters and their approximate 
  allowed ranges in the first row of Table~\ref{Tab:benchmark}. The range of the parameters 
  will be further clarified when discussing the high-energy resonance relevant for IceCube in 
 Sec.~\ref{Sec:IceCube}. In particular, we will see that the value of $\Delta m^2_{41}$
  is strictly related to that of  $|f|$.
    
\begin{table}[t!]
\centering
\resizebox{.99\textwidth}{!}{\begin{minipage}{\textwidth}
\caption{\label{Tab:benchmark}  
Benchmark values of parameters and their approximate allowed ranges. Note that $|f|$ and $\Delta m^2_{41}$ are
fully positively correlated since their ratio must be approximately  constant.
}
\begin{ruledtabular}
\begin{tabular}{lcccccccc}
Parameter & $f$ &  $\Delta m^2_{41}/\mathrm{eV}^2$ &  $s^2_{14}/10^{-2}$ & $s^2_{24}/10^{-3}$  & $s^2_{34}/10^{-3}$  &  $\delta_{13}/\pi$ & $\delta_{14}/\pi$ & $\delta_{34}/\pi$ \\
\hline
$ \mathrm{Benchmark\,\, value}$ & -20 & 60 & 2 & $1$ & $0$ & 1.5 & 0 &  -- \\
\hline
$ \mathrm{Preferred\,\, range}$ &  [-40, -10] & [30, 120] &  [1, 3]  & [0.5, 2] & $\lesssim 1$ & [1.1, 1.8] & [0, 0.3] $\cup$ [1.2, 2.0]  & [0, 2]\\
\end{tabular}
\end{ruledtabular}
\end{minipage}}
\end{table}

 \section{Impact on the 3-flavor atmospheric neutrino resonances at a few GeV}
\label{Sec:Res_6GeV}

\subsection{Explaining the excess of electron-like events in Super-Kamiokande}
 
We have shown the new sterile potential induces NSI-like couplings in low-energy atmospheric
neutrinos. Consequently, one may expect a modification of the standard 3-flavor MSW and parametric resonances,
which occur at a few GeV. Although these resonances are difficult to observe, a first
indication is coming from atmospheric neutrinos.
In fact, the preference for NO found in Super-Kamiokande, as stressed in~\cite{Super-Kamiokande:2023ahc},
is imputable to an excess of $\nu_e$-like events (without an excess of $\bar\nu_e$-like events) 
in the multi-GeV and multi-ring samples and their concurrent up-down asymmetry (see Figs.~6, 8 and 15),
which can naturally be traced to the resonantly enhanced muon-to-electron flavor conversion, occurring
in the neutrino (anti-neutrino) channel in NO (IO). Looking at the plots, the excess is bigger than expected in the 3-flavor framework, and this
plausibly leads to a statistical significance of the indication in favor of NO ($\Delta \chi^2 \simeq 6$), which is 
sensibly larger than the expected sensitivity ($\Delta \chi^2 \simeq 2$, see Fig.~16).
Concerning IceCube DeepCore, the official analysis~\cite{IceCubeCollaboration:2024ssx} 
does not provide information on the NMO preference. However, the preliminary study~\cite{ICThesis} using  a 9.28-years data sample,
indicates a preference for NO at $\Delta \chi^2\simeq 4.4$. In such a study it is emphasized that  
the origin of the preference can be traced to an excess of cascades events related to 
electron neutrinos (see Fig. 6.4), which as in Super-Kamiokande are expected to be
produced by the MSW resonant-like behavior for neutrino energies~$E \simeq$ 3-7 GeV. 
Also in this case, the indication has a statistical significance sensibly 
larger than the expected sensitivity (see Fig. 8.2), which is below $\sim 1.2\sigma$.

\begin{figure*}[t!]
\vspace*{0.05cm}
\hspace*{-0.1cm}
\includegraphics[height=7.5cm,width=7.87cm]{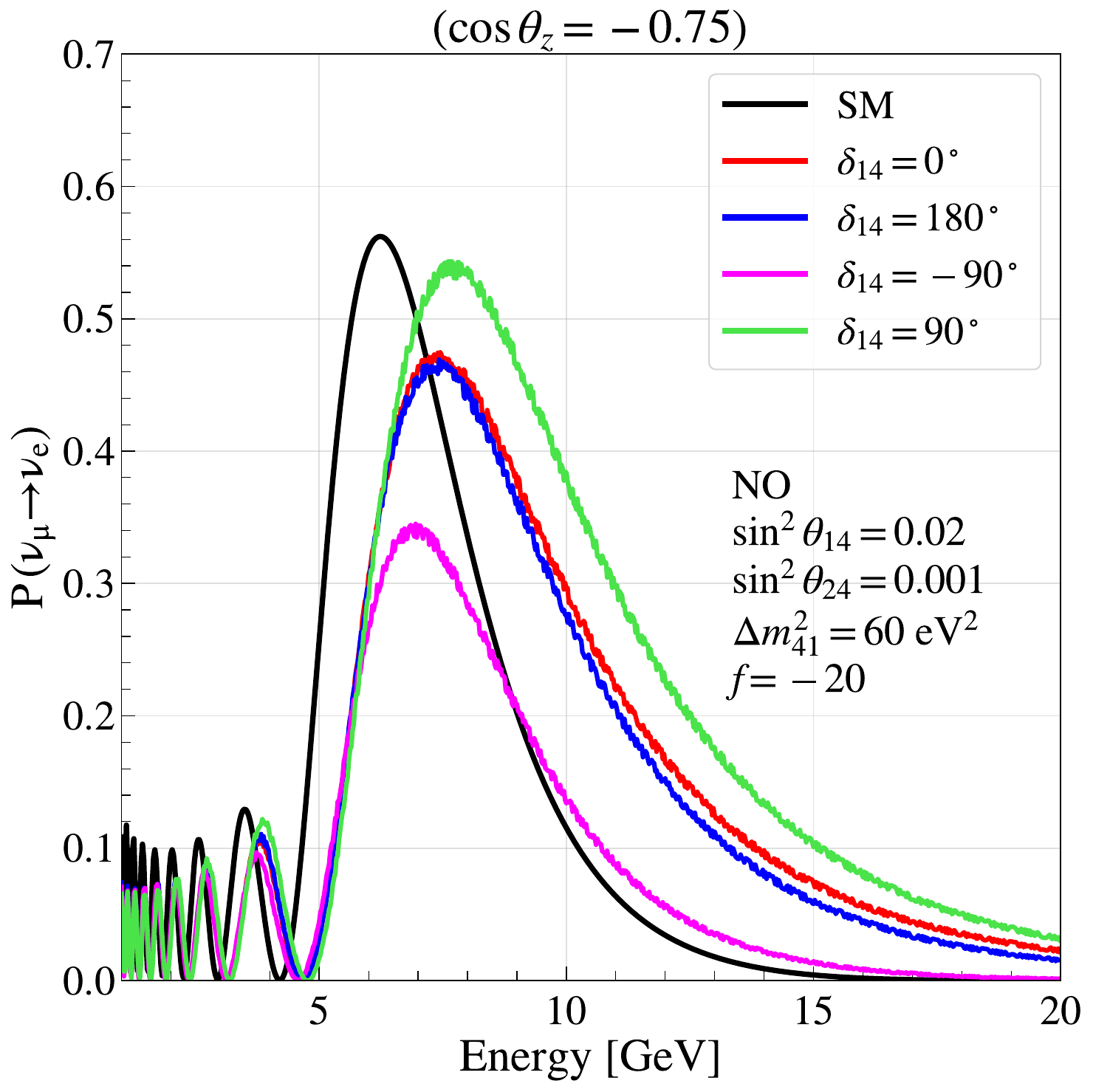}
\includegraphics[height=7.5cm,width=7.87cm]{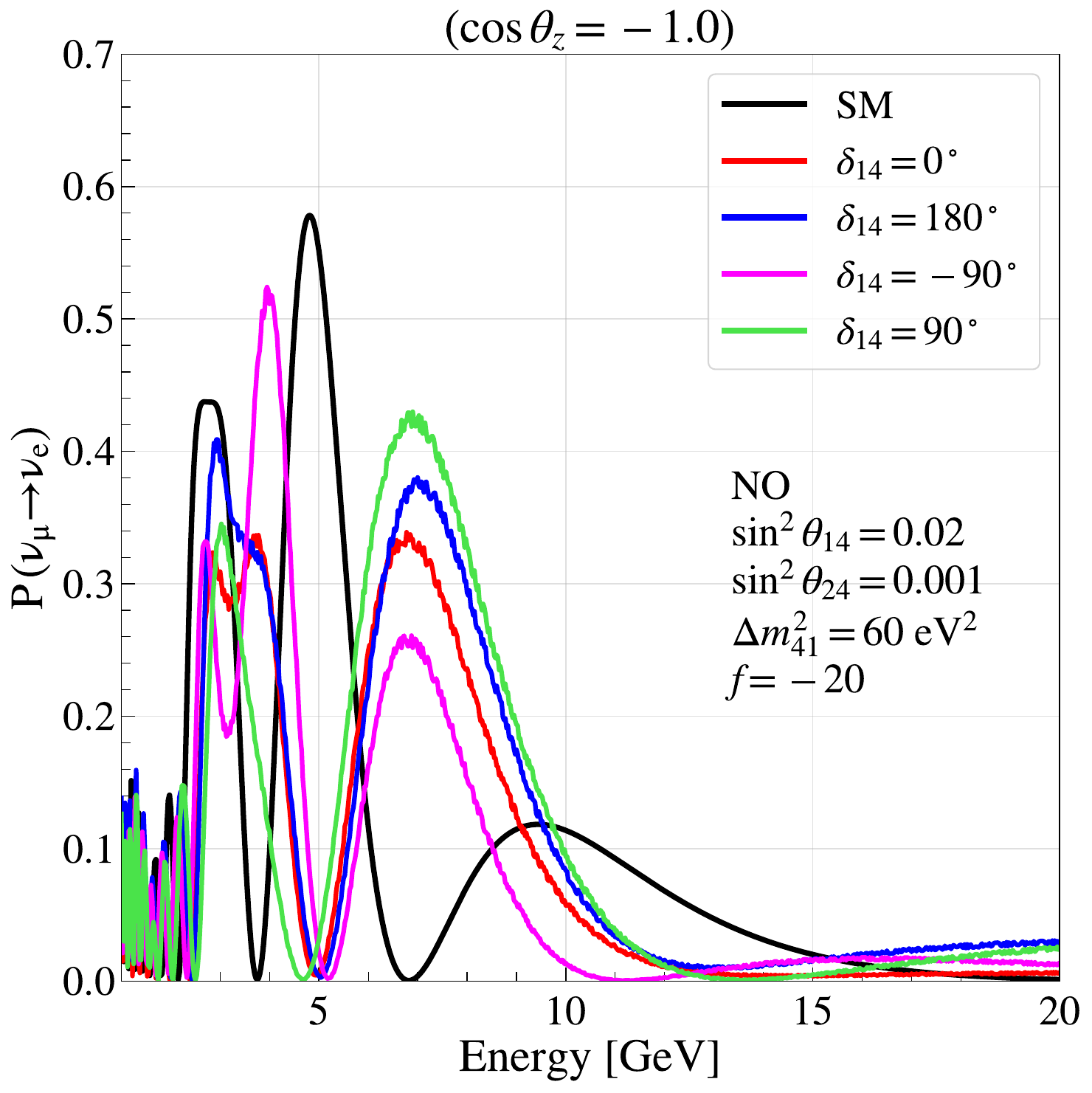}
\includegraphics[height=7.5cm,width=7.87cm]{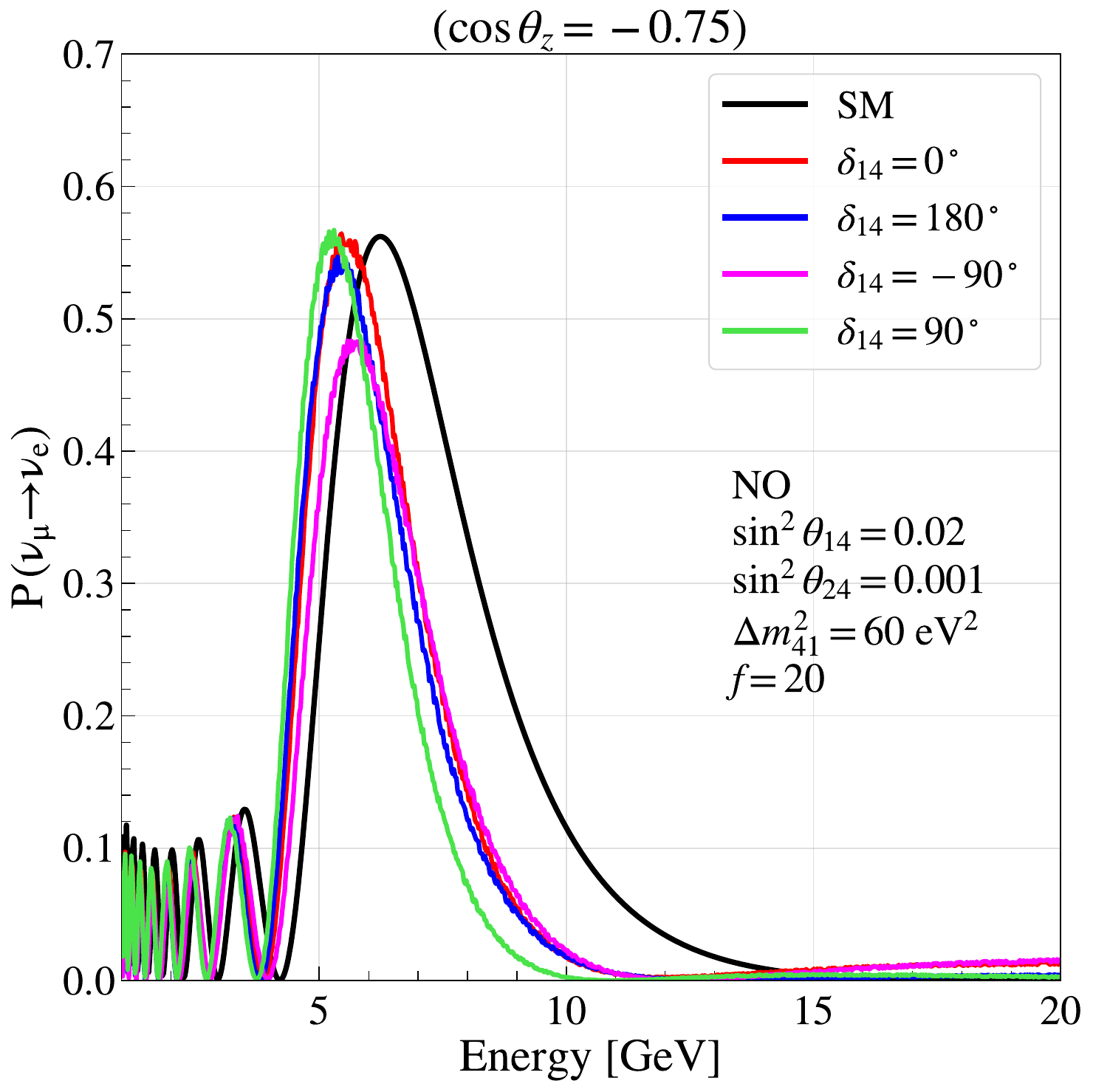}
\includegraphics[height=7.5cm,width=7.87cm]{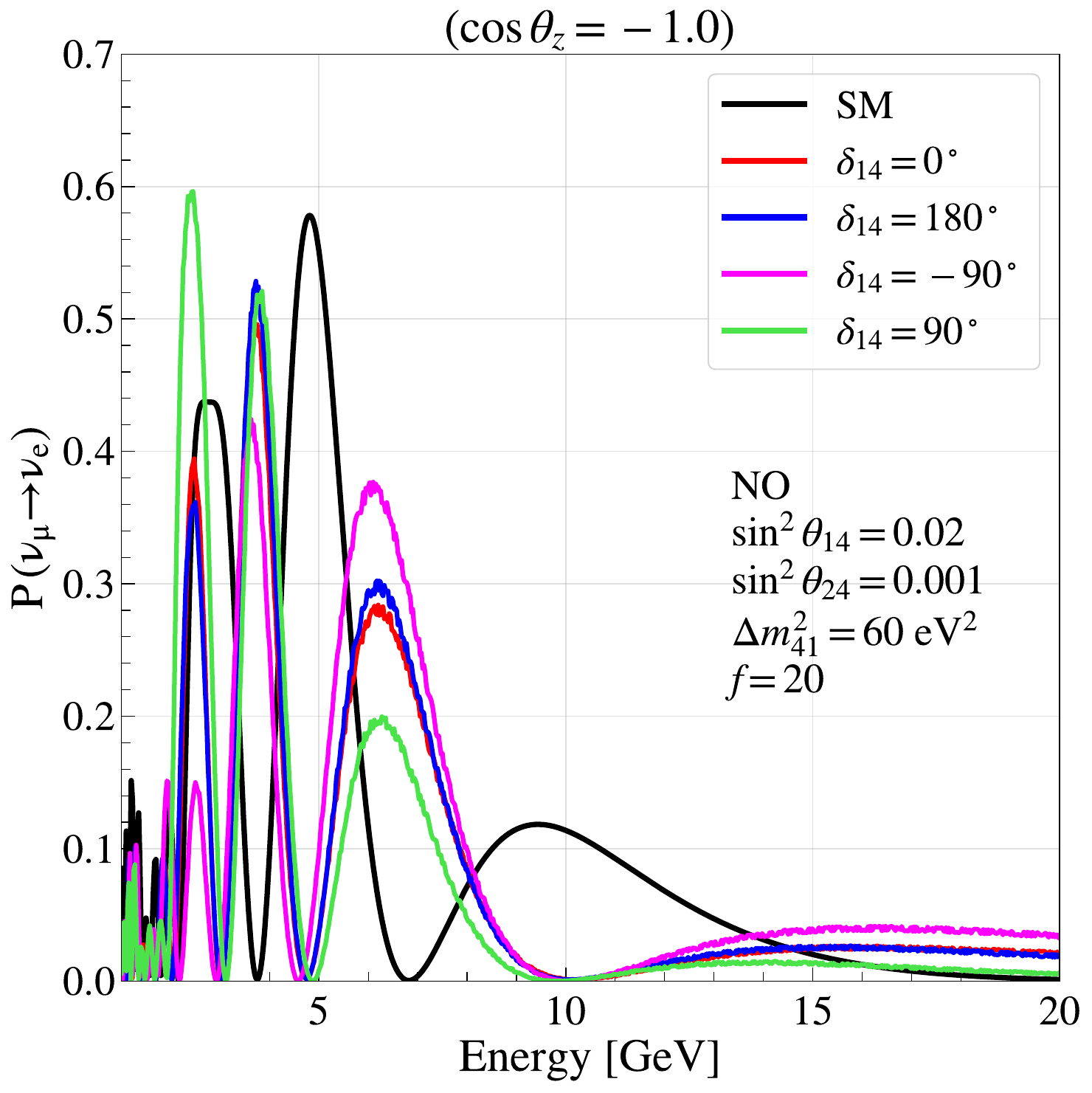}
\vspace*{-0.0cm}
\vspace*{0.0cm}
\caption{The plots represent the electron neutrino appearance probability for energies around a few GeV,
where the MSW and parametric resonances occur. The upper (lower) panels refer to a negative  
(positive) value of the parameter $f$ regulating the strength of the new potential.
The left panels illustrate a mantle-crossing trajectory with $\cos\theta_z = -0.75$,
in which case a MSW resonance is expected. The right panels correspond to a 
core-crossing trajectory with $\cos\theta_z = -1.0$, in which case three parametric enhancements occur.
The black profile depicts  the 3-flavor case. The colored curves represent the 3+1 model with the new 
matter potential with parameters $f = \pm20$, $\Delta m^2_{41} = $ 60 eV$^2$, $s^2_{14} = 0.02$, 
$s^2_{24} = 0.001$, $s^2_{34} = 0$. The CP-phase $\delta_{14}$ assumes the four values indicated in the legend.}
\label{fig_IceCube_prob_NMO}
\end{figure*}

While the excess of electron-like events is compatible with a statistical over fluctuation, it may constitute 
a manifestation of active-sterile neutrino oscillations with the novel potential. 
Here, we demonstrate that the proposed scenario can indeed lead to an enhanced excess of
$\nu_e$-like events at a few GeV. In order to illustrate this point, we show
in Fig.~\ref{fig_IceCube_prob_NMO} the plots of the transition
probability $P(\nu_\mu \to \nu_e$) in the neutrino channel.
The upper (lower) panels refer to the negative (positive) valued new potential with $|f| = 20$.
The left panels refer to a mantle-crossing trajectory while the right ones refer to  a core-crossing
 trajectory. In the first case, one expects a MSW resonance at  $\sim$ 6 GeV, while in the second case
 three parametric enhancements are predicted (see for example~\cite{Akhmedov:2006hb}).
The black profile corresponds to the 3-flavor scheme,
in which case we have fixed the parameters at the best fit values of the global analysis~\cite{Capozzi:2025wyn},
with the exception of $\delta_{13}$, which we have fixed at the best fit value obtained from
our joint fit of NOvA and T2K in the presence of the new matter potential, which is basically 
the same for both NMO ($\delta_{13} \simeq 1.5 \pi$). The colored profiles are obtained for the benchmark 
parameters of the model with the novel potential, $|f| = 20$, $\Delta m^2_{41} = 60$~eV$^2$, $s^2_{14} = 0.02$, $s^2_{24} = 10^{-3}$.
The different colors correspond to four different choices of the CP-phase $\delta_{14}$ indicated in the legend.
Focusing on the 6 GeV MSW resonance occurring for mantle crossing (left panels),
we can observe that all properties of the resonance (peak energy, height, width and tail) are markedly modified in the presence 
of the new physics. However, the effects are radically different for negative with respect to positive values of the potential. 
In the upper-left plot, corresponding to $f = - 20$, we see that the peak energy appreciably moves toward larger
energies with a shift $\Delta E \simeq +1.5$~GeV. The peak height strongly
depends on the value of the CP-phase $\delta_{14}$. 
In all cases, the width is approximately 
the double compared to the 3-flavor scheme and the tail extends to much higher energies.
This means that, for the same peak height, in the
presence of the new (negative) potential, the resonance
would lead to at least as twice the number of events compared with the 3-flavor case. Given the linear grow 
with the energy of the involved cross sections, the number of events can be even substantially larger.
In the left-lower plot, we can notice that in the case with positive potential ($f=+20$), the resonance
is shifted toward lower energies ($\Delta E \simeq -1$~GeV), and has smaller width. 
These findings, clearly indicate
that the new scenario goes in the direction indicated by data only for a negative valued potential.
The modifications of the parametric enhancements shown in the right panels are less trivial
and a numerical analysis is needed to properly evaluate the impact of the new effects.
However, due to the lower energy and lower area under the curve they should be
less important in the overall excess. We note that these findings on the resonance at a few GeV
 are in striking agreement with the independent indication provided by NOvA and T2K,
which prefer a negative potential.
We deem that the excess of electron-like events observed in Super-Kamiokande and IcecCube DeepCore 
may be easily explained by the scenario under consideration. Our educated guess is that Super-Kamiokande and
IceCube DeepCore could provide an indication in favor of the new model with a statistical level comparable
to that of the indication in favor of NO.

\subsection{Impact on the status of the current indication on the neutrino mass ordering}

The modification of the resonances at a few GeV naturally rises the issue of the robustness
 of the indication in favor of NO provided by global neutrino analyses in the 3+1 framework with pseudo-sterile
 neutrinos.
To understand this point, one must carefully consider the origin of the current preference for NO
in the 3-flavor scheme. The global fits incorporate several datasets, which overall in the combination lead to
the indication in favor of the NO. 
There are three distinct sources which provide the aforementioned  indication. First, there is the $\nu_\mu \to \nu_e$
appearance channel of NOvA and T2K. In the 3-flavor scheme the two LBL experiments jointly provide a weak indication
in favor of IO. In our analysis, we find that in the new scenario there is weak indication in favor of NO
(as also observed in analyses with ordinary NSI~\cite{Chatterjee:2020kkm,Chatterjee:2024kbn}).
In any case the hint from this source is currently very weak.
Second, there is a well known interplay 
between the LBL $\nu_\mu \to \nu_\mu$ disappearance channel and reactor neutrinos (currently dominated by Daya Bay and expected
to be dominated by JUNO in the near future). In fact, 
the joint estimates of the atmospheric mass-squared splitting $\Delta m^2_{31}$ from these two classes 
of experiments, provide a $\Delta \chi^2 \simeq 2.0$ in favor of the NO~\cite{Capozzi:2025wyn} (see also~\cite{Esteban:2024eli,deSalas:2020pgw}).%
\footnote{A recent exploratory analysis~\cite{Esteban:2026phq}
 of JUNO data in conjunction with NOvA and T2K provides an enhanced indication at $\Delta \chi^2 \simeq 5$ in favor of NO.}
 We have explicitly 
checked that the estimate of $\Delta m^2_{31}$ by NOvA and T2K is stable and independent
on the new matter potential. This can be easily understood because the NSI-like $\varepsilon_{\mu\mu}$
and $\varepsilon_{\mu\tau}$ couplings, which have more impact on the muon neutrino survival probability
that is sensitive to $\Delta m^2_{31}$, are negligibly small in our scenario. Similarly, Daya Bay
and other reactor experiments have basically no sensitivity to matter effects and also their
estimate of  $\Delta m^2_{31}$ is unaffected. As a consequence the interplay between LBL
and Reactor experiments remains intact in the presence of pseudo-sterile neutrinos, and their (small)
indication in favor of NO is unaltered. This is very important for the future searches. In fact,
in case of confirmation of the new 3+1 scenario here considered, the interplay of LBL experiments
with JUNO, which is expected to soon release relevant data, would provide a clean signature on
the NMO.   The third, currently dominant, piece of information on the NMO is  potentially more problematic,
because, as we have discussed in the subsection above, it comes from the SK 
atmospheric neutrino data~\cite{Super-Kamiokande:2023ahc},
which show a preference for NO with $\Delta \chi^2 \simeq 6$, which is strictly
related to the GeV resonant enhancements.
Based on our findings, the NO seems to be favored also in the presence of pseudo-sterile neutrinos,
because only in NO one can explain the observed excess of $\nu_e$-like events. One should
expect and increased statistical level of the indication in favor of NO since the new scenario offers
 a better fit of the experimental data. Hence, we can conclude, with the important caveat given below, that the status of the indication
in favor of NO in current global analyses is robust and the indication would be even enhanced by
LBL (which jointly prefer NO in the new scheme) and by Super-Kamiokande. 

In the discussion above, we have tacitly assumed that the solar lower octant ($\sin^2\theta_{12}<0.5$) 
solution is that chosen by Nature. However, we have  seen that in the case in which the new potential 
is proportional to the electron or proton number density, a degenerate solution exists for the dark octant
($\sin^2\theta_{12}>0.5$). In the presence of such a degenerate solution, one cannot distinguish between 
NO and IO. This happens because the 3-flavor scheme suffers from the so-called generalized  
mass ordering degeneracy~\cite{Coloma:2016gei}.
It implies that in the dark-octant solution ($\sin^2\theta_{12}\simeq 0.7$), all the existing experiments 
sensitive to the NMO prefer the alternate case (IO) with respect to the one (NO) preferred for the 
lower-octant solution ($\sin^2\theta_{12}\simeq 0.3$). In particular, also JUNO, in the presence of the
octant degeneracy is completely blind to the NMO~\cite{Fogli:2001wi}, and its combination to the disappearance channel of LBL
experiments is also ineffective in this respect.  
Therefore, if the existence of the dark-solution is taken into account (as it should be in 
the senario entailing pseudo-sterile neutrinos with a matter potential proportional to the electron or proton number
density), the NMO is completely undetermined by current data. However, we will see that the
high energy ($\sim 10$ TeV) atmospheric data offer a tool able 
to break the degeneracy because they are sensitive to a 4-flavor resonance which presents
different features in the two (otherwise) degenerate solutions.

\section{New three-level resonant conversion phenomena in high-energy atmospheric neutrinos}
\label{Sec:IceCube}

It is well known that neutrino flavor transitions in matter can give rise to resonant-like amplification. 
This distinctive behavior was first predicted in the context of solar neutrinos~\cite{Mikheev:1986gs,Mikheev:1986wj}.
Within the 3-flavor framework, 
it is extremely difficult  to observe the resonant phenomenon in terrestrial experiments, albeit
some hints are coming from multi-GeV atmospheric neutrinos, as we have discussed  in the previous section.
The situation is radically different when 
an extra light sterile neutrino species comes into play.
In fact, in the 3+1 scheme, as first evidenced in~\cite{Yasuda:2000xs}, 
high-energy ($E \simeq$ TeV) atmospheric neutrinos can undergo matter enhanced conversion.
As pointed out in~\cite{Nunokawa:2003ep}, neutrino telescopes have the unique potential to detect
such a resonant behavior.
More precisely, matter effects can result in the near complete disappearance 
of TeV-scale muon anti-neutrinos traveling through the Earth for a sterile neutrino with eV$^2$-scale mass-squared difference (see~\cite{Choubey:2007ji,Razzaque:2011ab,Barger:2011rc,Razzaque:2012tp,Esmaili:2012nz,Agarwalla:2012uj,Esmaili:2013vza,Lindner:2015iaa,Petcov:2016iiu,Miranda:2018buo,Cabrera:2024rgi,Brettell:2024zok} for further studies on this topic). 
The experimental search for the resonant signature has been pioneered by the IceCube Collaboration
with high-energy analyses exploiting neutrinos from $500$ GeV to $100$ TeV.
Starting from the eight-years data analysis~\cite{IceCube:2020tka}, the collaboration
has found a hint in favor of sterile neutrinos within the ordinary 3+1 scheme, 
which has been subsequently  confirmed and consolidated in 
the eleven-years data analyses~\cite{IceCubeCollaboration:2024nle,IceCubeCollaboration:2024dxk,IceCube:2024pky}.
This indication seems to come from the observation of a resonant-like behavior occurring at energies
of $\sim 10$ TeV. In the analysis~\cite{IceCubeCollaboration:2024nle}, performed under
the assumption of $\theta_{34}=0$, the best fit point is obtained for $\Delta m^2_{41} \sim$ 3.5 eV$^2$ and $s^2_{24} = 0.04$.
Such a best fit point is rather stable if $\theta_{34}$ is allowed to assume a non-zero value compatible
with the existing bounds, as shown in~\cite{IceCube:2024pky}.
Here, we will show that in the pseudo-sterile neutrino scenario the picture of the resonant phenomena
involving high-energy neutrinos becomes much more complex. In particular, we will see that 
together with the ordinary resonances involving a pair of active-sterile states
$(\nu_\alpha, \nu_s)$ ($\alpha = e,\mu, \tau$) one can have new resonances
in the active $(\nu_\alpha, \nu_\beta)$ ($\alpha, \beta = e,\mu, \tau$) sub-systems, 
mediated by the pseudo-sterile states. We will see that such new 
resonances are dynamically and algebraically different with respect to the conventional ones.

\subsection{Qualitative aspects of the conventional resonances and the bare level crossing diagram}
\label{SubSec:IceCube_qualitative}

As a first step in our study of the resonant behavior we clarify the framework behind the 
conventional resonances usually considered in the literature. This will allow us to  better
appreciate the unconventional features of the new amplification phenomena implied
by the new matter potential, which will be addressed in the following subsections.
A resonant behavior can take place if the matter potential is comparable to the wavenumber
$k_{41} = \Delta m^2_{41}/2E$ associated to the active-sterile oscillation frequency.
For an eV-scale sterile neutrino and matter density $\rho = 5.0$~g/cm$^3$ in the Earth, 
one gets the reference neutrino energy of $\sim10$ TeV as follows
\begin{eqnarray} 
\frac{|V_{NC}|}{k_{41} } 
&=& 
0.74 
\left(\frac{ \Delta m^2_{41} }{ 3.5~\mbox{eV}^2}\right)^{-1} \left(\frac{\rho}{5.0 \,\text{g/cm}^3}\right) \left(\frac{E}{10~\mbox{TeV}}\right)\,,
\label{a/Dm2}
\end{eqnarray}
where we have used as a benchmark the best fit value $\Delta m^2_{41} = 3.5~$eV$^2$ found in the 
IceCube analysis~\cite{IceCubeCollaboration:2024nle}. This value is roughly uncertain
by a factor of two as derived from the 90\% C.L. contours. 
Qualitatively, one expects that 
if  $|V_{NC}|$ is multiplied by a large factor $f$, the resonance energy will remain unaltered provided
that  $\Delta m^2_{41}$ is rescaled by the same factor (taking its absolute value).
In addition, one must note that  the sign of the rescaling factor $f$ would determine the channel
of the resonance. One expects the resonance in the neutrino (antineutrino) channel for $V_s<0$ 
($V_s>0$). 

\begin{figure*}[t!]
\vspace*{-0.0cm}
\hspace*{-0.1cm}
\includegraphics[height=5.5cm,width=5.9cm]{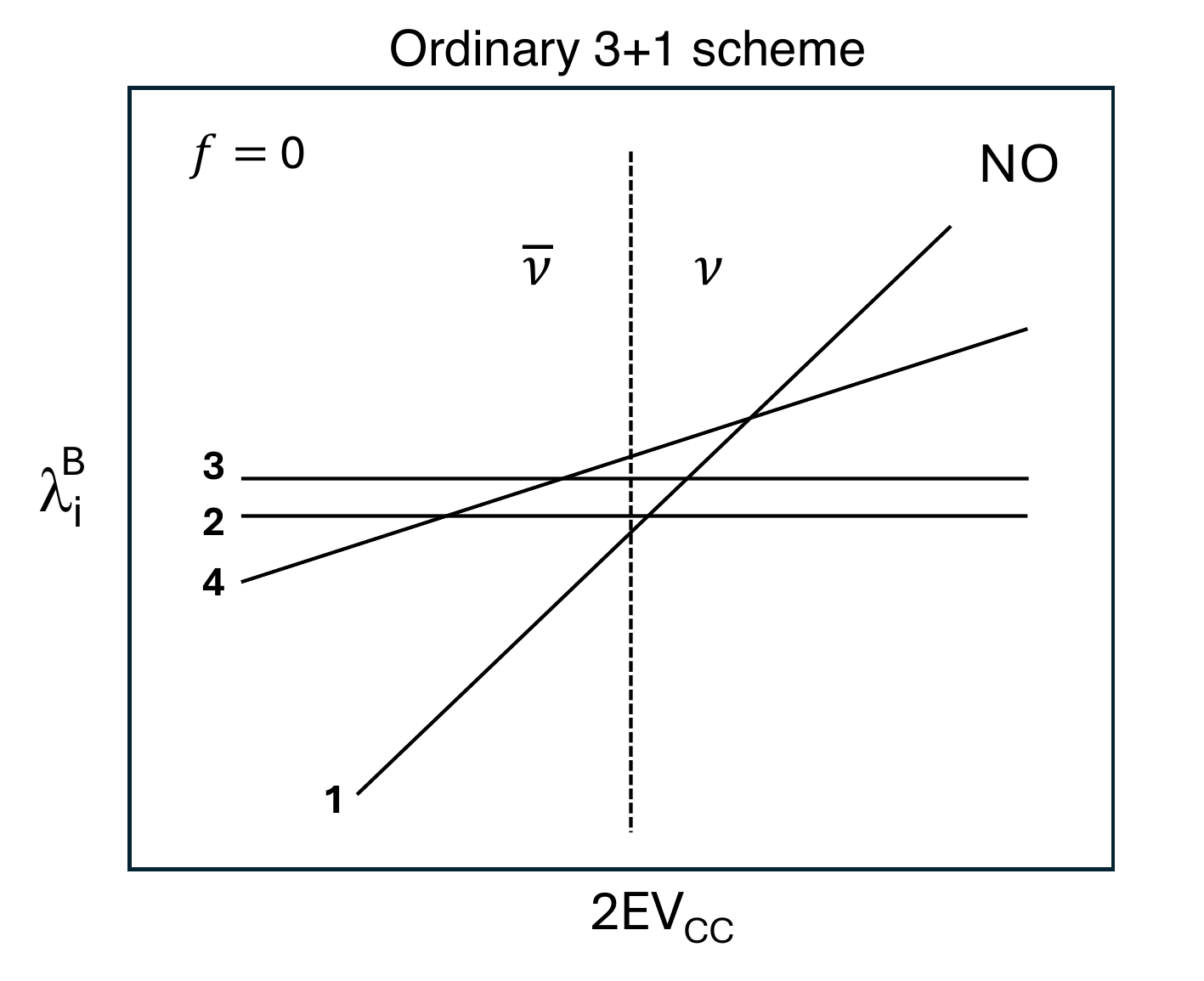}
\includegraphics[height=5.5cm,width=5.9cm]{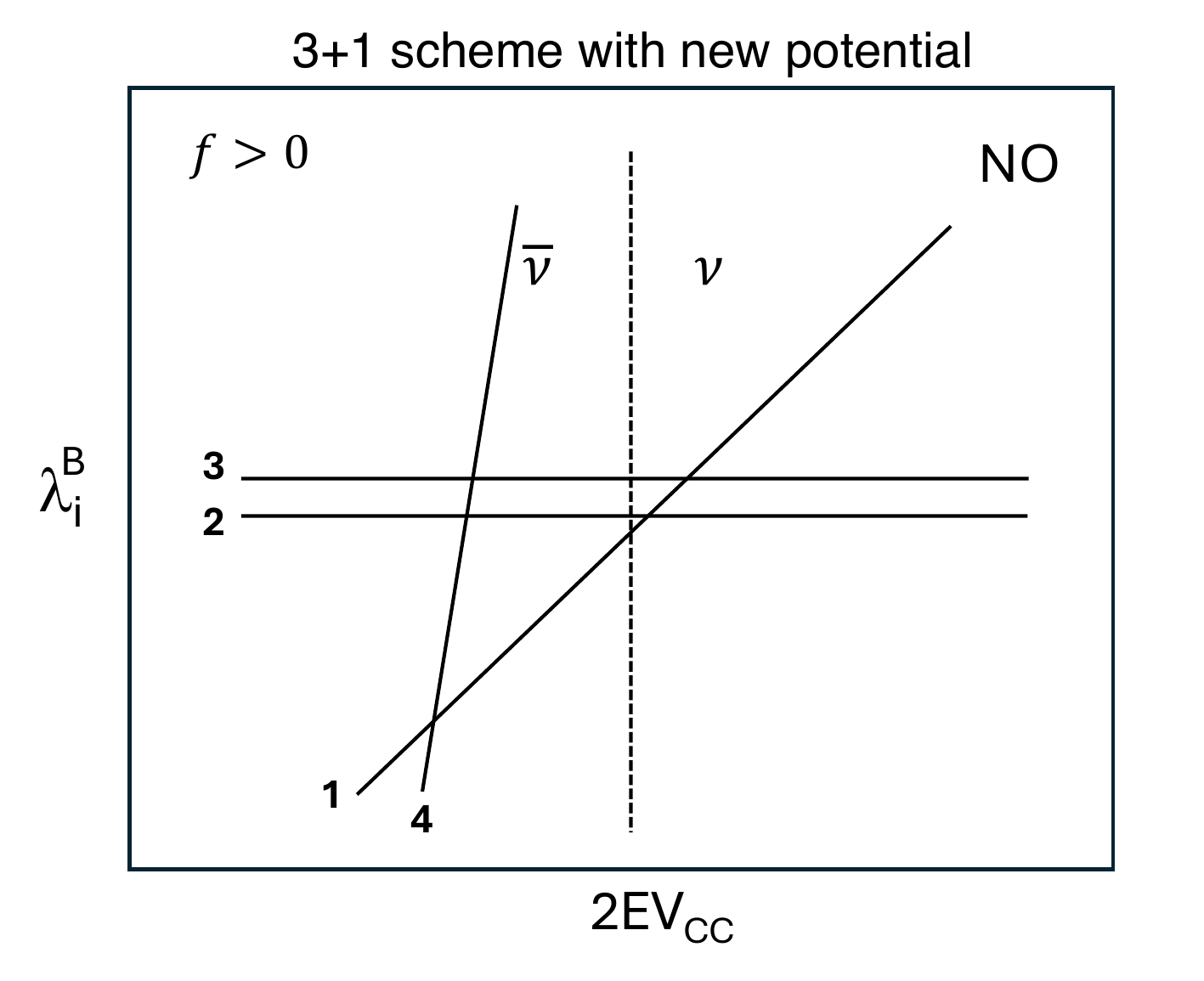}
\includegraphics[height=5.5cm,width=5.9cm]{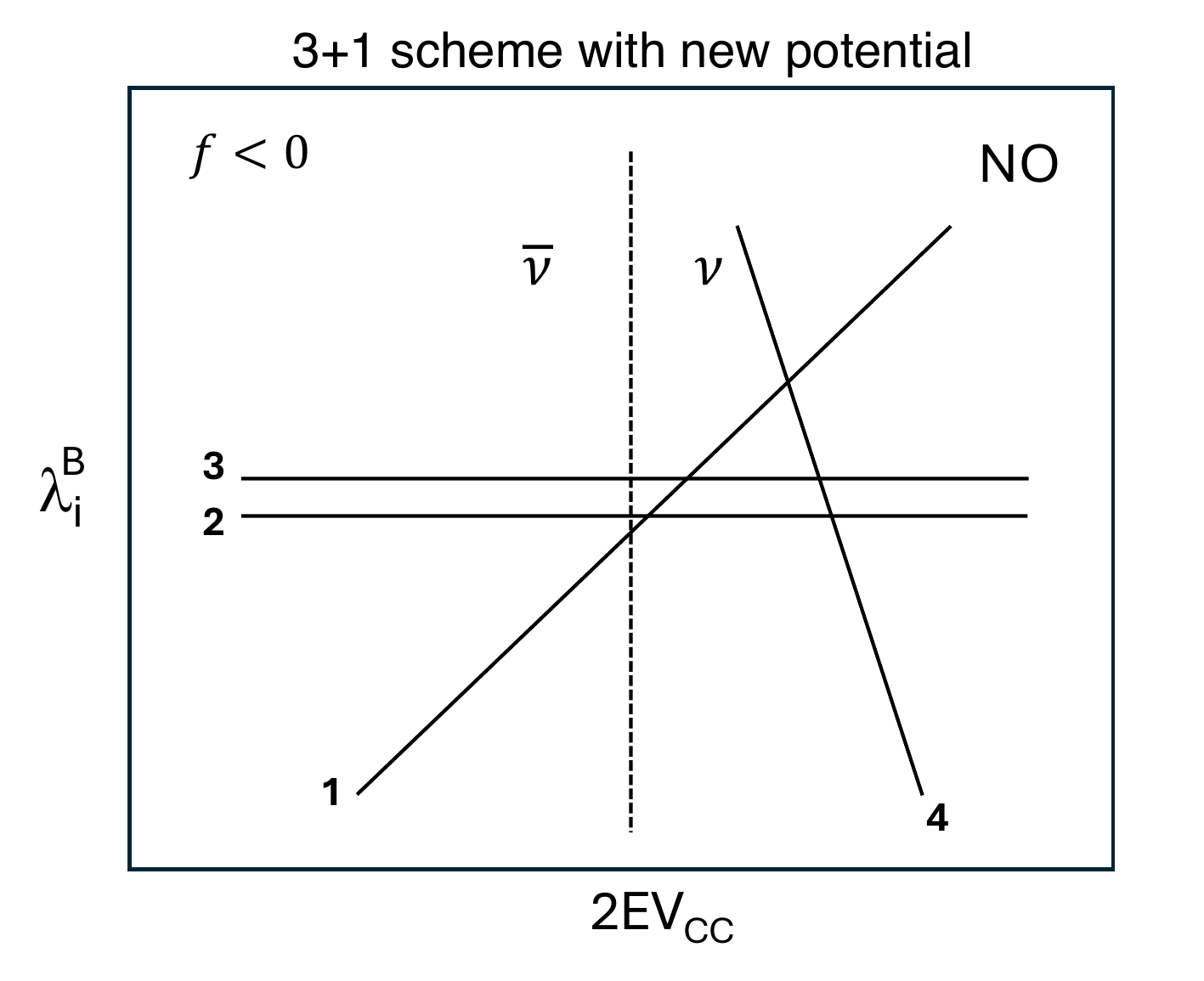}
\vspace*{-0.5cm}
\caption{The three panels represent the ``bare  level crossing diagram'' obtained by setting to zero all the
six mixing angles. The left panel refers to the 3+1 ordinary scheme. The central (right) panel refers to 
the 3+1 scheme with the new potential for a positive (negative) value of $f$ much larger than one. 
In all cases we assume NO for the three light states. The energy scale of the diagrams is qualitative as they
are envisaged to illustrate a topological structure.}
\label{fig_level_crossing}
\end{figure*}

 It is useful to define the ``bare'' Hamiltonian setting to zero all the six mixing angles,
 thus considering $U = \mathds{1}$ in the original Hamiltonian in Eq.~(\ref{Eq:H4nu}),
  obtaining 
\begin{eqnarray} \label{eq:H_3_ems_v1}
\arraycolsep=3pt
\medmuskip = 1mu
     H_{4\nu}^\mathrm{bare}  = K + V =
      \frac{1}{2E}
      \begin{bmatrix}
	 0   & 0  & 0	 & 0\\
	0 & \Delta m^2_{21} &  0  & 0\\
	0 & 0 & \Delta m^2_{31} & 0\\
	0& 0 & 0 &   \Delta m^2_{41}\\
    \end{bmatrix} +
    \begin{bmatrix}
	V_{\rm{CC}}   & 0  & 0	 & 0\\
	0 & 0 &  0  & 0\\
	0 & 0 &  0  & 0\\
	0& 0 & 0 &  \bar f V_{CC} \\
    \end{bmatrix}, \ 
\end{eqnarray}
which is diagonal. The matrix $2E H_{4\nu}^\mathrm{bare}$ has the (2,2) and (3,3) entries constant, 
 while the (1,1) and (4,4) entries present a linear dependency on the energy. The behavior
 of the eigenvalues of  $2E H_{4\nu}^\mathrm{bare}$, which we will denote with $\lambda^\mathrm{B}_{i}$
 ($i =1, 2, 3, 4$), can be represented in a diagram
 recently introduced in~\cite{Brettell:2024zok} to analyze the high--energy resonance in the ordinary 3+1 scheme.
 While in~\cite{Brettell:2024zok} the diagram is called with the name ``surface level crossing diagram'', 
 here we prefer to call it ``bare level crossing diagram'', in 
 contrast with the  ordinary level crossing diagram obtained when
 the mixing angles are taken into account in the full or ``dressed'' Hamiltonian.  
 In Fig.~\ref{fig_level_crossing} we show the bare level crossing diagram for
the ordinary 3+1 scheme (left panel) and for the 3+1 framework
with the new matter potential (central and right panel), where 
we have taken a large value $|f|\gg 1$, which is positive in the central panel
and negative in the right panel. In all panels we have considered the case of NO,
which implies that the 1-2 (solar) and 1-3 (atmospheric) crossings occur
for the neutrino channel.
 The vertical axis corresponds to zero potential/energy.%
 \footnote{When studying the resonant behavior in a constant density medium as in the present work, 
 in which case the potential is constant, the horizontal axis is proportional to the neutrino
 energy (positive for neutrinos, negative for antineutrinos). 
 In the context of flavor conversion of a neutrino with definite energy in an environment with varying density, the horizontal 
 axis is proportional to the matter potential which we will be a function of the time/position coordinate.} 
 The right part of the plane refers to neutrinos, 
while the left one corresponds to antineutrinos.  
The scales
are qualitative, as this diagram has the sole purpose to identify
the topological structure of the crossings among the four different energy levels.   
When the mixing angles are switched on to form the complete or ``dressed'' Hamiltonian,
the crossing points will become avoided crossings regions, 
where the phenomenon of eigenvalue repulsion~\cite{No-Crossing_1,No-Crossing_2} occurs mediated
by the off-diagonal entries which act as effective couplings among 
the different levels.

In the left panel we observe that in the ordinary  3+1 scheme, in addition to the
standard 3-flavor 1-2 and 1-3 crossings, one expects three crossings involving
the fourth state, a 1-4 crossing occurring in the neutrino channel and two crossings
(2-4 and 3-4) in the antineutrino channel. This crossing scheme is related to the
slope of the fourth state determined by the potential $0.5 V_{CC}$
appearing in the (4,4) entry of the bare Hamiltonian.
From the phenomenological point
of view only the 2-4 crossing in the antineutrino channel is relevant because
it induces $\bar \nu_\mu\to \bar \nu_s$ conversion observable in IceCube.
The 1-4 and 3-4 crossings are less relevant since IceCube is not sensitive
to   $\nu_e\to  \nu_s$ and $\bar \nu_\tau \to \bar \nu_s$ transitions due
to the flavor composition of the atmospheric neutrinos.
The central panel depicts the situation for a large positive value of the $f$ parameter,
which regulates the strength of the new potential. In this case the new potential
in the (4,4) entry $\bar f V_{CC} $ is positive as in the ordinary 3+1 scheme 
but the slope of the fourth state is much larger.
This implies that all the three crossings 
1-4, 2-4, 3-4 occur in the antineutrino channel. Conversely, for 
a negative value of $f$, as apparent in the right panel, all such three 
level crossings occur in the neutrino channel. 

One advantage of the bare level crossing diagram is its role in identifying 
the mixing angles in matter which regulate a given resonance~\cite{Brettell:2024zok}.
 Once the sterile vacuum mixing angles ($\theta_{14}$, $\theta_{24}$, $\theta_{34}$) 
are included to obtain the full Hamiltonian, three avoided crossings will emerge.
 These level repulsions will correspond to resonances%
 \footnote{The resonant enhancement in a constant density medium will appear as a dip/peak 
 when plotting the survival/transition probability as a function of the energy. 
 In the context of flavor conversion in a medium with varying density, the phenomenon of 
 adiabatic conversion (possibly corrected by non-adiabatic transitions)
 will occur when the neutrino traverses the resonance region.} 
originating respectively from
the [1-4, 2-4, 3-4] crossings, whose amplitude will be regulated
by the corresponding mixing angles in matter [$\theta_{14}^m$,
$\theta_{24}^m$, $\theta_{34}^m$], which will be amplified with
respect to their values in vacuum.%
\footnote{It should be underlined that for the realization of a resonant phenomenon,
an enhanced mixing angle in matter is a necessary but not sufficient condition. In fact, one also
needs a sufficiently large oscillating factor in the conversion probability,
which depends both on the dynamics (through the differences of the eigenvalues of the full Hamiltonian)
and a phenomenological parameter, which is the  baseline traveled by neutrinos.
It is a very lucky circumstance that the dimension of our planet allows the developments
of oscillating phases large enough to make such phenomena observable.}
 We will see that although these expectations will be fulfilled by the numerical simulations,
in addition to such conventional resonances involving one of the three active species and the sterile one,
we will encounter new unconventional resonances occurring among the active flavors 
$(\nu_e,\nu_\mu, \nu_\tau)$. The existence of such a kind of resonances is unexpected 
as it cannot be deduced from the bare level crossing diagram. We will see that 
they are generated by amplified second-order effects mediated 
by the sterile neutrino mixing and become appreciable 
only if $f$ is sufficiently large. In the following we will expound and explain this new 
amplification mechanism.

\subsection{Numerical diagonalization of the Hamiltonian in the case of constant density}
\label{SubSec:IceCube_numdiag}

As a second step towards the understanding the properties of the resonant behavior we perform a numerical
study considering a constant density profile with density equal to that of the Earth mantle. 
For the ordinary 3+1 scheme we assume $\Delta m^2_{41} = $ 3.5 eV$^2$ and $s^2_{24} = 0.04$, 
corresponding to the IceCube best fit point. For the new scenario
we take $f= -20$,  $s^2_{24} = 0.001$ and $\delta_{14} = 0$, which is our benchmark point.
In both the ordinary and new 3+1 scenarios we adopt the same value of  $s^2_{14}=0.02$
and set to zero $\theta_{34}$.
In the new scenario we rescale the value of the mass-squared difference by 
approximately the same factor $|f|\simeq 20 $ so as to maintain the resonance energy fixed at $\simeq 10~$TeV 
as indicated by IceCube. We find numerically that the best matching of the resonance energy is 
obtained for $\Delta m^2 = 60~$eV$^2$, which we adopt as the benchmark value.

\begin{figure*}[t!]
\vspace*{0.05cm}
\hspace*{-0.1cm}
\includegraphics[height=5.87cm,width=5.87cm]{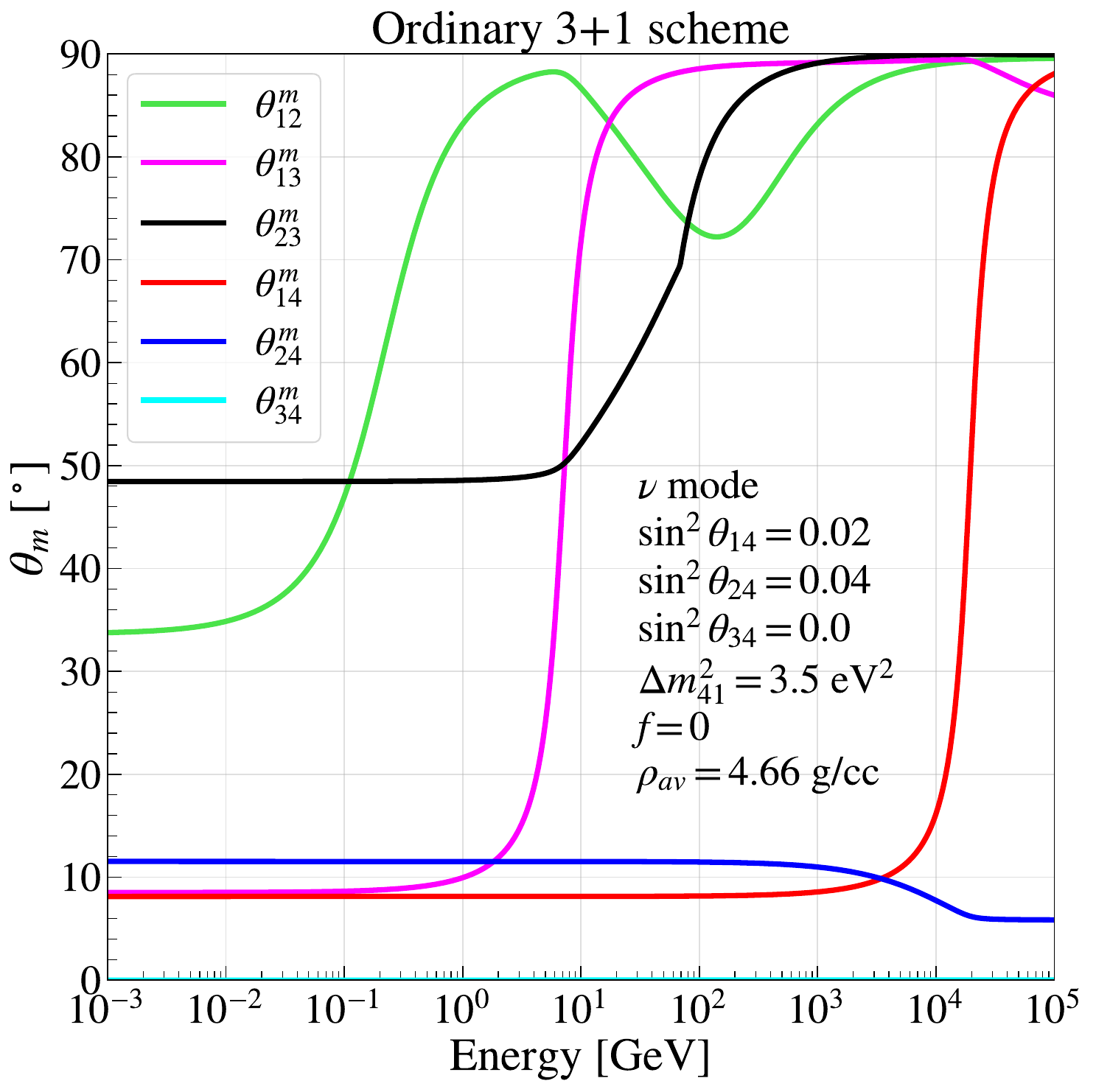}
\includegraphics[height=5.87cm,width=5.87cm]{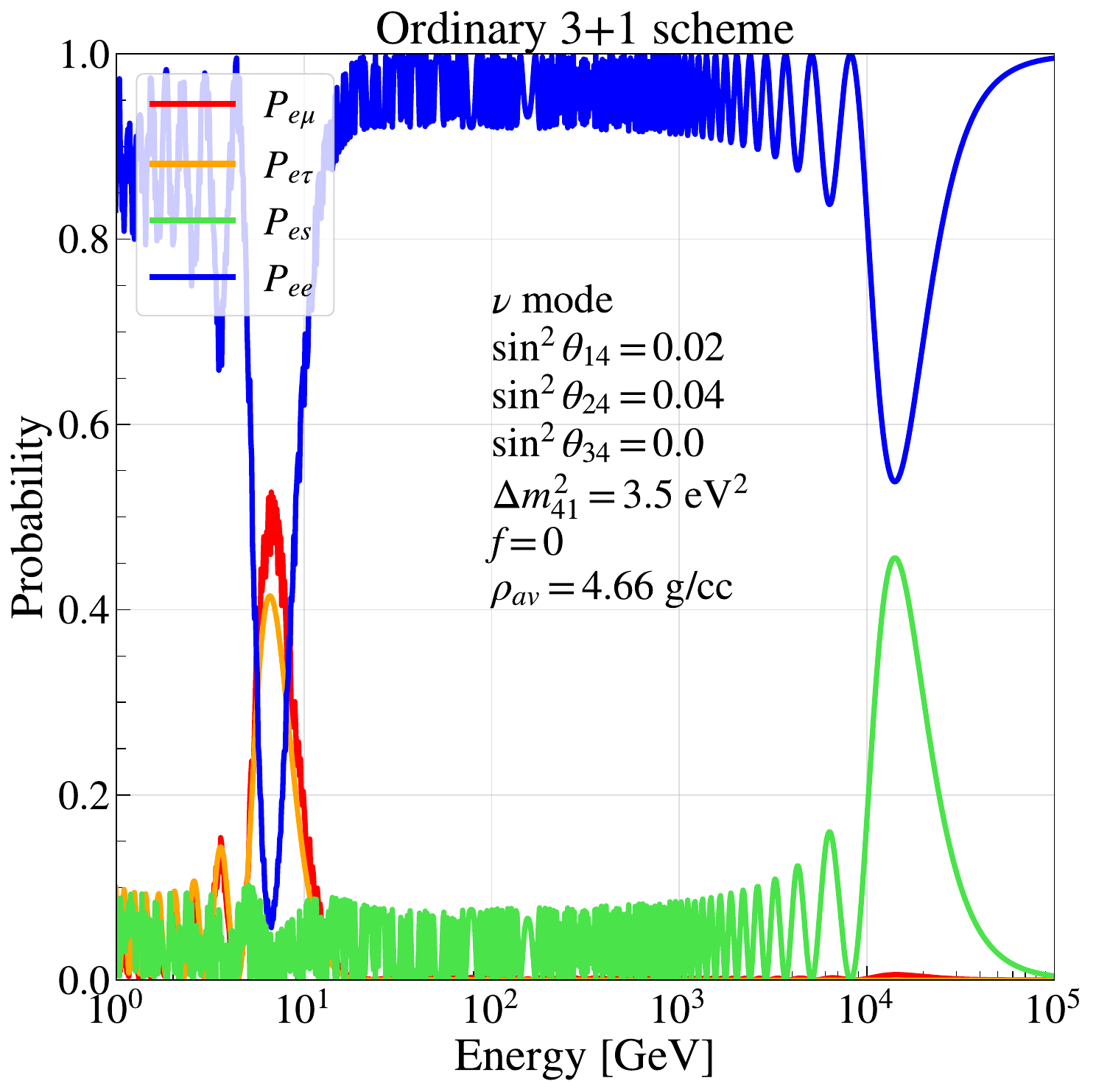}
\includegraphics[height=5.87cm,width=5.87cm]{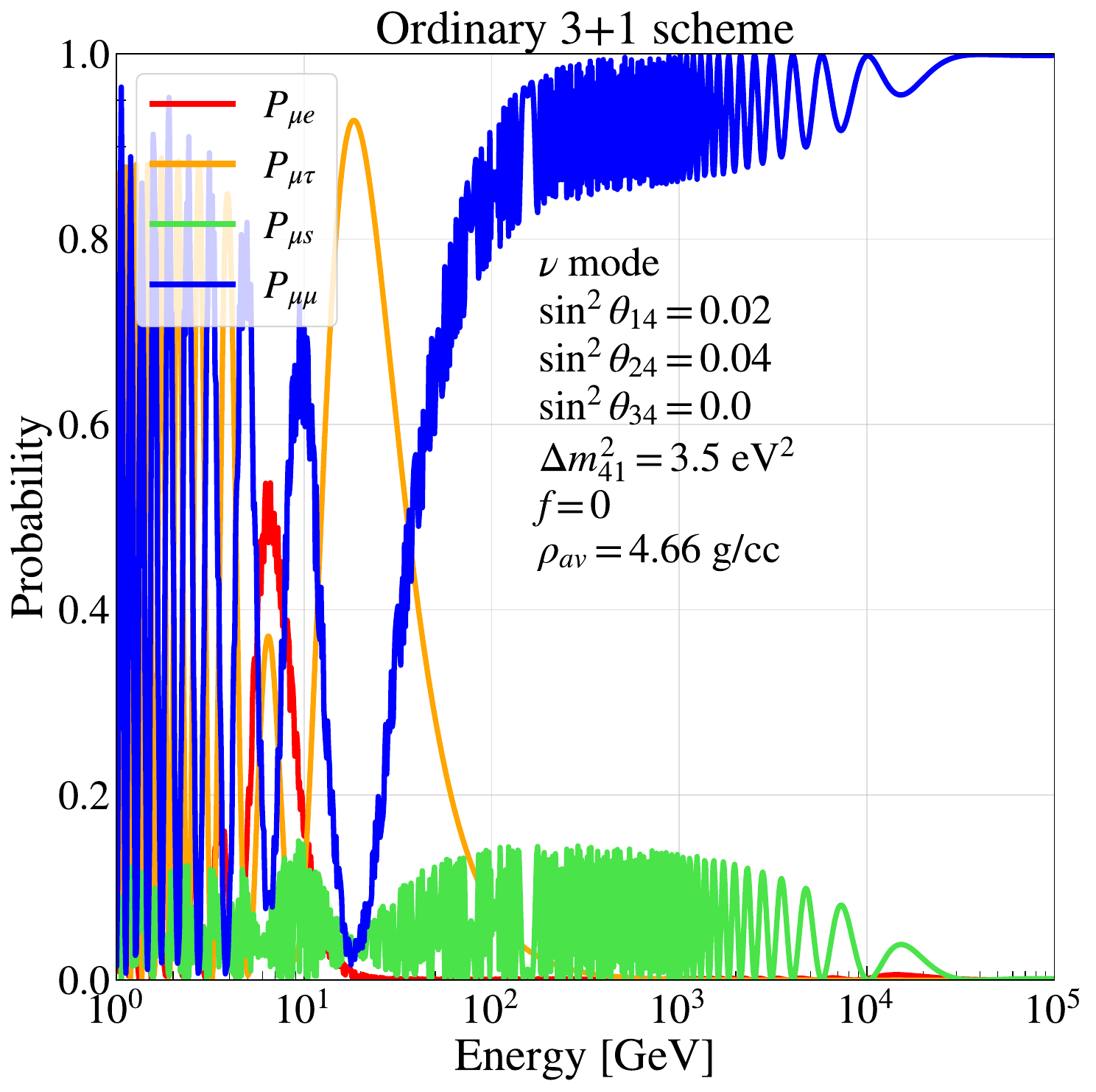}
\includegraphics[height=5.87cm,width=5.87cm]{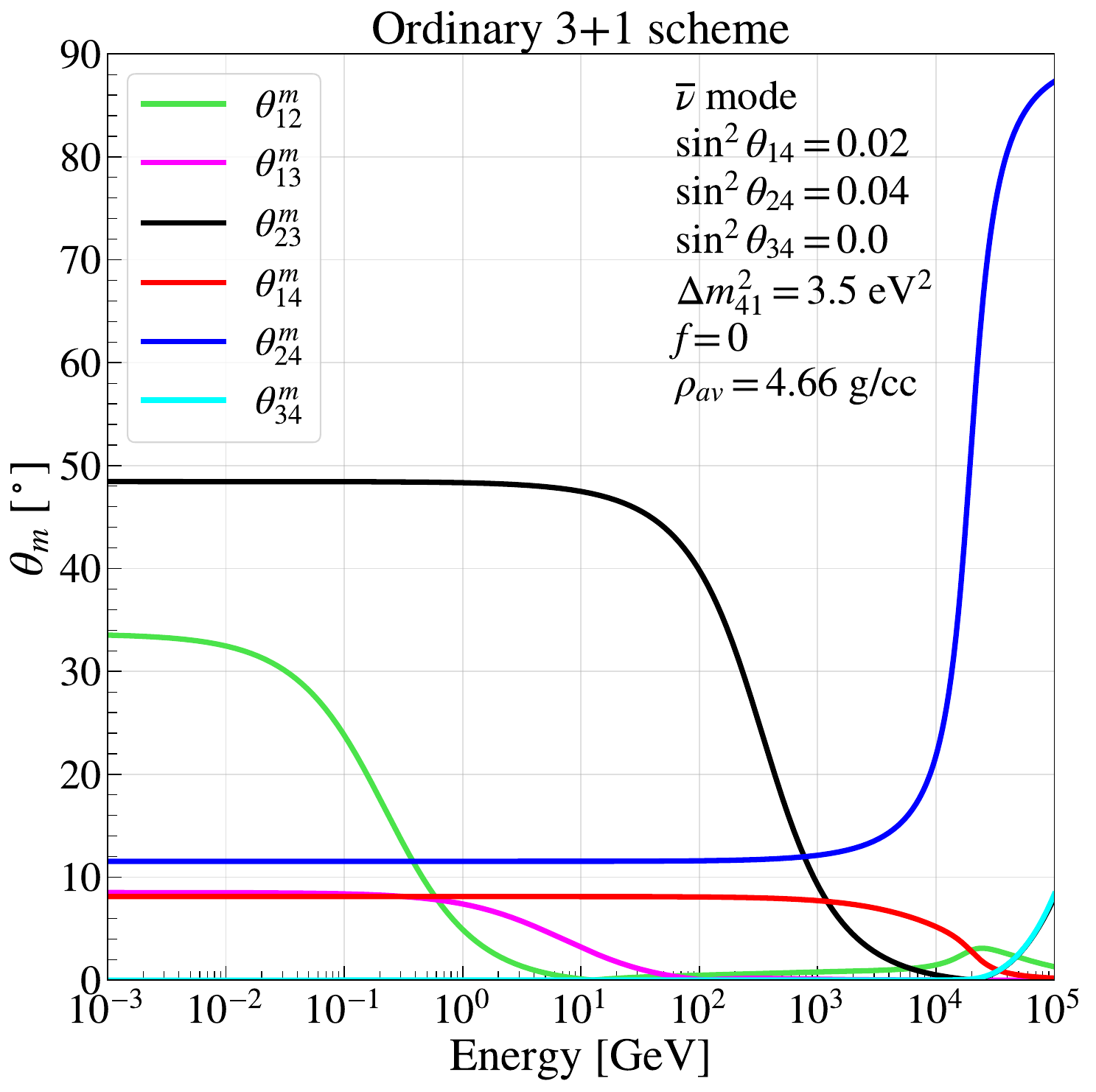}
\includegraphics[height=5.87cm,width=5.87cm]{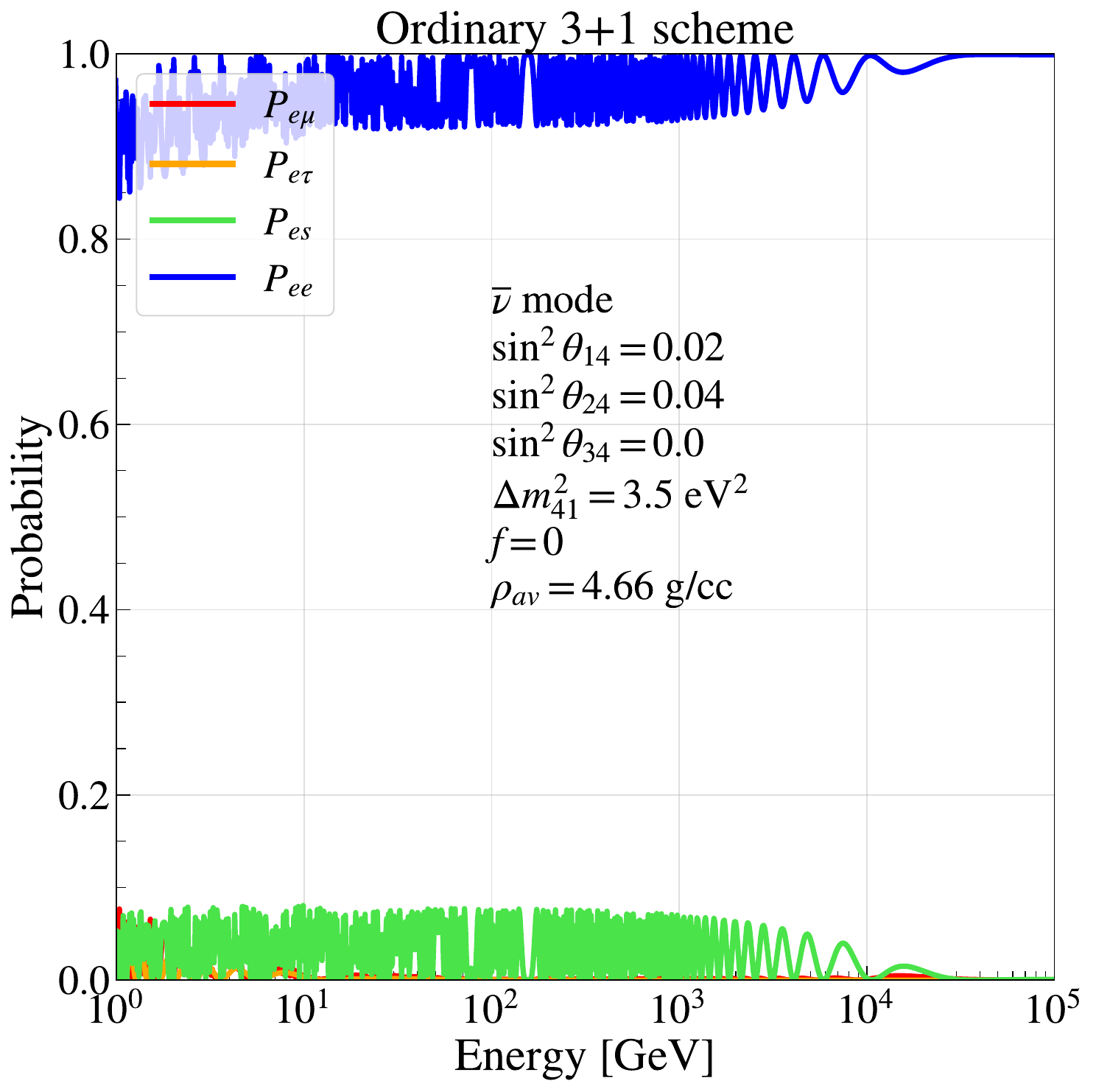}
\includegraphics[height=5.87cm,width=5.87cm]{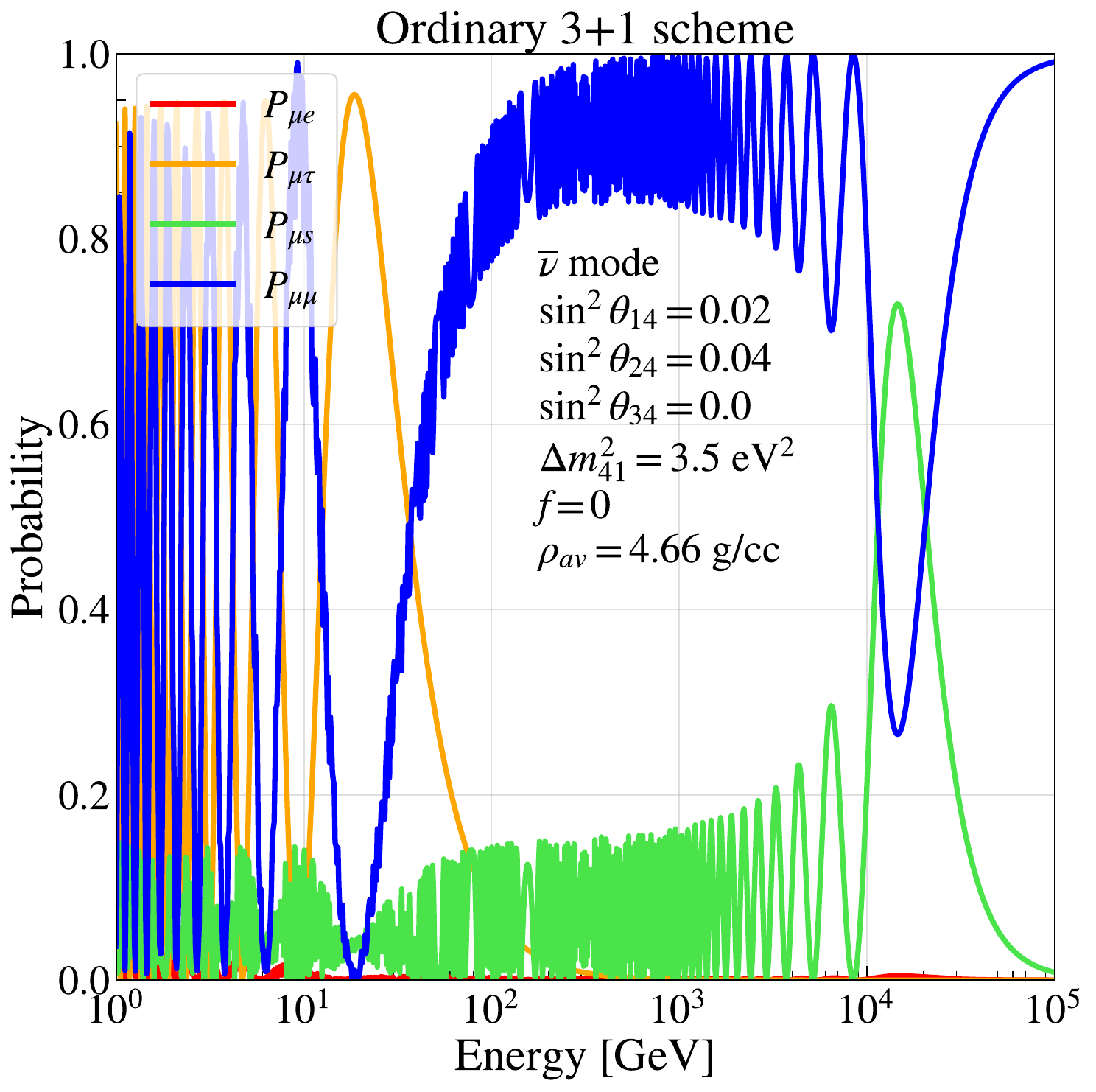}
\vspace*{0.2cm}
\caption{The six panels summarize the properties of the ordinary 3+1 scheme for constant density equal to Earth mantle value.
The upper (lower) panels refer to neutrinos (antineutrinos). The left panels represent the mixing angles in matter 
while the central and right ones depict the oscillation probabilities involving the electron and muon neutrino channel respectively.}
\label{fig_mix_3+1_ordinary}
\end{figure*} 

Figure~\ref{fig_mix_3+1_ordinary} 
refers to the ordinary 3+1 scheme, in which case the new matter potential is absent.
The upper (lower) panels refer to neutrino (antineutrino) channel. The left panels 
report the six mixing angles in matter (three active and three sterile) as a function of the neutrino energy.%
\footnote{Our findings on the behavior of the elements of the mixing 
matrix in matter as a function of the energy are in agreement with those obtained in~\cite{Xing:2018lob}
for the 3-flavor case and in~\cite{Zeng:2022rxm} for the ordinary 3+1 framework.}
The central panels depict the probabilities involving the electron neutrino channel,
while the right panels report the probabilities entailing the muon neutrino channel.
In the left-upper panel we recognize that around $E\simeq 15$ TeV there is a 1-4 resonance 
corresponding to a maximal value of the mixing angle $\theta_{14}^m$ in matter.
This resonance gives rise to enhanced conversion in the $(\nu_e,\nu_s)$ sub-system,
which reveals in the upper-central panel. Such a conversion is of scarse phenomenological relevance
because the flux of atmospheric electron neutrinos is very low, being one order of magnitude
smaller than that of muon neutrinos. We observe that in the range [0.1 TeV - 10 TeV] there are fast oscillations, which 
in average lead to an almost energy-independent suppression of the survival 
probability.  This suppression is regulated by the amplitude of the $\theta_{14}$ mixing angle in vacuum 
($s^2_{14} = 0.02$). The right-upper panel shows the probabilities
involving muon neutrinos, which, as expected do not present any resonant behavior.
Also for muon neutrinos, in the range [0.1 TeV - 10 TeV] there are fast oscillations
regulated by the amplitude of the $\theta_{24}$ mixing angle in vacuum, 
which in the ordinary 3+1 scheme is appreciably large ($s^2_{24} = 0.04$).
In the left-lower panel, corresponding to the antineutrino case, in the region around  $E\simeq 15$ TeV 
there is a 2-4 resonance, corresponding to a maximal value of the mixing angle $\theta_{24}^m$ in matter.
In this case, as confirmed by the lower-central panel, there is no resonance in the $(\nu_e,\nu_s)$ sub-system,
which is regulated by vacuum-like oscillations related to $s^2_{14}=0.02$.
In the right-lower panel we can observe the resonant conversion in the  $(\nu_\mu,\nu_s)$
sub-system, which is of phenomenological interest because the flux of atmospheric neutrinos
is mostly of the muon type. We notice that in the ordinary 3+1 scheme everything goes as expected
from the bare level crossing diagram (left panel of Fig.~\ref{fig_level_crossing}), with a 1-4 resonance in the  neutrino
channel and a 2-4 resonance in the antineutrino channel. The 3-4 resonance expected in the
antineutrino channel is not there simply because we have adopted a zero value for the
mixing angle $\theta_{34}$. This picture confirms  the usefulness of the bare diagram as a powerful
guide in identifying the relevant resonances. 

\begin{figure*}[t!]
\vspace*{0.05cm}
\hspace*{-0.1cm}
\includegraphics[height=5.87cm,width=5.87cm]{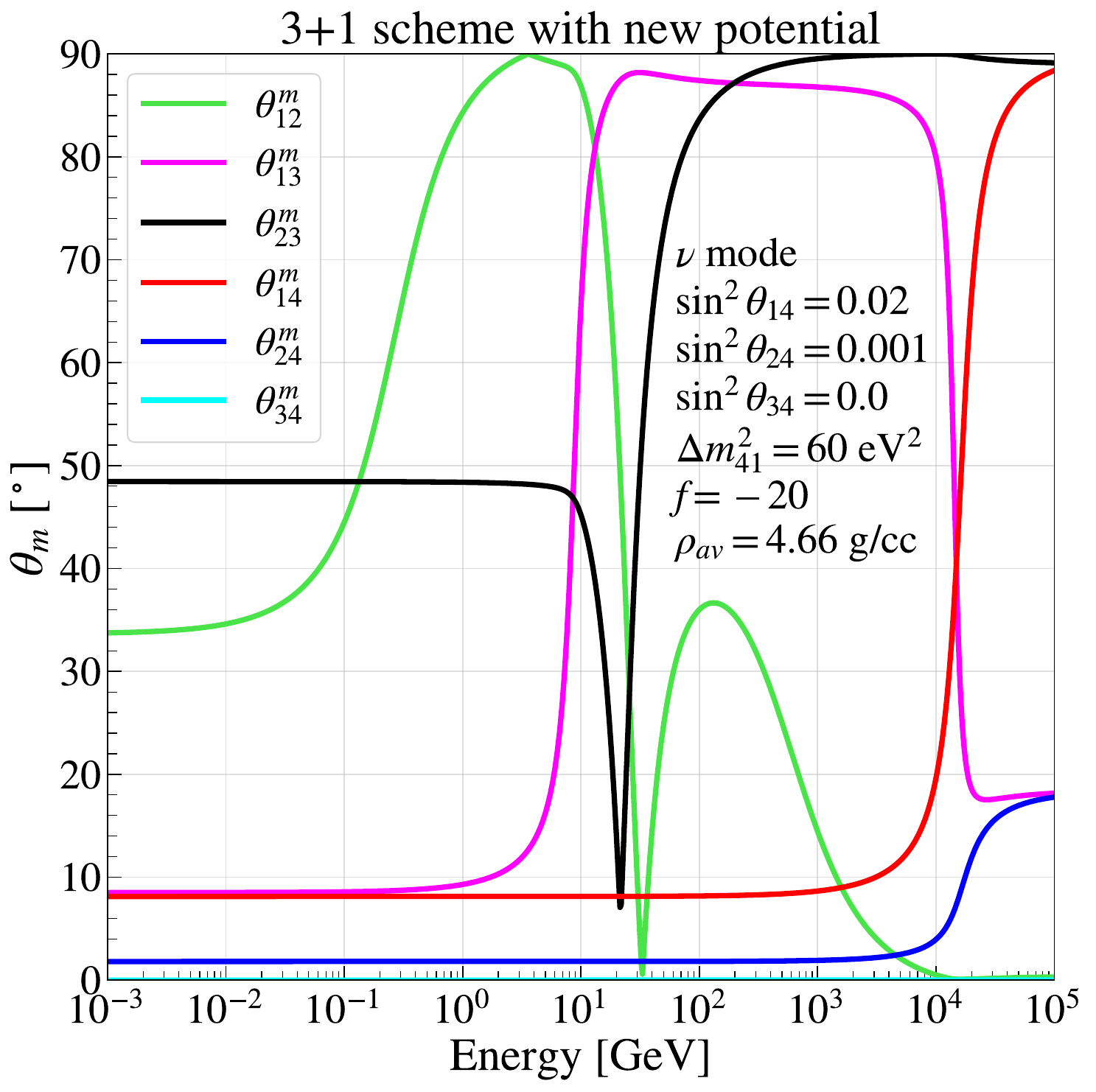}
\includegraphics[height=5.87cm,width=5.87cm]{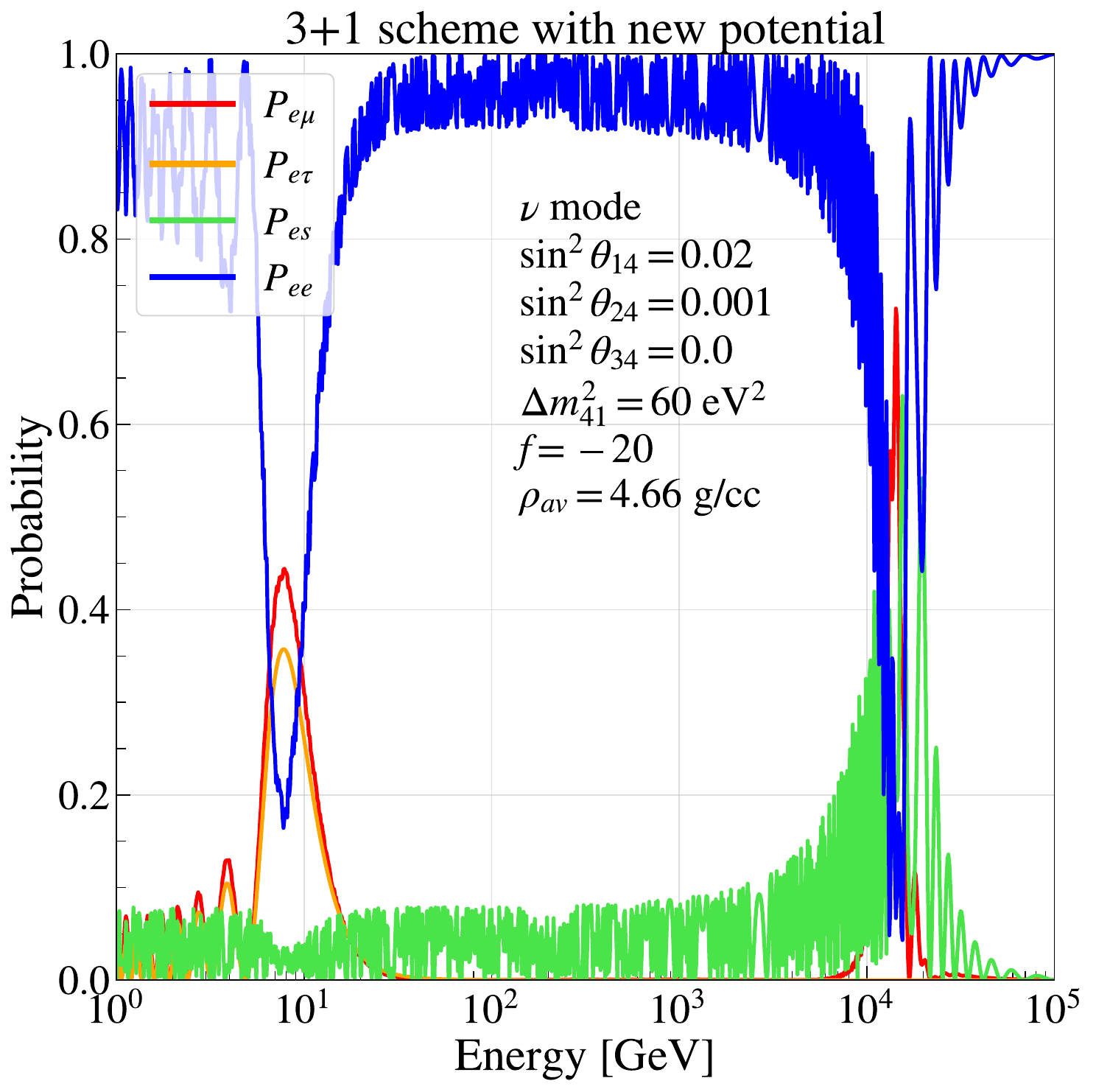}
\includegraphics[height=5.87cm,width=5.87cm]{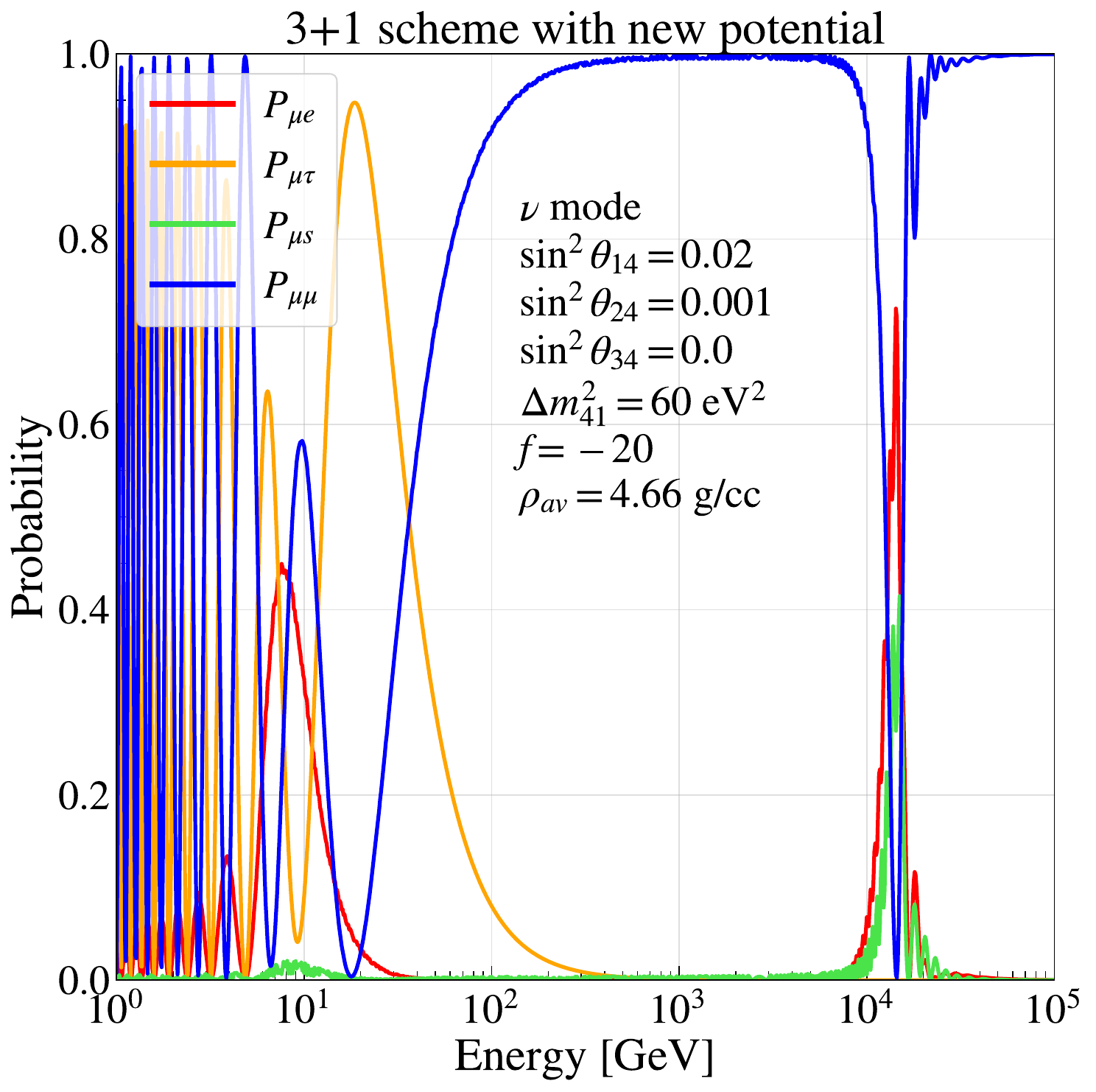}
\includegraphics[height=5.87cm,width=5.87cm]{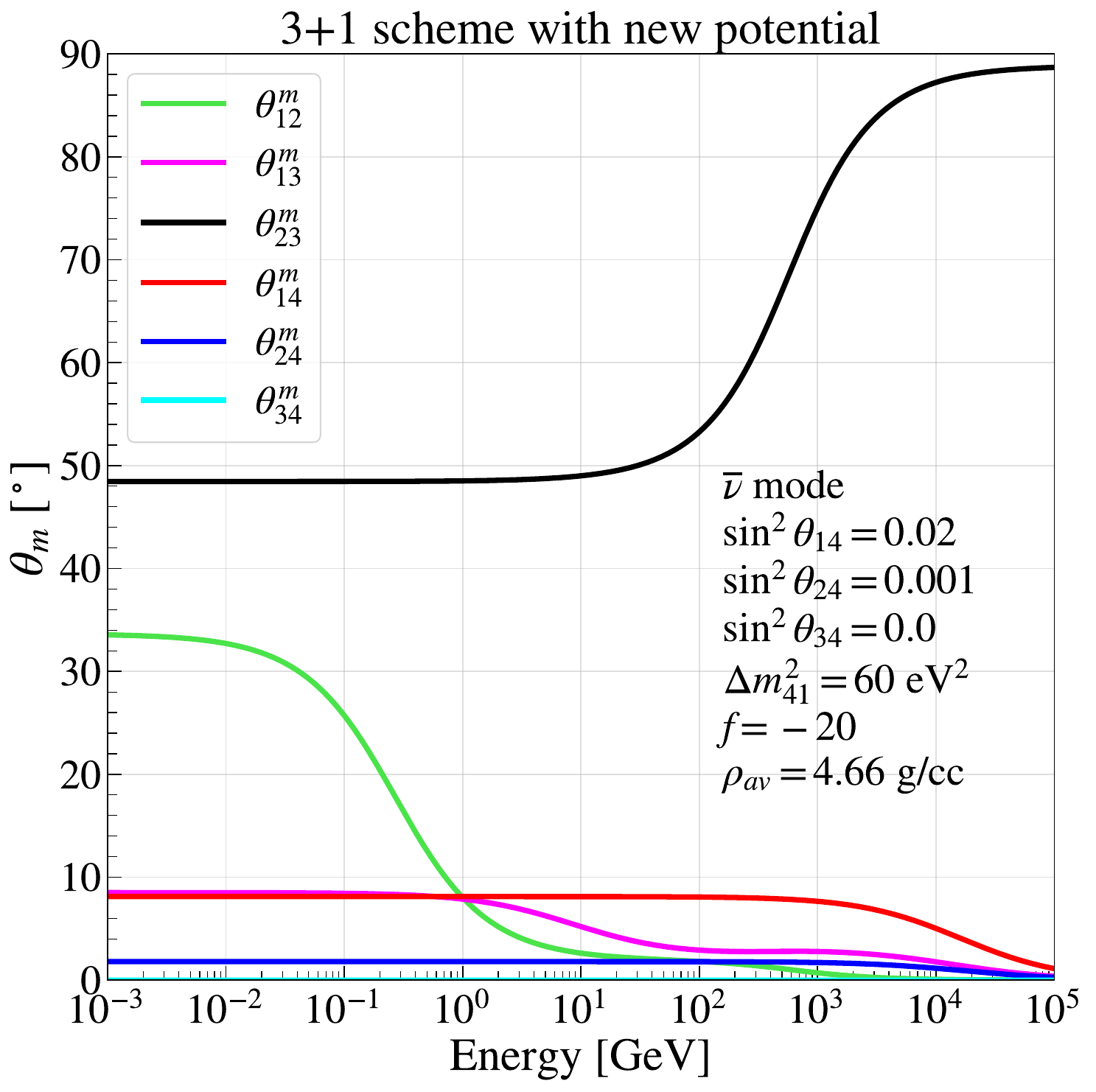}
\includegraphics[height=5.87cm,width=5.87cm]{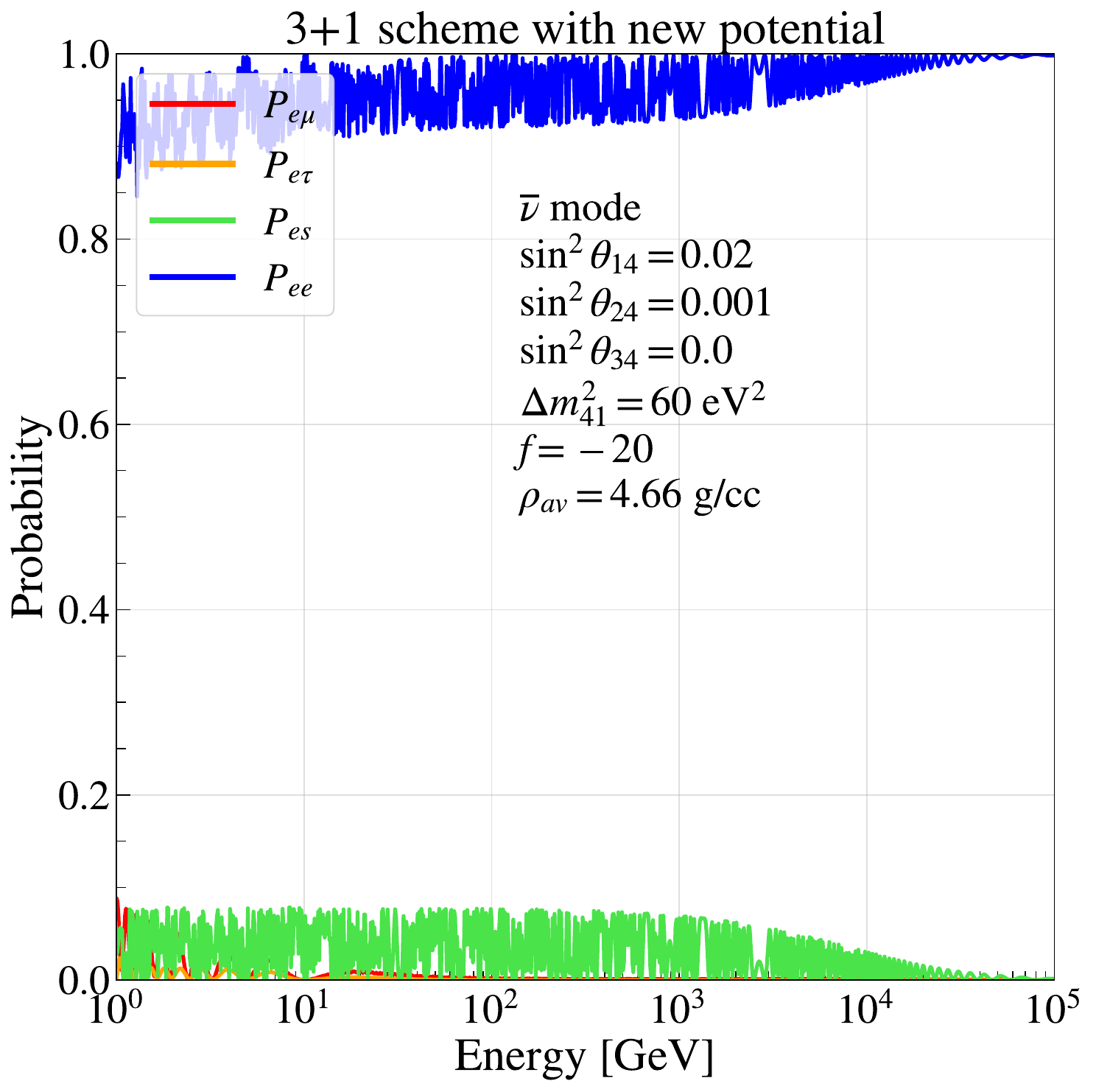}
\includegraphics[height=5.87cm,width=5.87cm]{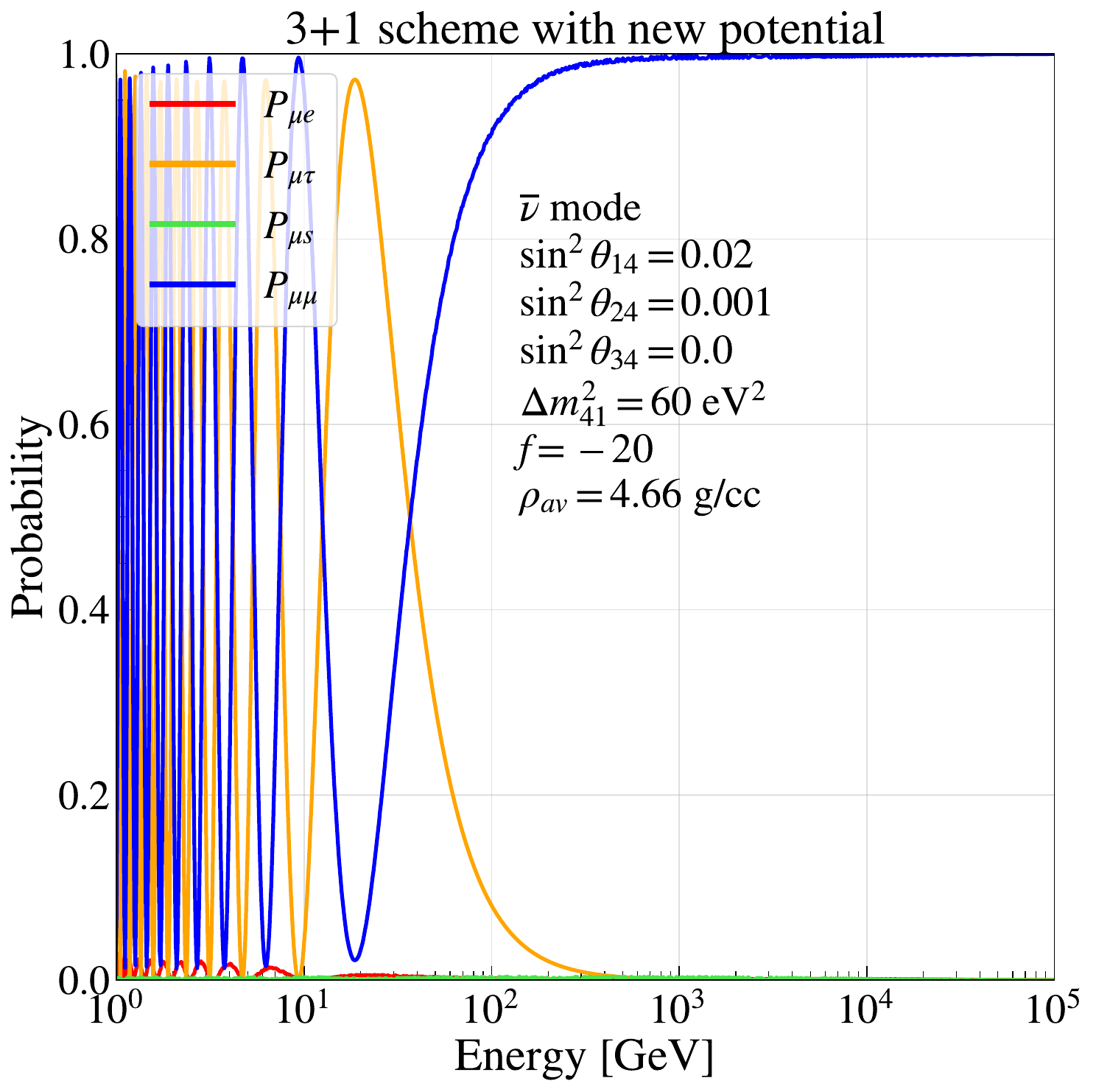}
\vspace*{0.2cm}
\caption{The six panels summarize the results obtained in the 3+1 scheme with the new potential for the benchmark case $f=-20$
for constant density equal to Earth mantle value. 
The upper (lower) panels refer to neutrinos (antineutrinos). 
The left panels represents the mixing angles in matter while 
the central and right ones depict the oscillation probabilities involving the electron and muon neutrino channel respectively.}
\label{fig_mix_3+1_pseudo_-20}
\end{figure*} 
 
Figure~\ref{fig_mix_3+1_pseudo_-20} shows analogous plots to Fig.~\ref{fig_mix_3+1_ordinary}
but refers to the 3+1 model with pseudo-sterile neutrinos.
The upper (lower) panels refer to the neutrino (antineutrino) channel.
 We use the benchmark parameters 
$f = -20$, $\Delta m^2_{41} = 60$ eV$^2$ and the mixing angles $s^2_{14} = 0.02$,
$s^2_{24} = 0.001$ and $\delta_{14} = 0$.
Looking at the mixing angles in matter in the left-upper panel, we observe that around $E\simeq 15$ TeV
there is a 1-4 resonance similarly to the ordinary 3+1 case and a 
2-4 resonance ($\theta_{24}^m$ enhanced albeit not maximal).
This pattern is expected based on the bare crossing diagram.
What is less expected is the 1-3 resonance ($\theta_{13}^m$ maximal), which,
as seen in both the upper-central and upper-right panels, drives amplified 
$(\nu_e,\nu_\mu)$ conversion. We notice that such a resonance is not predicted 
on the basis of the bare crossing diagram, where the only 1-3 crossing intervenes at
low energies and it is the well-known standard 3-flavor resonance occurring at  a few GeV. 
This circumstance suggests that the $(\nu_e,\nu_\mu)$ resonance is not the result of a usual two-level
avoided crossing.  In fact, as we will discuss in the next subsection, 
this resonance is related to a three-level behavior mediated by the pseudo-sterile
neutrinos.
We observe that $P_{e\mu} = P_{\mu e}$ as can be checked confronting the upper-central and upper-right panels,
as expected for the symmetric matter profile inside the Earth.  
In the central-upper and right-upper panels we can also observe the conventional
resonances in the  $(\nu_e,\nu_s)$ and $(\nu_\mu,\nu_s)$ channels.
In the right-upper panel we observe that  in the energy range around 1 TeV,  differently from the ordinary 3+1 scheme, 
the muon neutrino survival probability is  basically equal to one because the fast oscillations induced by the vacuum mixing
angle $\theta_{24}$ are suppressed by the very small value of this parameter 
($s^2_{24} = 0.001$). In the lower panels, representing
the antineutrino channel, we see that there is no resonance and that the behavior
of the probabilities is similar to the 3-flavor scheme apart from the fast oscillating
flat suppression of the electron survival probability in the lower central panel, 
which is a vacuum effect driven by the relatively large $s^2_{14}= 0.02$.

\subsection{Three-level dynamics as the origin of the resonance in the ($\nu_e, \nu_\mu$) sub-system}
\label{SubSec:IceCube_newres}

In the previous subsection we have seen that in the presence of the new potential a 
resonance in the ($\nu_e, \nu_\mu$) sub-system takes place. Here, we try to gain some insight
in this unexpected phenomenon. With this purpose, we set $\theta_{34}=0$ as in the previous subsection.
In the high-energy limit the 3-flavor parameters become irrelevant and it is easy to check that 
the Hamiltonian has the third row and third column exactly equal to zero. This means that
the $\nu_\tau$ state is unaffected by flavor conversion phenomena and remains unaltered 
during the evolution. Hence, we can extract from the four-flavor Hamiltonian the $3\times3$ sub-matrix 
with $(1,2,4)$ indexes, which regulates the dynamics of the 3-flavor sub-system $(\nu_e, \nu_\mu, \nu_s)$.
Neglecting the CP phase $\delta_{14}$, we obtain the effective 3-flavor Hamiltonian
\begin{eqnarray} \label{eq:H_3_ems_v1}
\arraycolsep=3pt
\medmuskip = 1mu
     H_{3\nu}^{(e \mu s)}  = 
     R_{24} R_{14} K^\prime R_{14}^T R_{24}^T + V^\prime \simeq
      \begin{bmatrix}
	 V_{\rm{CC}} + k s^2_{14}  & k   s_{14} s_{24}  & k  s_{14} 	\\
	\dagger & k  s_{24}^2 &  k  s_{24}  
	\\
	\dagger & \dagger
	&\bar f V_{CC} + k 
    \end{bmatrix}, \,
\end{eqnarray}
where $K^\prime = \mathrm{diag} (0,0, k\equiv k_{41})$ and $V^\prime = \mathrm{diag} (V_{CC},0, \bar f V_{CC} = V_S - V_{NC})$ 
are the 3-dimensional diagonal matrixes corresponding to the 4-dimensional matrixes
$K$ and $V$ from where we have omitted the third (equal to zero) entry.
In the second equality in Eq.~(\ref{eq:H_3_ems_v1}) we have taken the limit of small mixing angles
 which is of phenomenological interest.
The diagonalization of such an Hamiltonian in general will require the product of three orthogonal rotations 
defining the three mixing angles in matter $\theta_{24}^m$, $\theta_{14}^m$, $\theta_{12}^m$,
and can be cast in the form 
\begin{equation}
\label{eq:U_3nu_m}
U_{3\nu}^{m} =   R_{24}^m R_{14}^m R_{12}^m\,, 
\end{equation} 
where we have respected the same order of the three rotations in vacuum. 
Now, we observe that if  the mixing angle $\theta_{14} = 0$, the Hamiltonian  in Eq.~(\ref{eq:H_3_ems_v1})
has a block-diagonal form and is diagonalized by a simple rotation 
in the 2-4 plane, with $U_{3\nu}^{m} = R_{24}^m$ involving only the mixing angle $\theta_{24}^m$.
Physically, it happens that the $\nu_e$ flavor state evolves independently of
the $(\nu_\mu, \nu_s)$ sub-system. This system 
presents a resonance for  $[H(4,4) - H(2,2)] = 0$, i.e.
$k \simeq - \bar f V_{CC} \simeq -0.5 f V_{CC}$, which occurs in the neutrino
(antineutrino) channel for $f<0$ ($f>0$). Similarly, 
if the mixing angle $\theta_{24}=0$, the Hamiltonian has again
a block-diagonal form and is diagonalized by a simple rotation in the 1-4 plane
with $U_{3\nu}^{m} = R_{14}^m$ depending on the mixing angle $\theta_{14}^m$.
In this case, the evolution of the $\nu_\mu$ flavor state is decoupled from the   
$(\nu_e, \nu_s)$ sub-system, which presents a resonance for $[H(4,4) - H(1,1)] = 0$, 
i.e. $k (1-s^2_{14}) \simeq (1- \bar f) V_{CC}$, which for the values of the
parameters we are considering (small $s^2_{14}$ and large $\bar f$), 
presents only a minor shift with respect to the resonance in the $(\nu_\mu, \nu_s)$ sub-system.

In the most general case in which both mixing angles $\theta_{14}$ and $\theta_{24}$
are non-zero, the Hamiltonian in Eq.~(\ref{eq:H_3_ems_v1}) has not a block-diagonal structure anymore.
It is clear that, in order to obtain its diagonalization, 
one would need all the three mixing angles in matter, and in particular
a non-zero 1-2 mixing angle $\theta_{12}^m$ will emerge. 
What is less immediate to understand is that the 1-2 system may resonate,
with $\theta_{12}^m$ becoming maximal. With the purpose of shedding light on this point,
it is useful to come back to the rotated NSI-like Hamiltonian. Also in this case, taking the case
$\theta_{34}=0$, we can extract the $3\times 3$ submatrix,
\begin{eqnarray} \label{eq:Hdyn_3flavor}
\arraycolsep=3pt
\medmuskip = 1mu
    \bar H_{3\nu}^{(e \mu s)}  =  
   V_{\rm{CC}} \!   \begin{bmatrix}
	1 + \bar f_\e s^2_{14}  & \bar f_\e  \tilde s_{14} s_{24}  & (1- \bar f_\e)  \tilde s_{14}	\\
	\dagger &\bar f_\e s_{24}^2 & -\bar f_\e   s_{24}\, \\
	\dagger & \dagger & \bar f_\e + \frac{k_{41}}{V_{\rm{CC}}} 
	\end{bmatrix}
	=  
       V_{\rm{CC}} \!   \begin{bmatrix}
	1 + \varepsilon_{ee} & \varepsilon_{e\mu}  & \varepsilon_{es}	\\
	\dagger & \varepsilon_{\mu\mu} & \varepsilon_{\mu s} \\
	\dagger & \dagger & \delta(E) 
	\end{bmatrix}    
\end{eqnarray}
where for compactness we  have introduced
$\delta(E) = \bar f_\e + \frac{k_{41}}{V_{\rm{CC}}}$.
We see that in the 4-flavor dynamics the off-diagonal NSI-like couplings  $\varepsilon_{es}$ 
and  $\varepsilon_{\mu s}$  appear which involve one active state and the sterile one.
Now, we can note that the $\nu_\mu \to \nu_e$ transitions can be mediated
by the coupling $\varepsilon_{e\mu}$ or via the product $\varepsilon_{es}\varepsilon_{\mu s}$.  
In the first case the coupling appears directly in the Hamiltonian, while in the second case it 
represents a second-order effect.
In general, in the model we are considering,
both direct and indirect couplings are different from zero, and the first order and second order mechanisms 
will act simultaneously and interfere. However, one can note that
the second order effects are amplified by the new potential being 
$\varepsilon_{es}\varepsilon_{\mu s} \simeq \bar f_\e^2 s_{14} s_{24} $,  and dominate over the 
first order effects. 
These observations allow us to enucleate the core dynamics of the resonant $(\nu_e, \nu_\mu)$
conversion, which is captured by the simplified Hamiltonian 
\begin{eqnarray} \label{eq:H_Toy}
\arraycolsep=3pt
\medmuskip = 1mu
    \bar H_{3\nu}^{\prime (e \mu s)}  =  
       V_{\rm{CC}} \!   \begin{bmatrix}
	1  & 0  & \varepsilon_{es}	\\
	0 & 0 & \varepsilon_{\mu s}  \\
	\dagger & \dagger & \delta(E) 
	\end{bmatrix} 
\end{eqnarray}
 where we have set to zero all the other couplings and for simplicity we consider real couplings (neglecting the CP-phases).
  Of course the neglegted couplings will modify the picture but are less fundamental. 
The system has two natural resonances corresponding to  $\delta = 0$ and 
 $\delta =1$. When $\varepsilon_{es}=0$ the ($\nu_\mu, \nu_s$) sub-system resonates for
 $\delta =0$, while when $\varepsilon_{\mu s}=0$ the  ($\nu_e, \nu_s$) sub-system  
 resonates for $\delta = 1$. When both couplings are non-zero as implied in our model,
 the second order effects, far from the regions of the two well localized resonances 
 [($0 \lesssim \delta \lesssim 1$) for the internal region and ($\delta \lesssim 0$, $\delta \gtrsim 1$) for the external region],
are expected to provide an effective NSI-like coupling $\propto \varepsilon_{es}\varepsilon_{\mu s} /\delta$. 
This perturbative approach, for small values of the couplings is expected to hold also
in between the two resonances, where $\delta \simeq 1/2$, which remain well separated.
For increasing value of the couplings these corrections will blow up in the region
between the resonances making necessary a full 3-level treatment. 
One expects that the two resonances will get coupled for increasing 
 values of the off-diagonal couplings giving rise to a more complex resonant behavior. 
 In order to illustrate this point, in our full model we fix $s^2_{14} = s^2_{24}=0.005$ 
 and consider the two values of $f = -4$ and $f=-20$. These two choices make our full model 
 very close to the toy model described by the Hamiltionian in Eq.~(\ref{eq:H_Toy}) and correspond
 to two values of $\varepsilon_{es} \varepsilon_{\mu s} \simeq 0.07$ and $0.45$ respectively.
 In the first (second) case, one expects a small (large) perturbation of the unperturbed Hamiltonian.

\begin{figure*}[t!]
\vspace*{0.05cm}
\hspace*{-0.1cm}
\includegraphics[height=5.87cm,width=5.87cm]{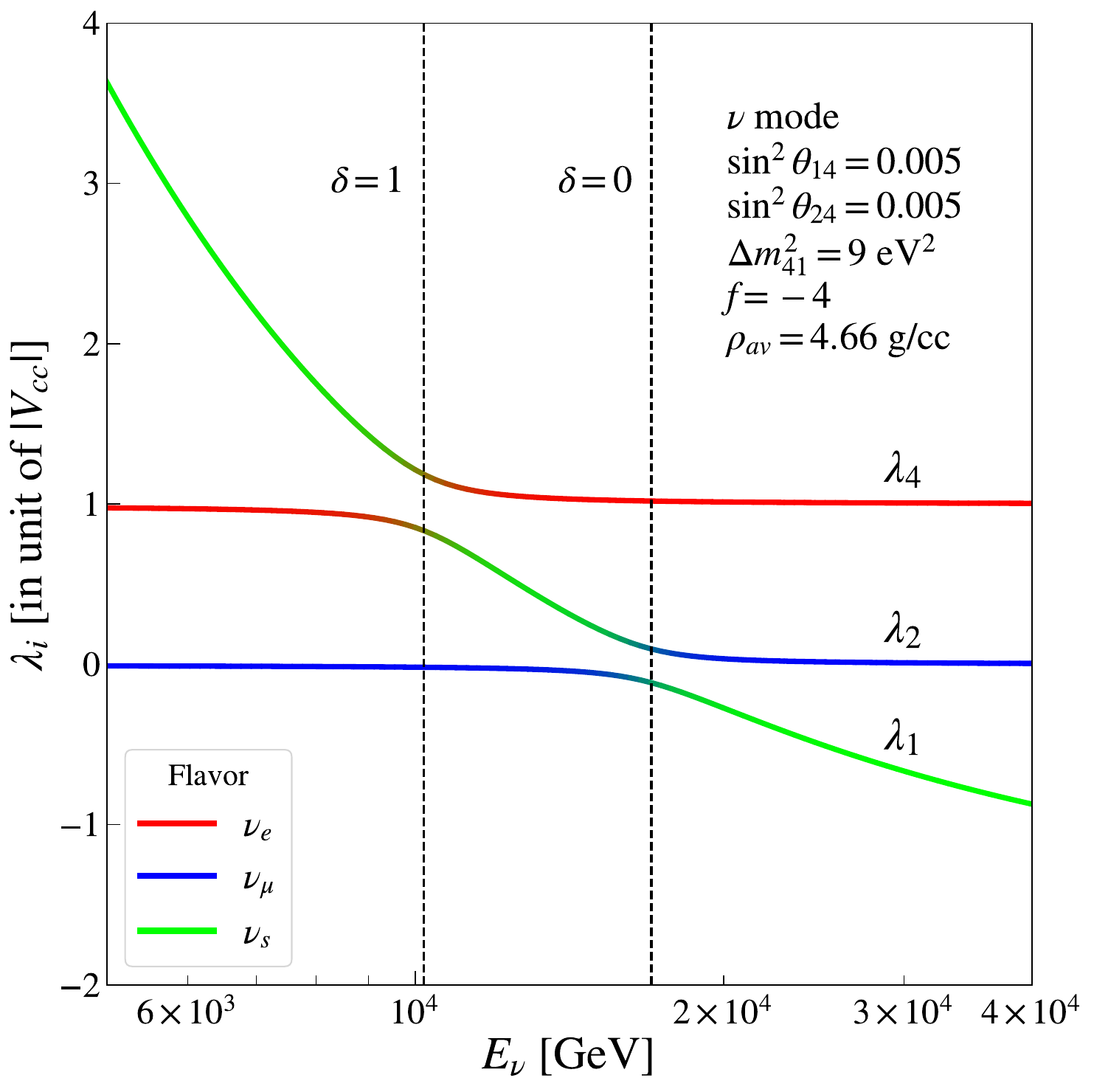}
\includegraphics[height=5.87cm,width=5.87cm]{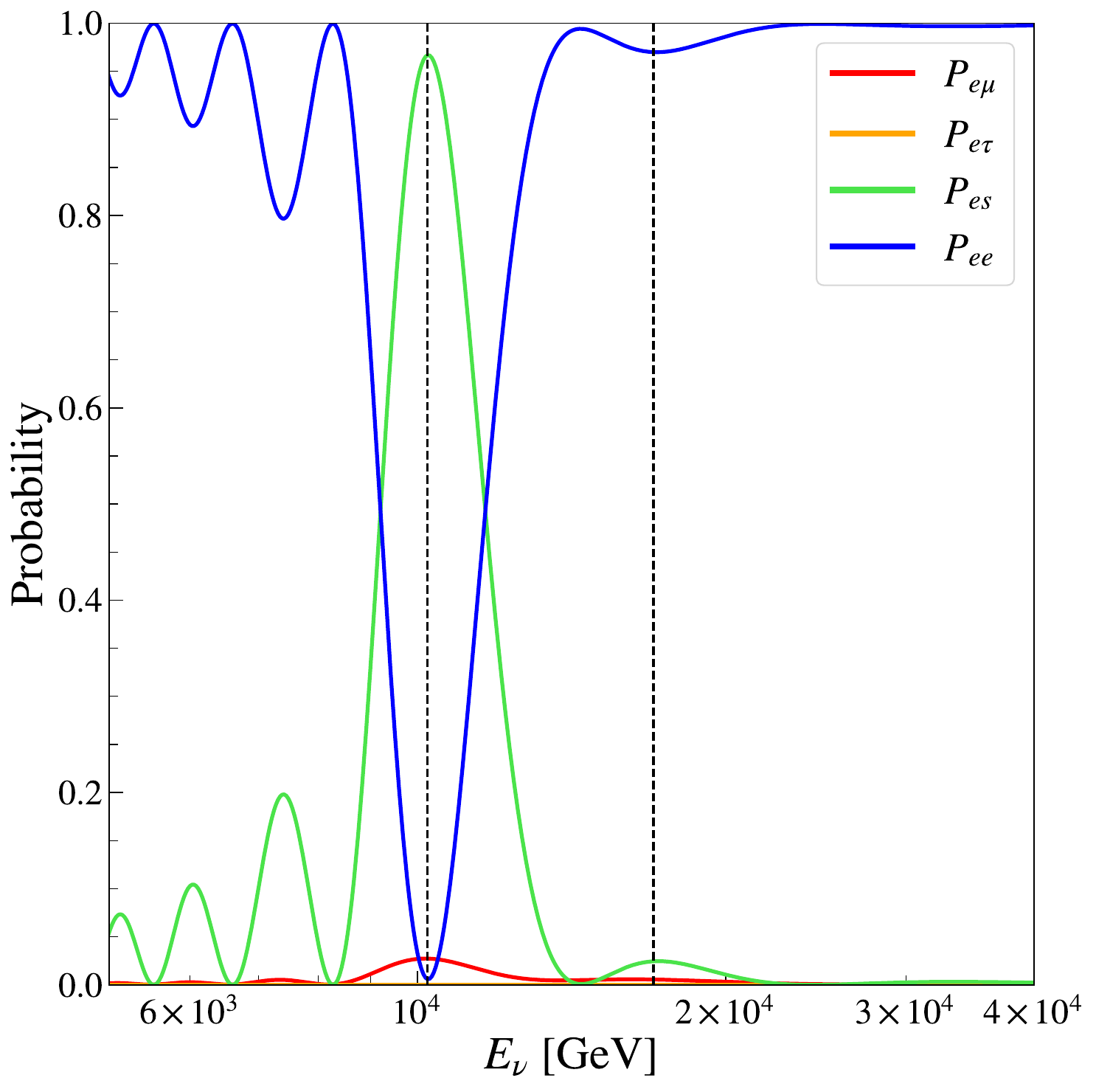}
\includegraphics[height=5.87cm,width=5.87cm]{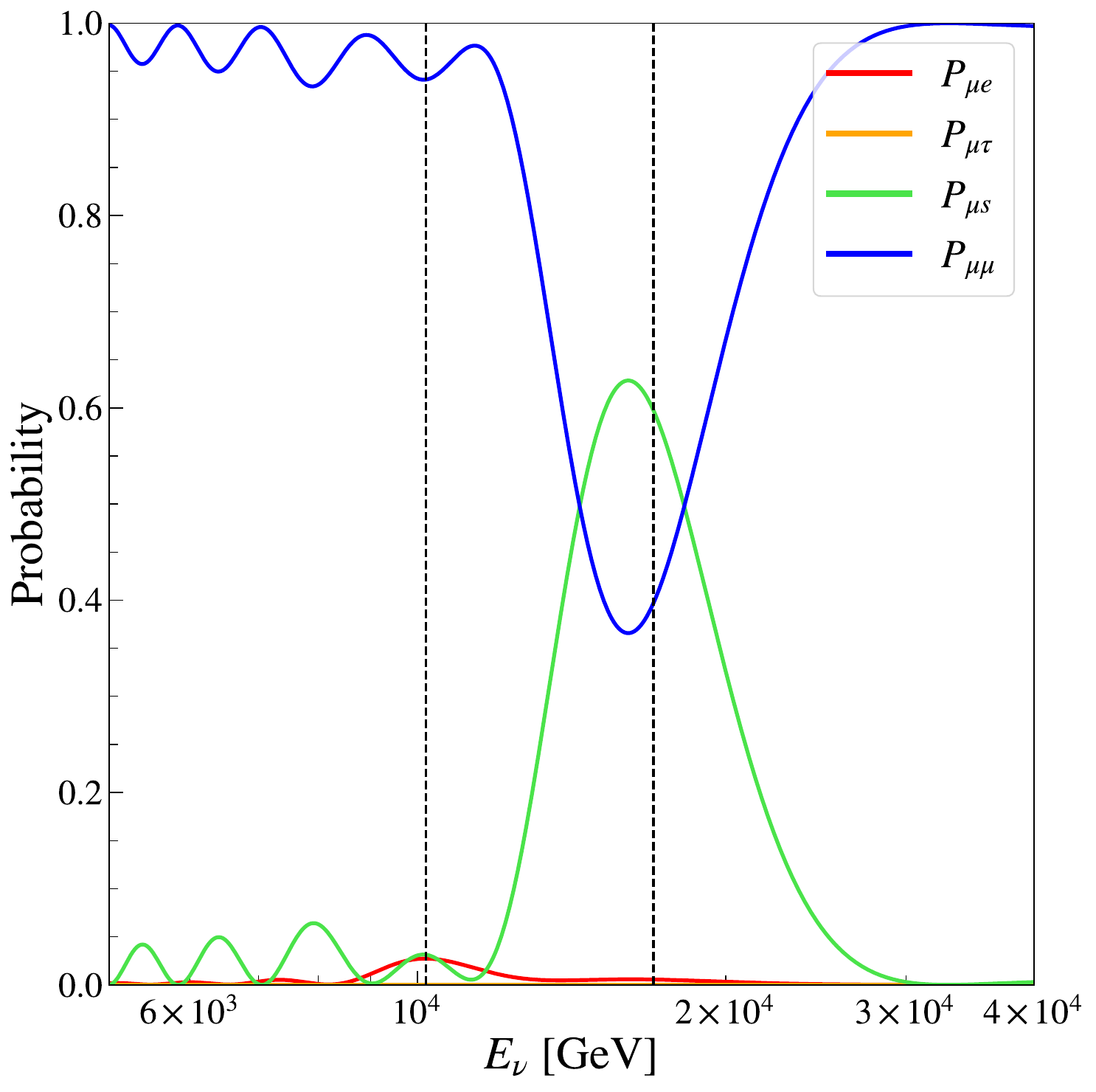}
\includegraphics[height=5.87cm,width=5.87cm]{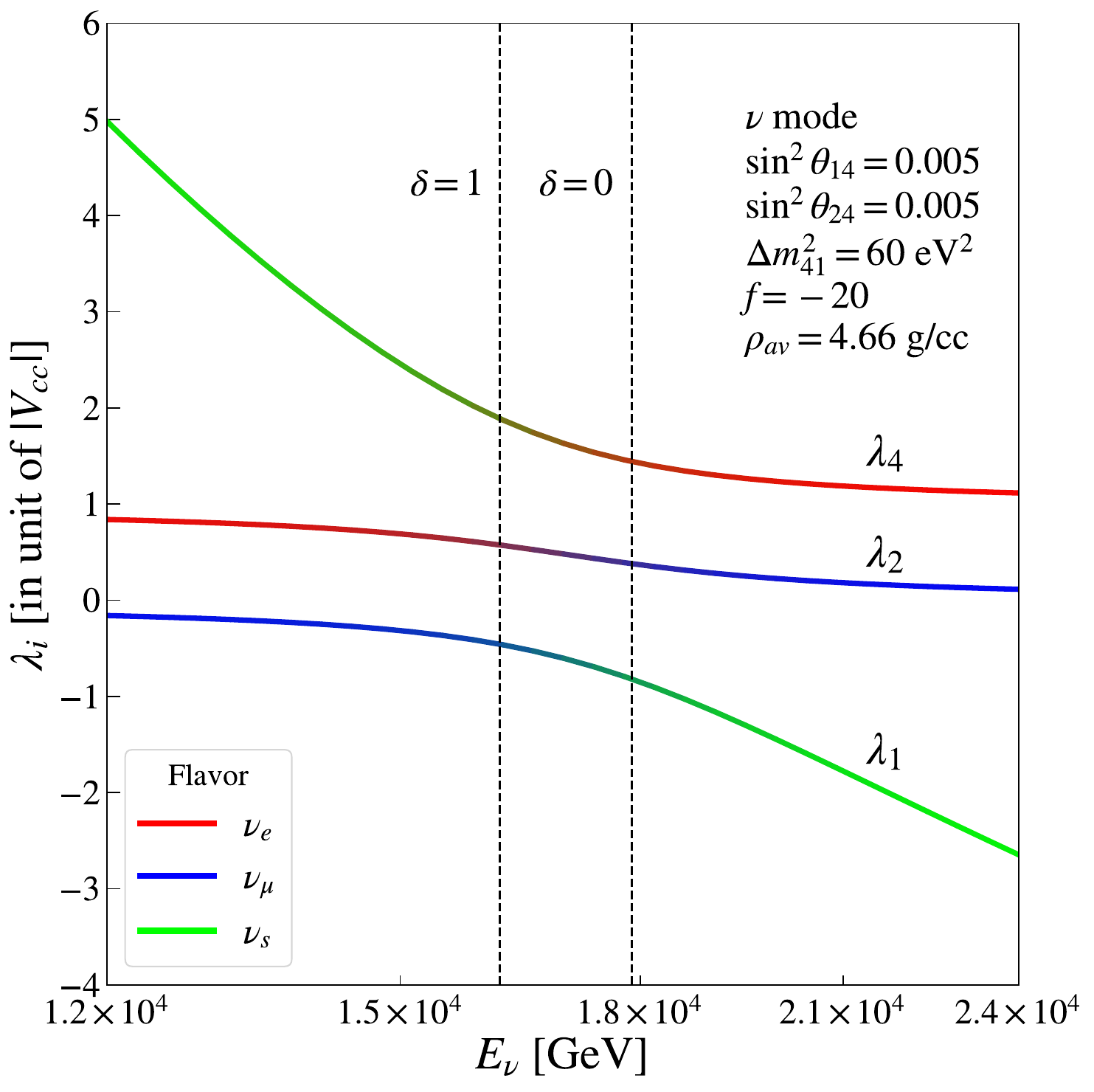}
\includegraphics[height=5.87cm,width=5.87cm]{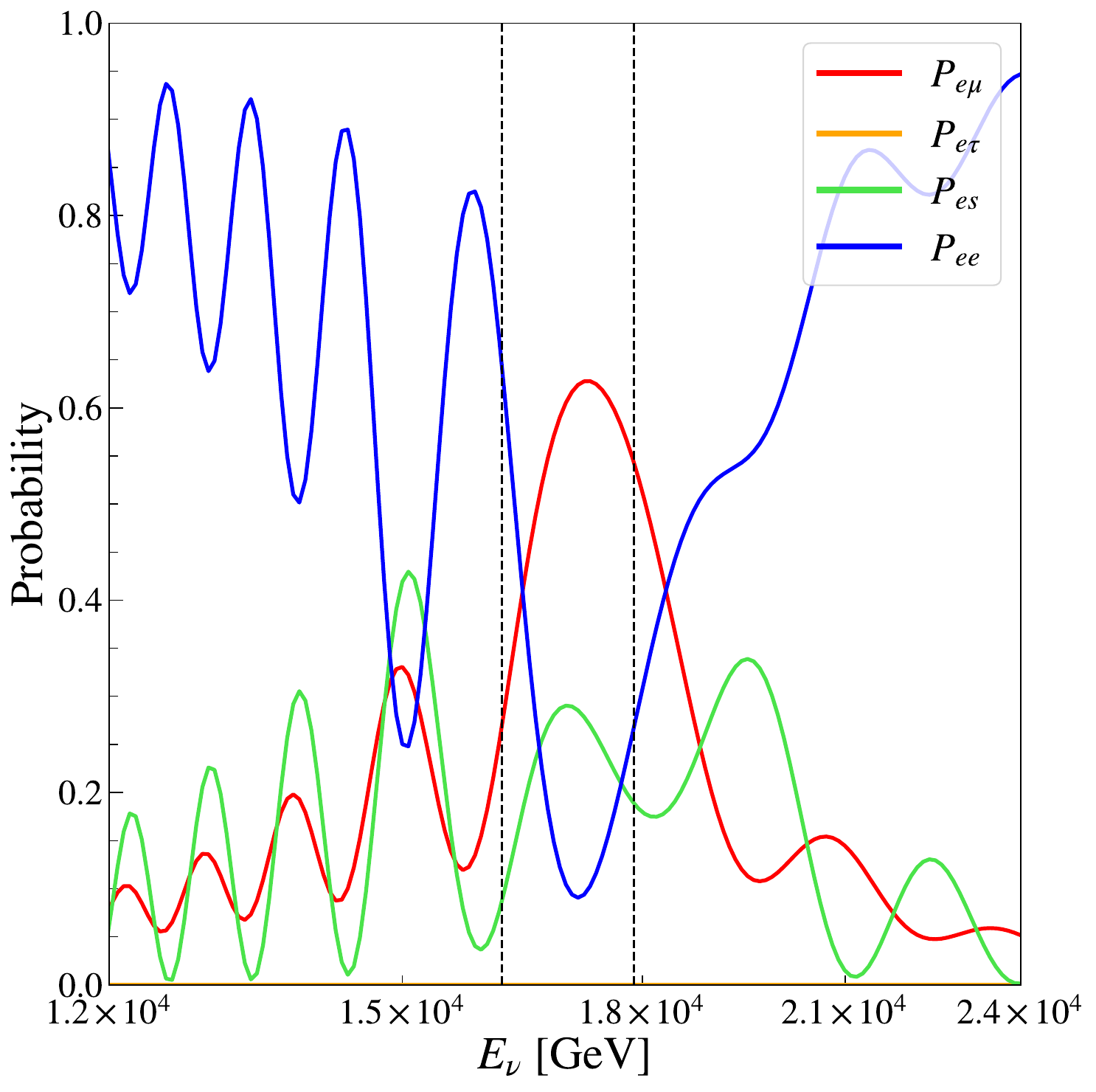}
\includegraphics[height=5.87cm,width=5.87cm]{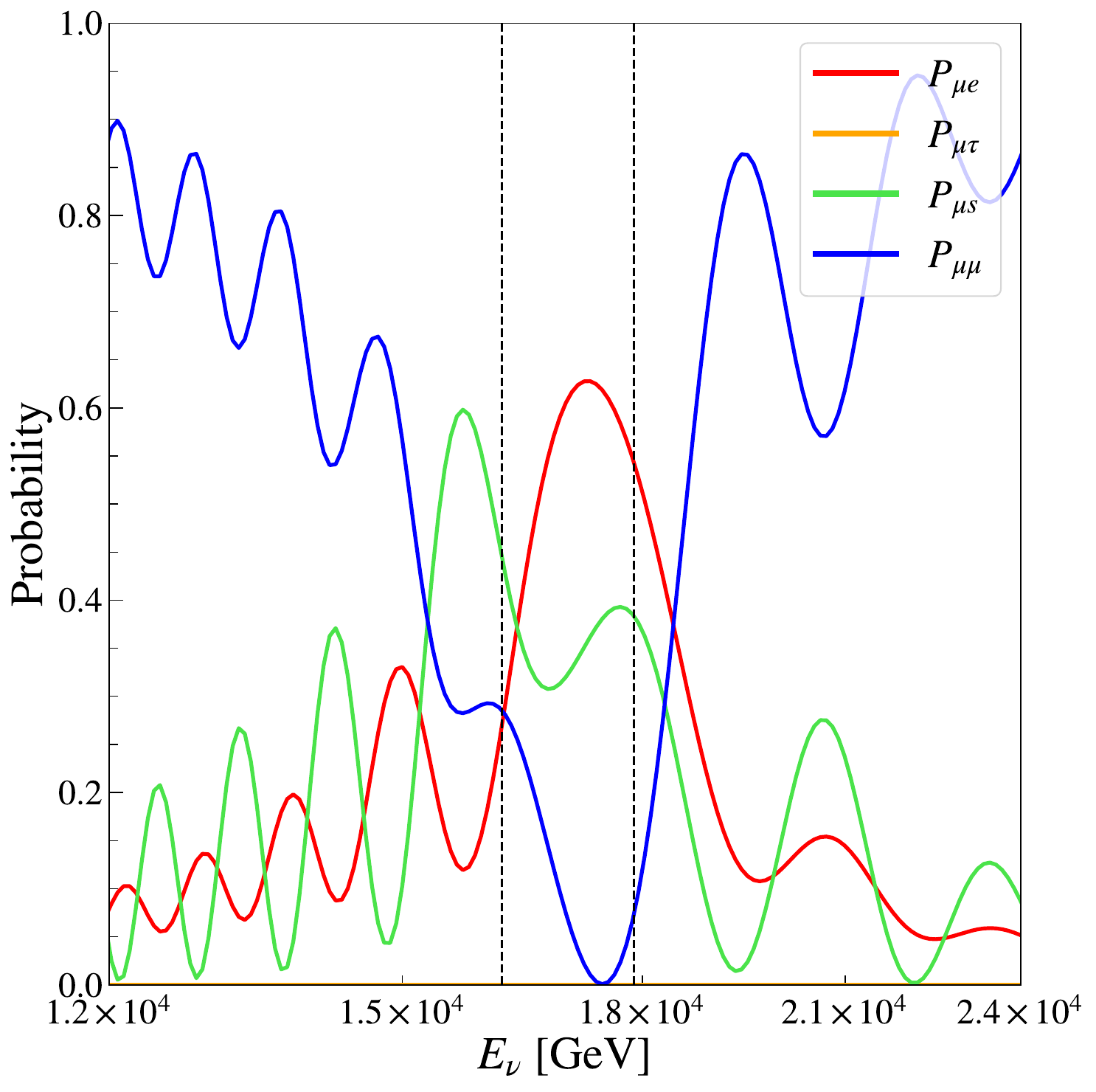}
\vspace*{0.2cm}
\caption{The six panels summarize the behavior of the model in the resonance region
highlighting the transition of the system from a two-level regime to a three-level one.
In the upper (lower) panels we have taken two different values of the $f$ parameter (respectively $f=-4$ and $f=-20$).  
The left panels report the behavior of the eigenvalues. The central (right) panels represent the
oscillation probabilities relevant for the electron (muon) neutrino channel.}
\label{fig_multilevel_3+1_ordinary}
\end{figure*} 

Figure~\ref{fig_multilevel_3+1_ordinary} summarizes the behavior of the model in the resonance region
highlighting the transition of the system from a regime in which the two resonances are decoupled to one
in which they get coupled. In the upper (lower) panels we have taken two different values of the $f$ parameter (respectively $f=-4$ and $f=-20$).  
The left panels report the energy eigenvalues in units of $|V_{CC}|$. 
The varying color along the curves represents the flavor composition of the eigenvectors in matter $\nu_i^m$ $(i =1,2,4)$. 
Thanks to such colors we can clearly visualize the relevant flavor transitions occurring in the resonance region. 
For the smaller value of $f=-4$ (left-upper panel)
we observe the presence of two distinct resonances corresponding to two separate avoided crossings,
 the first one involving the ($\nu_e, \nu_s$) system at $\delta = 1$ and
the second one involving the ($\nu_\mu, \nu_s$) system at $\delta = 0$. 
The central and right panels represent the
oscillation probabilities relevant for the electron and  muon neutrino channels respectively.
The upper-central panel and the upper-right
panels confirm that in the case of $f=-4$ there are only two resonances involving ($\nu_e,\nu_s$) and  ($\nu_\mu,\nu_s$) transitions.
For large values of $f=-20$ (left-lower panel) the two avoided crossings merge to form a single 
triple-avoided crossing involving the full ($\nu_e, \nu_\mu, \nu_s$) system. From the colors of the curves, we see that 
in the region between $\delta = 1$ and $\delta =0$, each eigenvector in matter is a superposition of at least
two flavors, and we can deduce that the transitions will involve all the three sub-systems, and in particular the
 ($\nu_e, \nu_\mu$) one. This is confirmed by the lower-central and 
lower-right panels, where we see that a resonance in the  ($\nu_e, \nu_\mu$) channel coexists with the two resonances in the ($\nu_e, \nu_s$) and
 ($\nu_\mu, \nu_s$) systems.  Therefore, for increasing values of the off-diagonal couplings $\varepsilon_{es}$ and $\varepsilon_{\mu s}$,
 the system undergoes a transition from a regime where it is possible to describe it in terms of two separate 2-level systems
 to a regime where a genuine 3-level dynamics is at play. 
 Finally, we note that the differences between the eigenvalues are related to the scale of the
  matter potential $V_{CC}$. Such large differences correspond to large phase factors
   $\phi_{ij} = (\lambda_i - \lambda_j) L \equiv \tilde{\Delta} m^2_{ij}L/2E$ ($L$ being the
    baseline and with $\tilde{\Delta} m^2_{ij}$ we have indicated the mass-squared differences in matter)
 in the transition probabilities, which reveal as a fast oscillating behavior evidenced in the central and right panels of 
 Fig.~\ref{fig_multilevel_3+1_ordinary}. The full analytical treatment of the 3-level system is
 beyond the scope of the present work and can  be performed using for example the method recently introduced in~\cite{Denton:2019ovn}. 
 However, the resulting formulas are not transparent since one cannot avoid to resolve a cubic secular equation
 in order to calculate the analytical expressions of the eigenvalues.

A closing technical remark is in order. In the simplified 3-flavor framework
we have considered in this sub-section, we have seen that a matter 1-2 mixing angle 
$\theta_{12}^m$ emerges naturally. In the 4-flavor results presented in sub-section~\ref{SubSec:IceCube_numdiag}
a non-zero 1-3 mixing appears (see Fig.~\ref{fig_mix_3+1_pseudo_-20}). The explanation for this apparent inconsistency 
is the following. In the full 4-flavor system the $\nu_\tau$ flavor is present, although
at high energies, for $\theta_{34}=0$ it does not participate to the flavor conversion.
If one considers the flavor composition of the eigenvectors in matter $\nu_i^m$,
one would find that at energies below the resonance, at around $E \simeq 1$ TeV,
one has that the second eigenstate coincides with the $\tau$ flavor state ($\nu_2^m \simeq \nu_\tau$).
This can be checked by considering the expressions of the elements of 
the mixing matrix in matter $U^m$ and is related to the fact that neutrinos (not antineutrinos)
undergo the two standard resonances (solar and atmospheric) at low energies.
This implies that the dynamics of the $(\nu_e, \nu_\mu)$
system involved in the new resonance is regulated by the $(\nu_1^m, \nu_3^m)$ 
eigenvectors in matter. At the algebraic level this is achieved by replacing a rotation in
the 1-2 plane (related to $\theta_{12}^m$), with the product of a rotation in the 1-3 plane 
 (related to $\theta_{13}^m$) and a rotation in the 2-3 plane with $\theta_{23}^m = \pi/2$.
 In fact, from the left-upper panel in Fig.~\ref{fig_mix_3+1_pseudo_-20}  it can be seen that for energies $E\gtrsim 1$ TeV, 
the mixing angle $\theta_{23}^m$ assumes a value equal to  $\pi/2$.
 More precisely, one must note that in the full diagonalization, the 3-flavor mixing matrix
$U_{3\nu}^m = R_{23}^m R_{13}^m R_{12}^m$ in matter lies at the external position,
so this matrix performs the last step in the diagonalization process, after which the Hamiltonian acquires
the diagonal form $H^{4nu}_\mathrm{diag} = \mathrm{diag} (\lambda_1, \lambda_2, \lambda_3, \lambda_4)$.
One can check that the action of $U_{3\nu}^m = R_{23}(\pi/2) R_{13}(\theta_{13}^m) R_{12}(0)$
would produce the swapping of $\lambda_2 \leftrightarrow \lambda_3$. 
 We have also checked that  in the high-energy regime, 
the value of $\theta_{12}^m$ obtained in the simplified 3-flavor system is
identical to the value of $\theta_{13}^m$ obtained in the 4-flavor case. 
Basically, this is an algebraic subtlety needed  to ensure that the relevant 
dynamics intervenes between the two physical states involved in the level repulsion. 
We have also checked
that in the antineutrino channel, where the $(\nu_e, \nu_\mu)$ resonant amplification
occurs for positive values of $f$, one has $\nu_3^m \simeq \nu_\tau$, and in that case no
2-3 level swapping is needed and one simply has a rotation in the 1-2 plane with 
$U_{3\nu}^m = R_{23}(0) R_{13}(0) R_{12}(\theta_{12}^m) \equiv R_{12}(\theta_{12}^m)$.

\subsection{Impact of non-zero $\theta_{34}$ and factorization of the 4-level dynamics}
\label{SubSec:IceCube_theta_34}

\begin{figure*}[t!]
\vspace*{0.05cm}
\hspace*{-0.1cm}
\includegraphics[height=5.87cm,width=5.87cm]{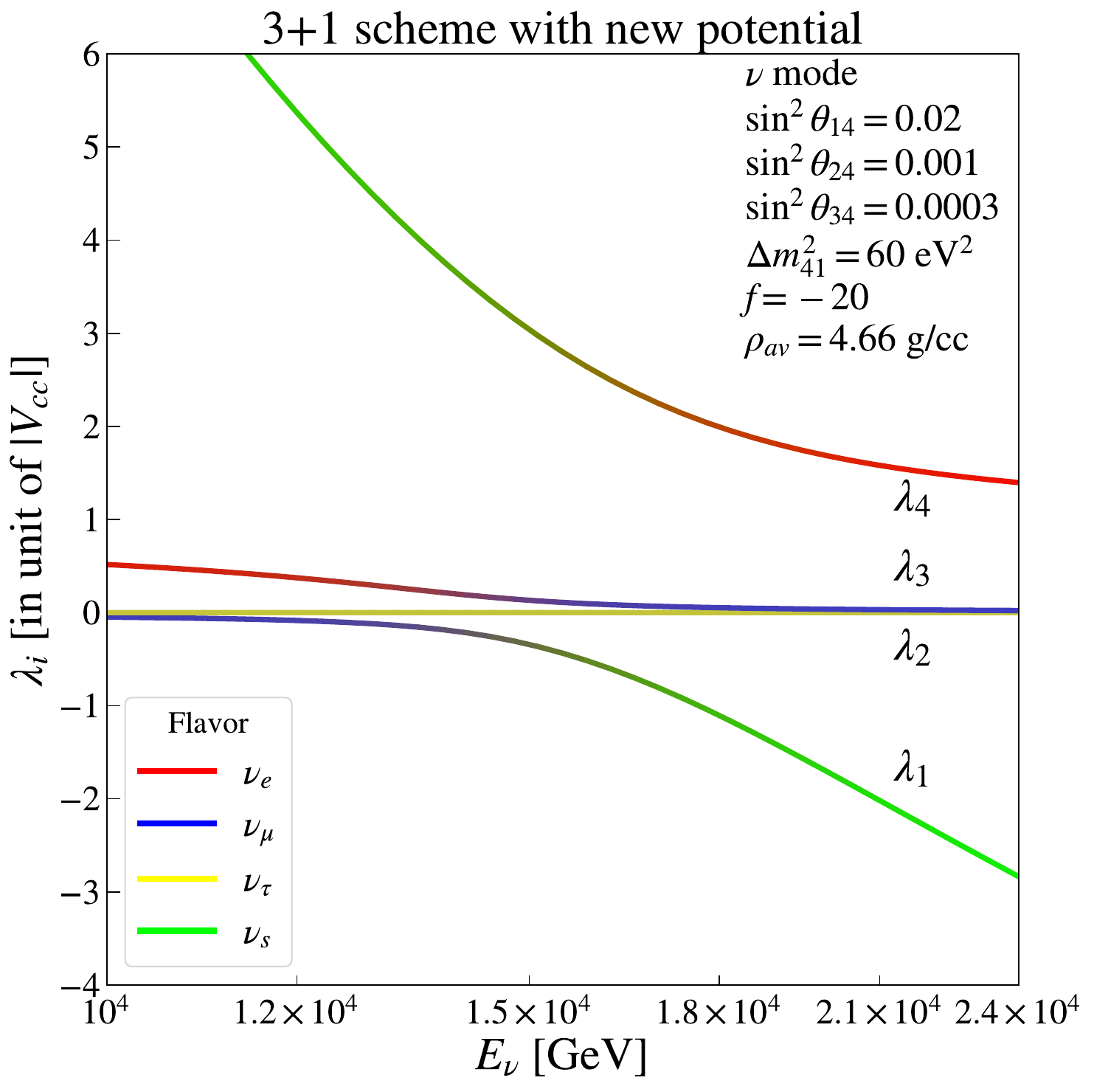}
\includegraphics[height=5.87cm,width=5.87cm]{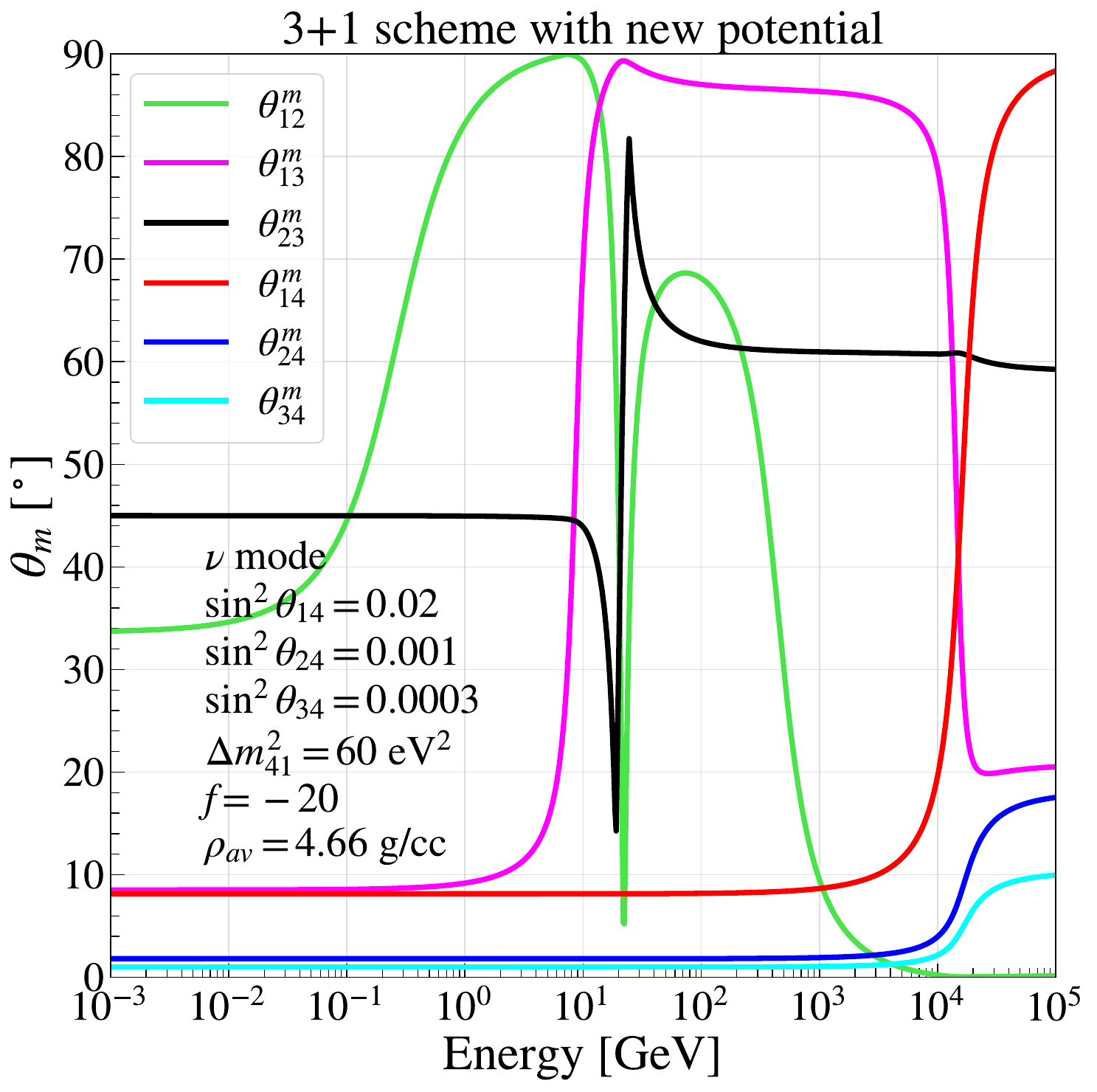}
\includegraphics[height=5.87cm,width=5.87cm]{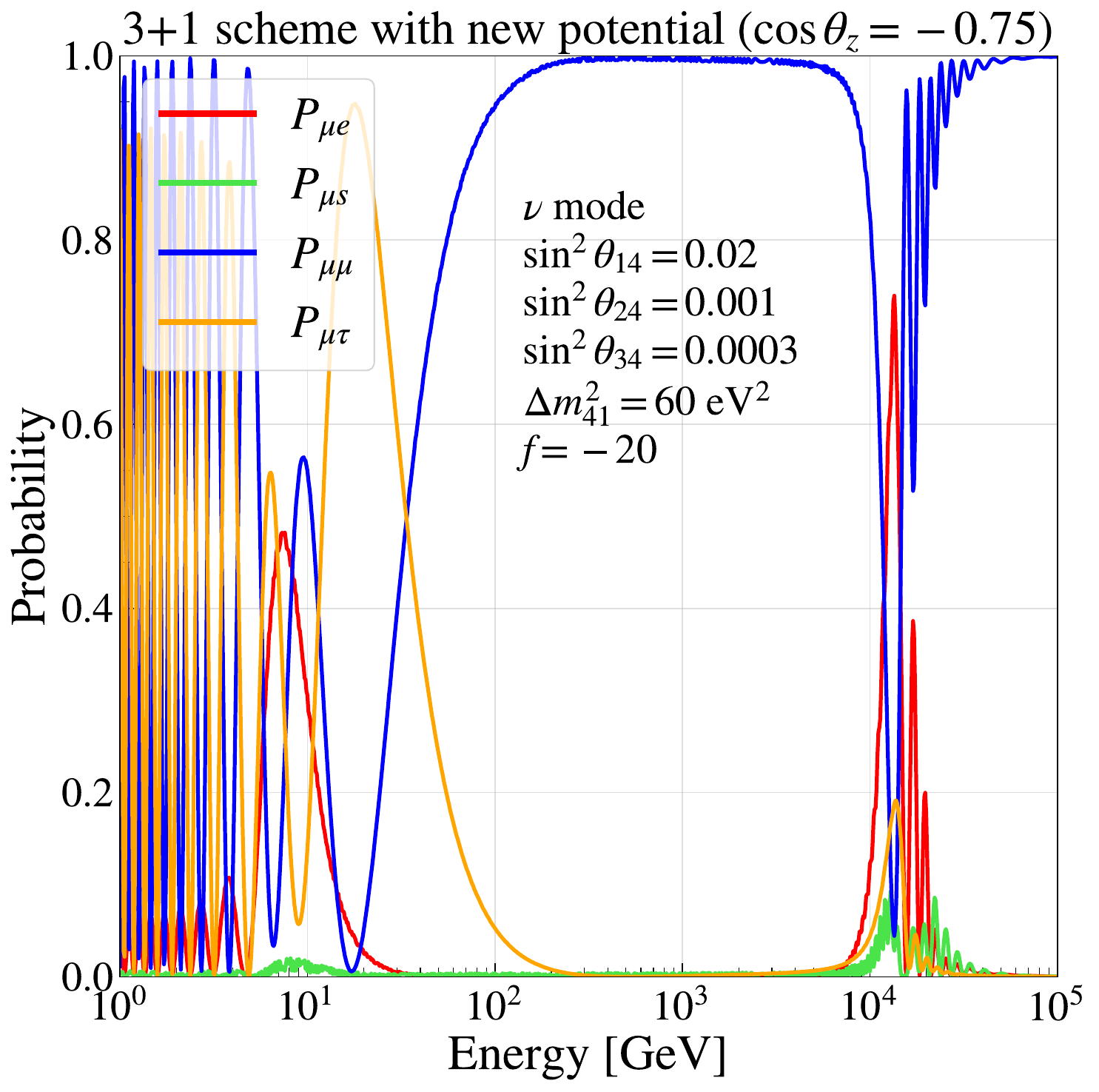}
\vspace*{0.1cm}
\caption{The three panels summarize the results obtained in the 3+1 scheme with the new potential for a non-zero value of $\theta_{34}$.
The left panel represents the behavior of the eigenvalues in the high-energy resonance region. The central
panel displays the mixing angles in matter for constant density equal to the average mantle value. The  
right panel reports the oscillation probabilities relevant for the muon neutrino channel for 
a mantle-crossing trajectory using the PREM profile.}  
\label{fig_mix_3+1_pseudo_-20_theta_34}
\end{figure*}

Until now, for definiteness, we have restricted our study to $\theta_{34} =0$, in which case the $\nu_\tau$ 
is not involved in the flavor transitions at  high energies. However, in general, when such a  mixing angle is non-zero,
the resonant flavor conversion phenomena are expected to involve also the $\nu_\tau$ flavor.
When considering  a non-zero $\theta_{34}$ one 
has to respect the low-energy constraints on the NSI-like couplings $\varepsilon_{e\tau}$, 
$\varepsilon_{\mu\tau}$ and $\varepsilon_{\tau\tau}$ which are related to such an angle. 
In particular, one has to respect the bound on the coupling 
$\varepsilon_{\mu\tau} =\bar f s_{24} \tilde s_{34}^*$, which is the most stringent one being
 $|\varepsilon_{\mu\tau}| \lesssim 3\times 10^{-3}$. For our benchmark values $f= - 20$,  $s^2_{24}=0.001$,
one has $s^2_{34}\lesssim3\times 10^{-4}$. Even for such a low values we find observable effects at the 
level of the oscillation probabilities. In Fig.~\ref{fig_mix_3+1_pseudo_-20_theta_34} we show the plots 
corresponding to  $s^2_{34} = 3\times 10^{-4}$. 
The left panel represents the four eigenvalues in the resonance region. We note that 
in our previous eigenvalues plots, only three eigenvalues where shown since we illustrated
the effective 3-flavor Hamiltonian. In such a context the fourth eigenvalue was zero by construction
and was not shown. However, we can observe that in the more general case with non-zero $\theta_{34}$ 
the situation is very similar to the case of zero $\theta_{34}$.
In fact, one of the eigenvalues ($\lambda_2$) is constant and equal to zero. 
In order to understand this feature, let's consider the full 
rotated Hamiltonian for small values of the mixing-angles
\begin{eqnarray} \label{eq:Hdyn_4}
\arraycolsep=3pt
\medmuskip = 1mu
    \bar H_{4\nu}  \approx  
   V_{\rm{CC}} \!   \begin{bmatrix}
	1 + \bar f_\e s^2_{14}  & \bar f_\e  \tilde s_{14} s_{24}  & \bar f_\e \tilde s_{14} \tilde s_{34}^* & (1- \bar f_\e)  \tilde s_{14}	\\
	\dagger &\bar f_\e s_{24}^2 & \bar f_\e s_{24} \tilde s_{34}^* & -\bar f_\e   s_{24}\, \\
	\dagger & \dagger  &\bar f_\e s_{34}^2 & -\bar f_\e  \tilde s_{34}\,\\
	\dagger & \dagger & \dagger & \bar f_\e + \frac{k_{41}}{V_{\rm{CC}}} 
	\end{bmatrix} \,.
\end{eqnarray}
In order to simplify the treatment, it is useful to extend our effective 3-flavor
toy Hamiltonian in Eq.~(\ref{eq:H_Toy}) to the 4-flavor case, defining
\begin{eqnarray} \label{eq:H_Toy_4nu}
\arraycolsep=3pt
\medmuskip = 1mu
    \bar H_{4\nu}^{(e \mu\tau s)}  =  
       V_{\rm{CC}} \!   \begin{bmatrix}
	1  & 0  & 0 &  \varepsilon_{es}	\\
	0 & 0 &  0  & \varepsilon_{\mu s}  \\
	0 & 0 &  0 & \varepsilon_{\tau s}  \\
	\dagger & \dagger & \dagger & \delta(E)\,.
	\end{bmatrix} 
\end{eqnarray}
where we consider the real NSI-couplings $\varepsilon_{es}, \varepsilon_{\mu s},\varepsilon_{\tau s}$
setting to zero the CP-phases and neglect all the other couplings. In the limit in which the three off-diagonal 
couplings  $\varepsilon_{es}, \varepsilon_{\mu s},\varepsilon_{\tau s}$
are equal to zero, the system presents one 1-4 crossing for $\delta =1$, and two coinciding 
crossings at $\delta =0$ involving the 2-4 and 3-4 states. The structure
of the Hamiltonian implies that in the ($\nu_\mu, \nu_\tau$) sub-space
only a linear combination will be coupled to the sterile state. 
In fact, defining $\varepsilon_{as} = \sqrt{(\varepsilon_{\mu s}^2 + \varepsilon_{\tau s}^2)}$
and introducing the orthogonal normalized states
\begin{eqnarray} 
\arraycolsep=3pt
\medmuskip = 1mu
\label{eq:H_states_+}
     | \nu_+ \rangle =  \frac{\varepsilon_{\mu s}  | \nu_\mu \rangle +  \varepsilon_{\tau s}| \nu_\tau \rangle}{\varepsilon_{as}} 
                             =  \frac{s_{24}  | \nu_\mu \rangle +  s_{34}| \nu_\tau \rangle}{\sqrt{s^2_{24}+ s^2_{34}}}\,, \\
\label{eq:H_states_-}
     | \nu_- \rangle  =  \frac{-\varepsilon_{\tau s}  | \nu_\mu \rangle +  \varepsilon_{\mu s}| \nu_\tau \rangle}{\varepsilon_{as}}
    			 	=  \frac{- s_{34}  | \nu_\mu \rangle  + s_{24}| \nu_\tau \rangle}{\sqrt{s^2_{24}+ s^2_{34}}}\,,      
\end{eqnarray}
in the new basis ($\nu_e,\nu_+, \nu_-, \nu_s)$ the Hamiltonian takes the form%
\footnote{The 4-flavor Hamiltonian is factorized by an orthogonal rotation in the $(\nu_\mu, \nu_\tau)$
plane with an angle defined by $\cos \alpha = \varepsilon_{\mu s}/\varepsilon_{as}$
and $\sin\alpha = \varepsilon_{\tau s}/\varepsilon_{as}$.}
\begin{eqnarray} \label{eq:H_Toy_4nu_prime}
\arraycolsep=3pt
\medmuskip = 1mu
    \bar H_{4\nu}^{(e + - s)}  =  
       V_{\rm{CC}} \!   \begin{bmatrix}
	1  & 0  & 0 &  \varepsilon_{es}	\\
	0 & 0 &  0  & \varepsilon_{as}  \\
	0 & 0 &  0 & 0  \\
	\dagger & \dagger & 0 & \delta(E)
	\end{bmatrix}\,.
\end{eqnarray}
Therefore, the 4-level system is factorized
in a non-trivial part involving the three-level system ($\nu_e,\nu_+, \nu_s)$ regulated by the Hamiltonian
\begin{eqnarray} \label{eq:H_Toy_3nu_prime}
\arraycolsep=3pt
\medmuskip = 1mu
    \bar H_{3\nu}^{(e + s)}  =  
       V_{\rm{CC}} \!   \begin{bmatrix}
	1  & 0  &   \varepsilon_{es}	\\
	0 & 0 &   \varepsilon_{as}  \\
	\dagger &  \dagger & \delta(E)
	\end{bmatrix}\,,
\end{eqnarray}
and a trivial part involving the ``spectator'' state $\nu_-$, which has constant zero energy.%
\footnote{In the language of multi-level systems, in particular in atomic physics and
quantum optics, the spectator state is dubbed as a ``dark state'' (see~\cite{Zhao:2025rjg}).}
This toy model provides a very good description of the real 4-flavor system because all the couplings
involving directly the $\nu_\tau$ flavor (for example $\varepsilon_{\mu \tau}$) are very small. 
It can be shown that the full 4-flavor model, because of the particular structure of the 
Hamiltonian, preserves exactly the basic properties of the toy model. In fact,
it can be checked that in the full Hamiltonian~(\ref{eq:Hdyn_4}) the second and third columns (or rows)
are linearly dependent and therefore it has rank 3. It happens that the state $\nu_-$
defined in Eq.~(\ref{eq:H_states_-}) is an exact eigenvector of the Hamiltonian~(\ref{eq:Hdyn_4}) with zero eigenvalue.
Therefore, the system can be factorized with the same rotation used for the toy model. 
 The effective 3-flavor Hamiltonian in Eq.~(\ref{eq:H_Toy_3nu_prime})
will induce (possibly resonant) transitions within the ($\nu_e,\nu_\mu$) and ($\nu_e,\nu_\tau$) sub-systems
mediated by the second-order coupling $\varepsilon_{es} \varepsilon_{as}$ 
of the $\nu_e$ with the $\nu_+$ state.
It is important to notice that the 4-dimensional Hamiltonian Eq.~(\ref{eq:H_Toy_4nu_prime}) 
will also induce ($\nu_\mu,\nu_\tau$) transitions mediated by the interference of the survival amplitudes 
of the spectator state $\nu_-$ (which has trivial dynamics) and of the state $\nu_+$ (which has non-trivial dynamics). 
Hence, we can conclude that in the presence of a non-zero $\theta_{34}$,
the core dynamics of the 4-flavor system remains that of a 3-level system,
allowing at the same time the $\nu_\tau$ flavor to fully participate to the resonant conversion
phenomena.

The central panel in Fig.~\ref{fig_mix_3+1_pseudo_-20_theta_34} representing
the running of the mixing angles in matter shows that in the high-energy resonant region,
the mixing angle $\theta_{34}^m$ becomes appreciable and has a similar behavior 
to $\theta_{24}^m$. In fact, we have checked that these two angles are identical if their 
vacuum value is the same, as expected from the $\mu-\tau$ symmetry of the Hamiltonian. 
Also we observe that the mixing angle $\theta_{23}^m$ assumes values different from $\pi/2$ 
and that it does not have a resonant like behavior. This angle basically regulates the relative 
amplitude of the 2-4 and 3-4 resonances. 
The right panel represents the transition probabilities
involving the muon neutrino channel. We can observe that, albeit not very pronounced, there  is a resonance
involving the ($\nu_\mu, \nu_\tau$) sector. Similarly, we find (not shown) that there is a comparable resonance
in the ($\nu_e, \nu_\tau$) sector. Both such resonances can become more pronounced for values of $\theta_{34}$
larger than our benchmark choice. Hence, we see that in the most general case, all the active flavors are involved
in the resonant conversion mediated by the sterile neutrino potential, as expected on the basis of our analytical treatment.

\begin{figure*}[t!]
\vspace*{0.05cm}
\hspace*{-0.1cm}
\includegraphics[height=6.87cm,width=7.87cm]{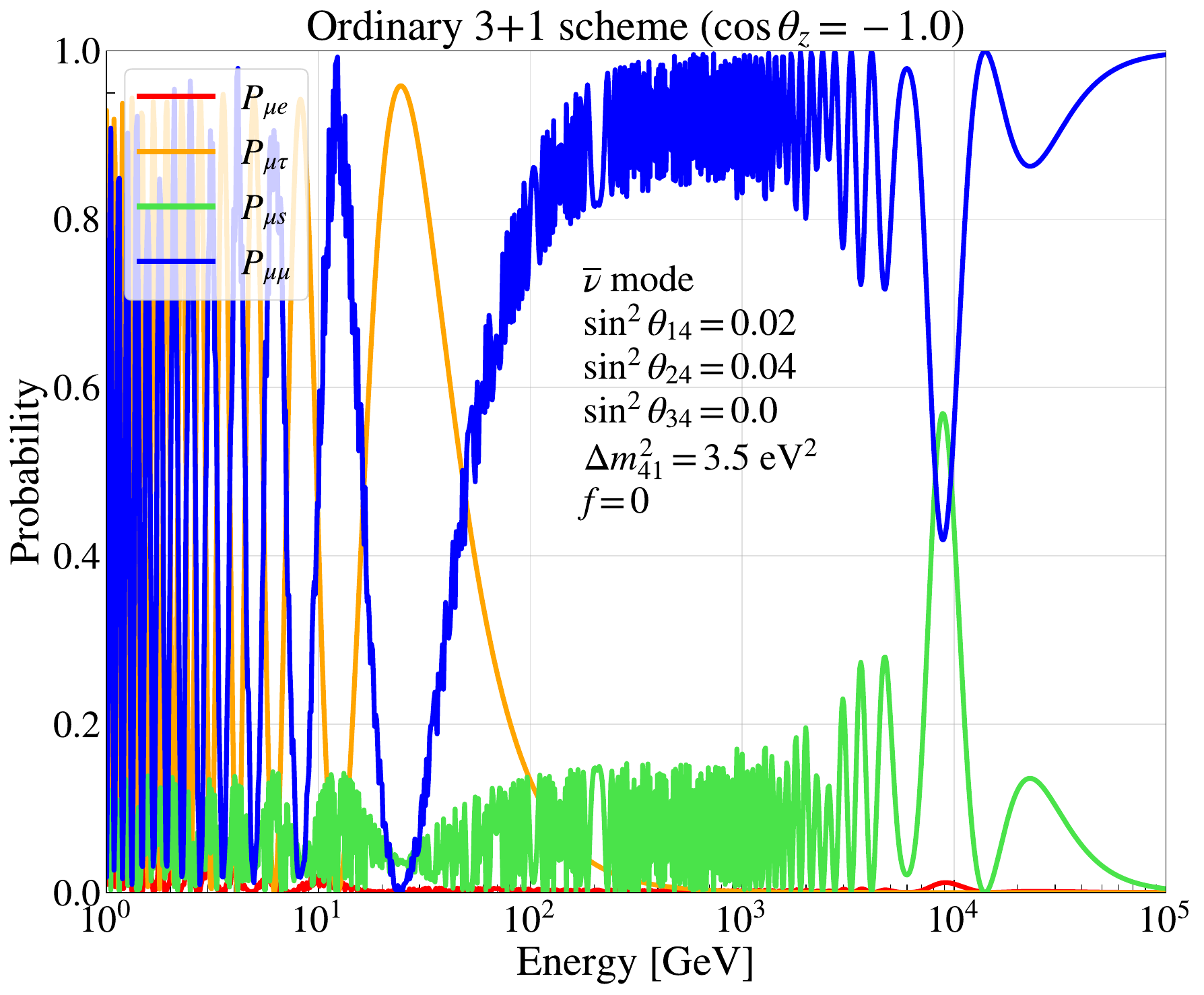}
\includegraphics[height=6.87cm,width=7.87cm]{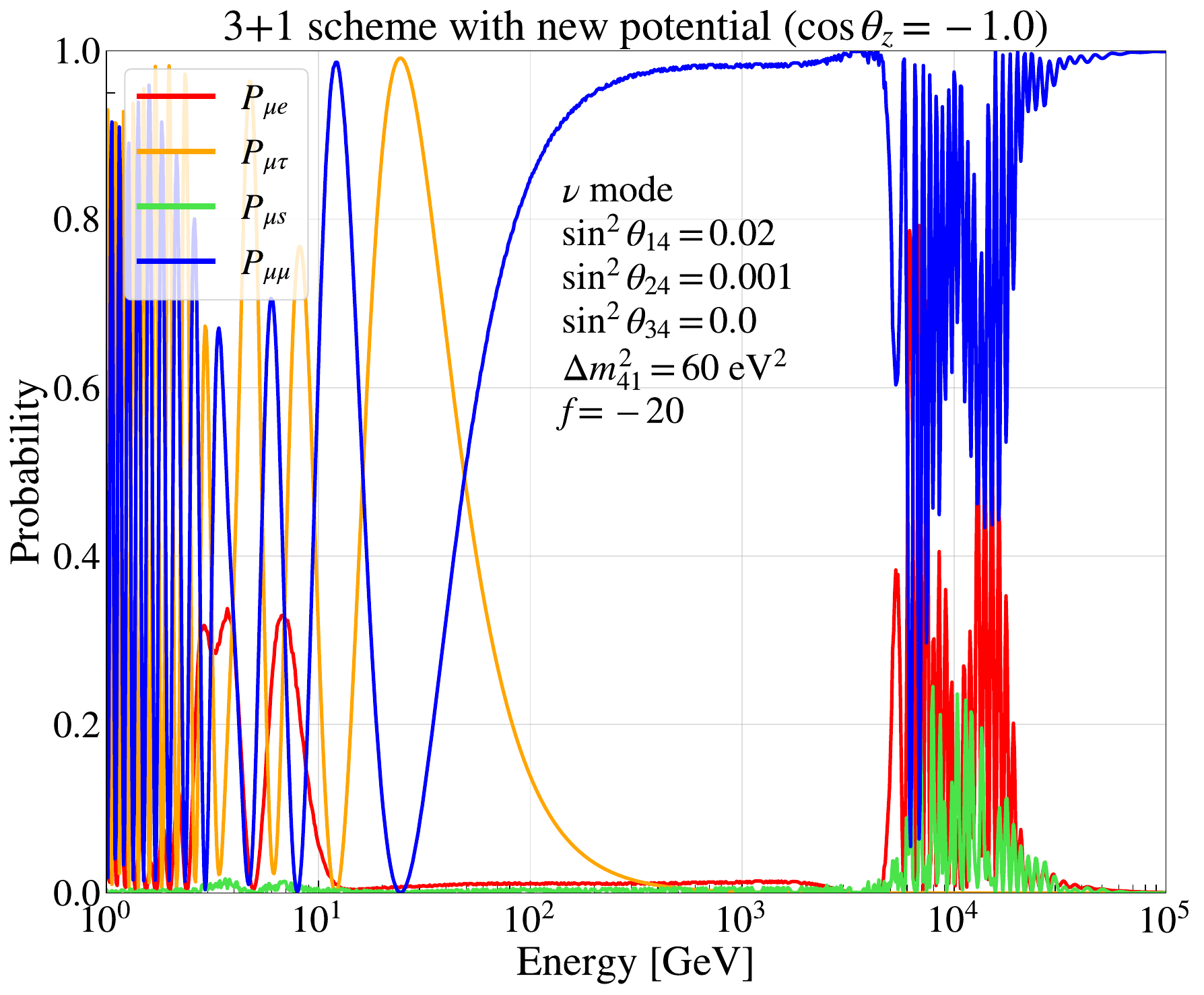}\\
\vspace*{0.5cm}
\includegraphics[height=6.87cm,width=7.87cm]{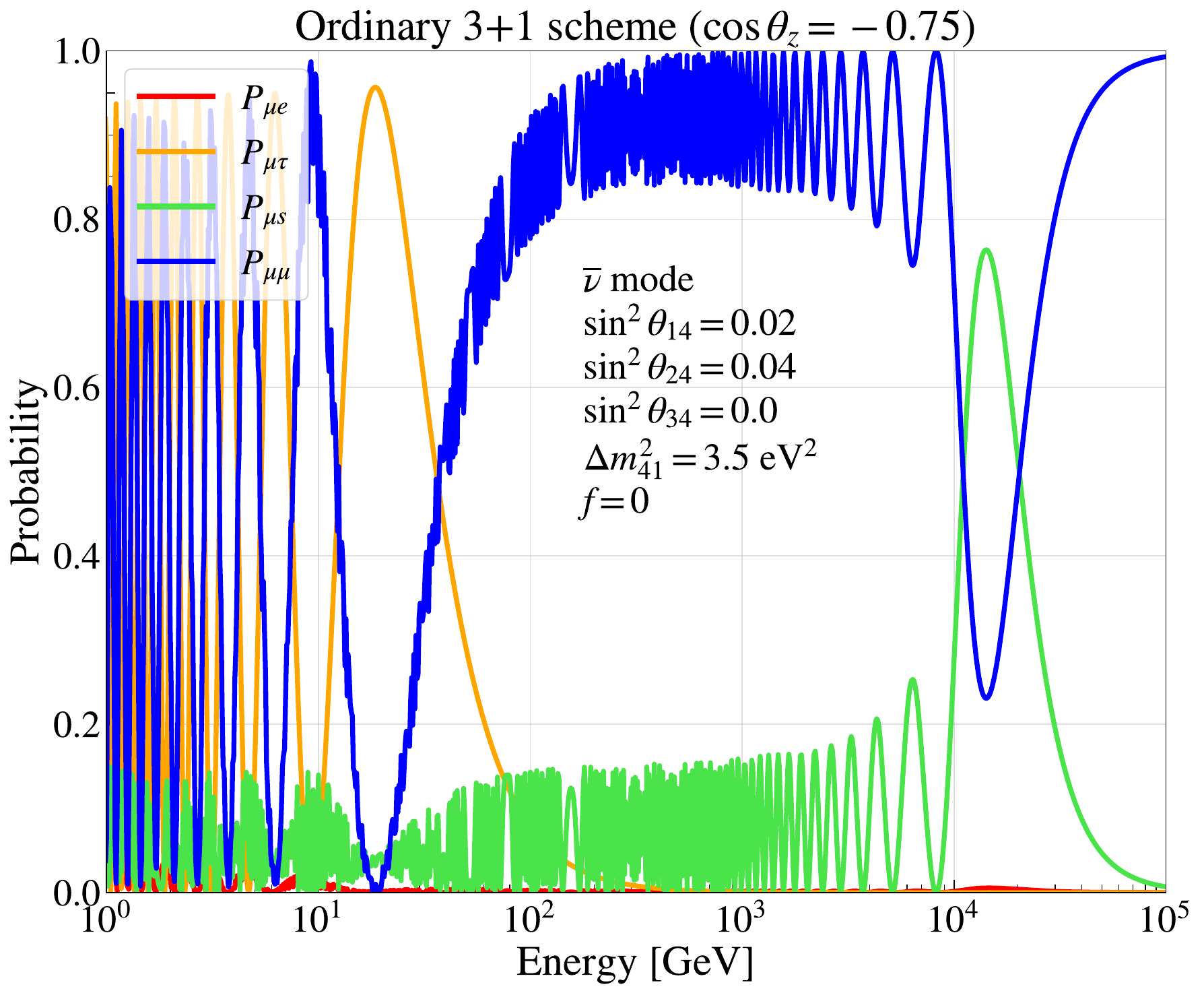}
\includegraphics[height=6.87cm,width=7.87cm]{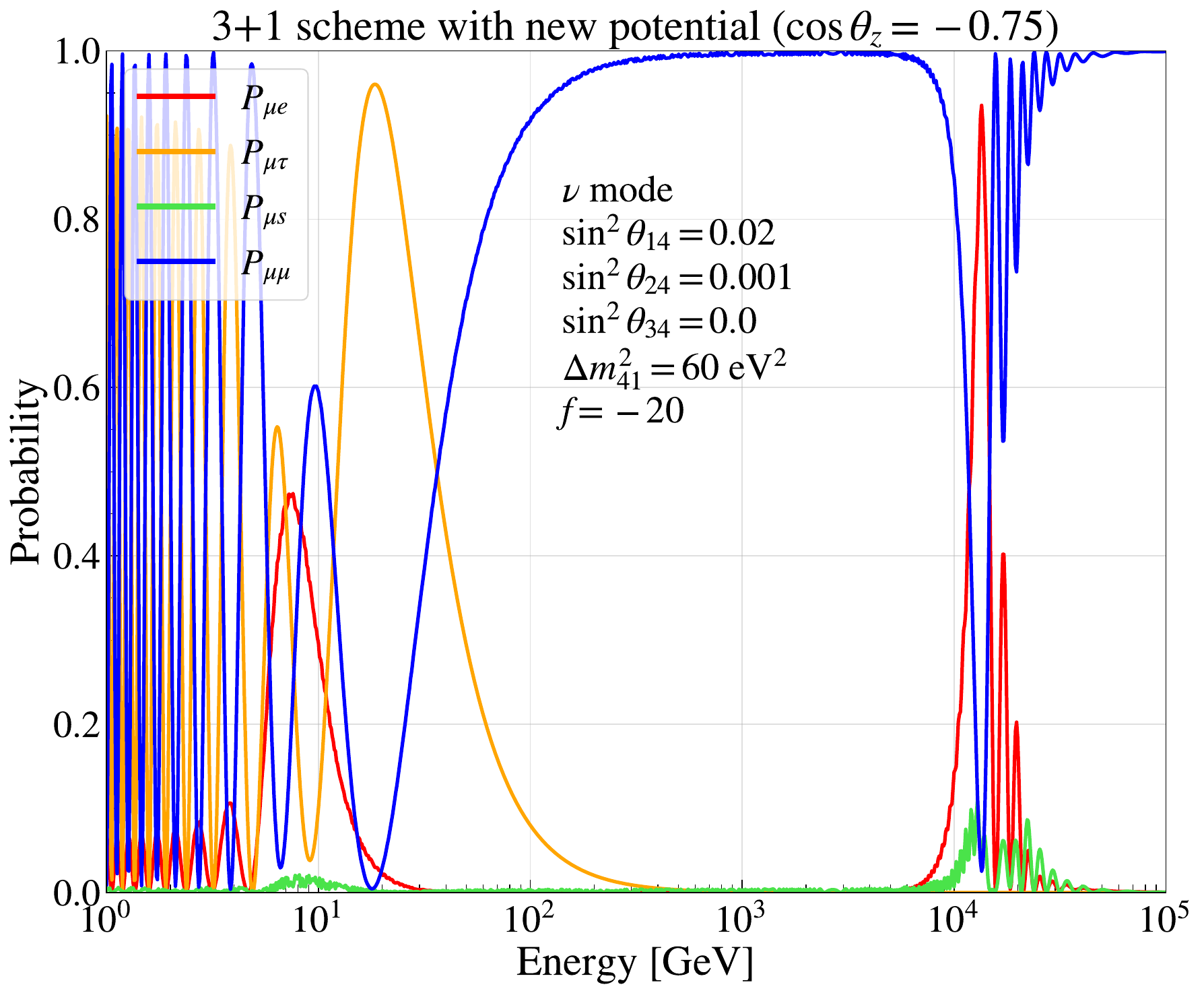}
\vspace*{0.0cm}
\caption{Oscillation probabilities for the ordinary 3+1 scheme (left panels) and the 3+1 scheme with new potential (right panels).
The left panels represent antineutrino oscillations probabilities, while the right panels represent  neutrino oscillation probabilities. 
The upper panels refer to a radial core-crossing trajectory with $\cos\theta_z = -1.0$, while the lower panels refer to a mantle-crossing
trajectory with  $\cos\theta_z = -0.75$. In the left panels we assume $\Delta m^2_{41} = $ 3.5 eV$^2$ and $s^2_{24} = 0.04$, $s^2_{14} = 0.02$.
corresponding the IceCube best fit point in the ordinary 3+1 scheme. In the right panels we take $f= -20$, $\Delta m^2_{41} = $ 60 eV$^2$, $s^2_{14} = 0.02$, 
$s^2_{24} = 0.001$ and $\delta_{14} = 0$, which is our benchmark point in the scenario with the new potential.}
\label{fig_IceCube_prob_1}
\end{figure*} 

\subsection{Full numerical results with the PREM density model and parametric enhancement}
\label{SubSec:IceCube_num}

As a final step in our exploration of the resonant phenomenon, we consider some numerical results using the realistic PREM model
of the Earth density.  Figure~\ref{fig_IceCube_prob_1} displays a few 
representative examples of the oscillation probabilities focusing on the muon neutrino channel,
which is phenomenologically relevant.
The left panels refers to the ordinary 3+1 scheme ($f=0$)
while the right panels represent the 3+1 scheme with the new potential for the benchmark 
value $f=-20$.  All the parameters are fixed at the benchmark values adopted in the previous 
subsection.
We stress that the left (right) panels represent antineutrino (neutrino) oscillations probabilities.
The upper panels refer to a radial core-crossing trajectory with $\cos\theta_z = -1.0$ where $\theta_z$ is the zenith angle,
while the lower panels 
refer to a mantle-crossing trajectory with  $\cos\theta_z = -0.75$. The right plots confirm
that in the presence of the new potential there is a strong resonance in the 
$\nu_\mu \to \nu_e$  channel (red color) as expected from the results obtained for the 
constant density case. However, while for the mantle-crossing trajectory the probabilities
are very similar to the case of constant density (compare the right-lower panel to the right-upper panel in Fig.~\ref{fig_mix_3+1_pseudo_-20}),
for the core-crossing  trajectory the probabilities have a radically different behavior,
presenting a very large width with fast oscillations. In addition, the muon survival probability
presents two distinct dips. This behavior is the sign that huge
parametric effects are at play induced by the layered structure of the Earth density.
While IceCube cannot resolve the detail of the fast oscillations because of the limited energy resolution of the
detector, it may be able to observe the large width of the resonance and possibly identify the two dips.

\begin{figure*}[t!]
\vspace*{0.05cm}
\hspace*{-0.1cm}
\includegraphics[height=7.87cm,width=8.87cm]{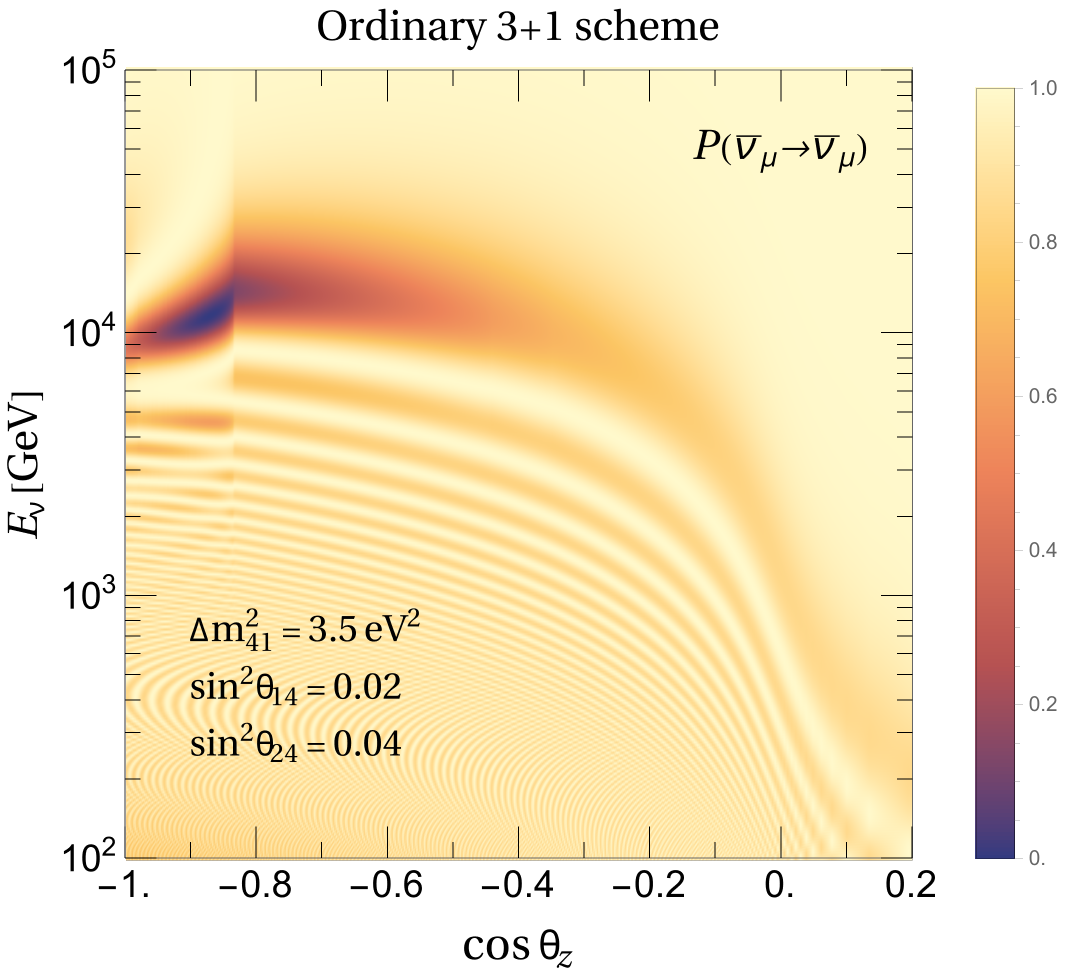}
\includegraphics[height=7.87cm,width=8.87cm]{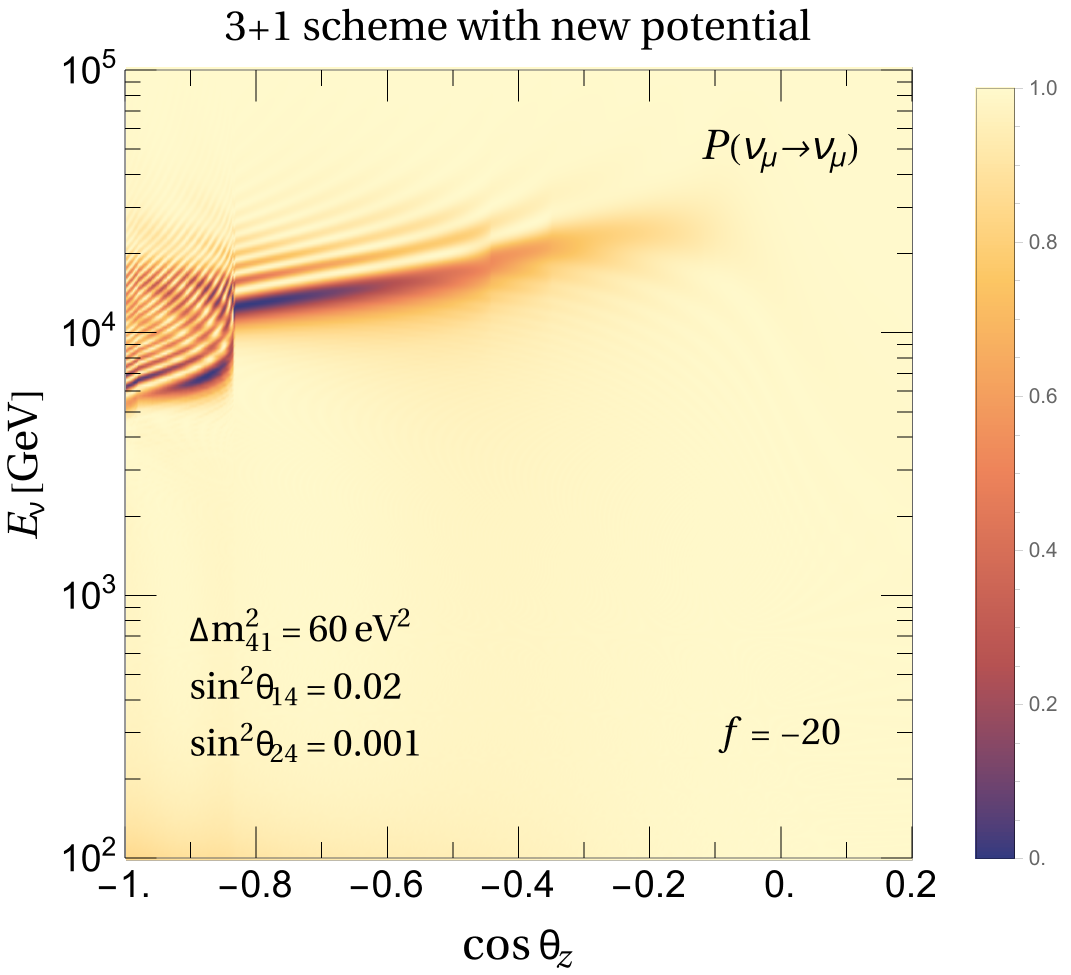}
\vspace*{-0.3cm}
\caption{Oscillograms for the ordinary 3+1 scheme (left panel) and the 3+1 scheme with new potential (right panel).
The left panel refers to muon antineutrinos, while the right panel refers to muon neutrinos.}
\label{fig_oscillogram_mm}
\end{figure*}

The resonant behavior can be further illustrated with the help of the so-called oscillograms,
which are 2-dimensional iso-contour plots of the muon neutrino disappearance probability
as a function of neutrino energy $E$ and cosine of the zenith angle $\cos(\theta_z)$.
In the left panel of Fig.~\ref{fig_oscillogram_mm} we show the oscillogram for the ordinary 3+1
case, with the same parameters used in Fig.~\ref{fig_IceCube_prob_1}. The plot refers to antineutrinos,
for which one expects  the resonance to occur in the ordinary 3+1 scheme and the oscillation pattern is
well understood. The resonance enhancement of the $\bar{\nu}_\mu\to\bar{\nu}_s$ oscillation manifests 
as antineutrinos pass through the Earth's core and/or mantle.  For trajectories crossing the core, that is
$\cos\theta_z \lesssim -0.8$, the enhancement is described
by a parametric resonance~\cite{Razzaque:2011ab,Esmaili:2013vza}
at the energy $\simeq (9~{\rm TeV})~(\Delta m_{41}^2/3.5~{\rm eV}^2)$.
For the mantle-crossing trajectories ($\cos\theta_z \gtrsim -0.8$) the enhancement
is a MSW resonance at the energy $\simeq(14~{\rm TeV})~(\Delta m_{41}^2/3.5~{\rm eV}^2)$. 
The right panel displays the oscillogram that we obtain for the 3+1 scenario with
the novel potential, where the resonance occurs in the neutrino channel. 
From this plot we see that, as expected, there is a resonant behavior at the same energies involved
in the ordinary 3+1 scheme. We still observe a well defined separation indicative
of the transition between core-crossing and mantle-crossing trajectories. We find that the 
depth and the width of the resonance (both for core and mantle-crossing trajectories)
depends on the value of $s^2_{24}$ and the width can be noticeably influenced by $s^2_{14}$,
becoming larger (smaller) for higher (lower) values of this parameter. The oscillogram confirms
the presence of fast oscillations for core-crossing trajectories, which are imputable to parametric effects.

We could not perform a numerical study of the IceCube resonance beyond the level of probability plots and oscillograms because 
reproducing the IceCube analysis is unfeasible from outside the collaboration.
Hence, it remains to be seen if the quality of the fit of the IceCube data to 
the new resonance is similar to that of the usual one. We note that the reconstruction 
of the neutrino energy is subject to a substantial degradation due to the 
limited energy resolution of the detector and what appears as a clear 
resonance at $E_\nu = 10$~TeV in the oscillogram, looking by eye at the IceCube 2D-histograms
of events, is translated into a deficit of events around the muon energy of $E_\mu \sim 1$~TeV
and to an arch-like pattern.
So, it is plausible that the distinction between the ordinary resonance and the one predicted
in the model under investigation should be difficult. However, an accurate 
official simulation performed by the IceCube collaboration is needed to clarify the issue.
We also notice that IceCube could distinguish the new resonance from the ordinary one looking at 
the appearance of electron neutrinos, which should manifest as cascade events. However, according
 to a recent study~\cite{Smithers:2021orb} 
 (see also the early study~\cite{Esmaili:2013cja})
performed in the 3+1 scheme with a large non-zero  $\theta_{34}$,  where $\nu_\tau$ appearance
is strongly enhanced giving rise to $\nu_\tau$ events (indistinguishable from the $\nu_e$ ones),
the sensitivity of IceCube to cascades is currently limited,
albeit it could improve in the future by considering an expanded data sample larger (up to ten-years)
than the one-year sample presently used in the analysis. Therefore, cascade events are
a signature of the model under consideration provided that there is enough sensitivity to
detect them. 

An important remark is in order concerning the compatibility of the benchmark parameters of the model
with the existing data. In Sec.~\ref{Sec:LBL_ATM} we have seen that with the choice of a
non-negligible value $s^2_{14} = 0.02$,  solar and LBL data indicates $O(10)$ values for the 
$f$ parameter regulating the strength of the new potential.  In the present section, we have also
seen that in order to maintain fixed the energy of the resonance at the value hinted 
at by IceCube ($E\simeq 10$ TeV), the new mass-squared must be rescaled 
by the factor $|f|$ with respect to the best fit value obtained in the ordinary 3+1 scheme, leading to 
the choice $\Delta m^2_{41} = 60\,$eV$^2$. Here, one must notice that for such large values 
of the mass-squared splitting there are strong constraints on $s^2_{14}$ established by
KATRIN~\cite{KATRIN:2025lph}. Looking at the exclusion regions in~\cite{KATRIN:2025lph}
(see Fig.~3), one can observe that in the window $\Delta m^2_{41} \simeq [30, 80]$ eV$^2$, 
the constraints are less restrictive compared with those provided for lower/higher values. 
This is confirmed by a very recent analysis combining KATRIN with SBL disappearance data~\cite{Giunti:2026uaf}.
In particular,  for $\Delta m^2_{41} = 60\,$eV$^2$ KATRIN allows values of $s^2_{14} \lesssim 0.025$. 
Therefore, our benchmark parameters are fully compatible with KATRIN. 
We observe that values of $\Delta m^2_{41}$ (and consequently of $f$)
a factor of two smaller or larger than our benchmark choice would be disfavored by KATRIN
since the bound on $s^2_{14}$ becomes stronger. It is interesting to observe that 
because of the connection between $f$ and $\Delta m^2_{41}$,
KATRIN indirectly constrains the new potential and it provides an independent
indication that $f$ should be $O(10)$.  New data with increased statistics
expected to come from KATRIN  should be able to push the sensitivity around the benchmark 
value of our model possibly confirming/disconfirming it.

As a closing observation, we stress that the estimated $O(10)$ value
 for the parameter $f$ is strictly related to our assumption to fix an appreciable
value of $s^2_{14} \simeq 0.01-0.03$. We have checked that basically all manifestations 
of the model where matter effects are crucial, such as the explanation of the LBL discrepancy
between NOvA and T2K, the production of an excess of $\nu_e$-like events in Super-Kamiokande 
multi-GeV atmospheric neutrinos and the multi-TeV resonance in IceCube, are intact also for 
very small values of $s^2_{14}$ provided that the parameters $f$ and $s^2_{24}$ are appropriately
rescaled (considering much larger values for $f$ and much smaller values  for $s^2_{24}$),
in order to keep fixed the $\varepsilon_{ee}^\e$ and $\varepsilon_{e\mu}^\e$ couplings at their
benchmark values identified in Sec.~\ref{Sec:LBL_ATM} [see Eqs.~(\ref{eq:eps_equivalent_2})-(\ref{eq:eps_equivalent_1})].
However, one must notice that a realization of the scenario 
with a very small value of $s^2_{14} \ll 0.01$ would be of no relevance for the explanation of 
the SBL electron neutrino disappearance findings such as the Gallium and Reactor anomalies.

\section{Effective two-level resonance in IceCube for $\theta_{12}$ in the dark octant}
\label{Sec:Dark}

Until now we have assumed that the solar mixing angle lies in the lower octant ($\sin^2\theta_{12} <0.5$).
In fact, we have seen that if the new potential is proportional to the neutron number density, the combination of solar
neutrino data with NOvA and T2K, excludes the higher (or dark) octant  ($\sin^2\theta_{12} > 0.5$) solution.
However, if the new potential is proportional to the electron or proton number density, the
degeneracy persists in the combination of solar and LBL experiments. This is a consequence
of the so-called generalized mass-ordering degeneracy~\cite{Coloma:2016gei}, which implies that 
the oscillation physics is invariant under the combined transformation%
\footnote{In the presence of off-diagonal NSI-like couplings the symmetry requires $\varepsilon_{\alpha\beta} \to
\varepsilon_{\alpha\beta}^*$, which in our model is obtained by the replacement of the CP-phases 
$\delta_{14} \to 2\pi - \delta_{14}$  and $\delta_{34} \to 2\pi - \delta_{34}$.}
 $\varepsilon_{ee} \to -2 - \varepsilon_{ee}$,  $\theta_{12} \to \pi/2 - \theta_{12}$, 
$\Delta m^2_{31} \to -\Delta m^2_{32}$ (i.e. NO $\to$ IO), $\delta_{13} \to \pi - \delta_{13}$.
Therefore, in this case the dark degenerate solution cannot be discarded. 
According to the degeneracy theorem, the 3-flavor phenomenology will not allow the distinction
between the two neutrino mass orderings. For example, in the dark octant solution,
the 6 GeV resonance in Super-Kamionande in the neutrino 
channel will occur in IO instead of NO. Also, 
the current preference for NO of reactor experiments
when combined with LBL experiments, would become a preference for IO. 
All in all, in the 3-flavor framework, in the presence of large NSI-like couplings (possibly close to $\varepsilon_{ee}\simeq -2$), 
there is no way to determine the NMO if the NSI are proportional to the electron or proton number density.

It is natural to ask if in the 3+1 scheme with the new matter potential, it is possible to distinguish
between the two degenerate solutions. The answer is positive essentially for a very simple reason.
Differently from what happens for the NMO in the 3-flavor scheme, the sign of $\Delta m^2_{41}$
is fixed to be positive. The case of a negative $\Delta m^2_{41}$ is excluded because the light
neutrino states would have a mass incompatible with cosmology, especially in our scenario where
$\Delta m^2_{41}$ can be much bigger than the usual $\sim 1$\,eV$^2$ scale more often considered in the literature.
This renders less dangerous a possible 4-flavor generalization of the 3-flavor degeneracy theorem.
If we consider the IceCube resonance at 10 TeV, the 3-flavor parameters are irrelevant, so
apparently it is blind to the octant of $\theta_{12}$ and to the NMO. On the other hand, 
one must observe that 
the coupling $\varepsilon_{ee}^\e = \bar f_\e s^2_{14}$ is relatively small in the
lower-octant solution (since $f \simeq -20$), while it becomes close to $-2$ in the dark-octant
solution for which one needs much larger (negative) values of $f$. 
Since the value of $f$ is different and much larger in the dark-octant solution, one may
expect a different behavior in the two degenerate solutions. This means, that 
the study of the IceCube resonance, breaking the generalized 3-flavor degeneracy,
can allow us to pin down indirectly the correct octant of the solar mixing angle and the NMO.
In fact, we find that the degeneracy is maximally broken since for the dark-octant solution
there exists a $(\nu_e, \nu_\mu)$ resonance in the antineutrino channel,
which has no counterparts for the ordinary solution. 

 In oder to understand this behavior we report below
the expression of the Hamiltonian regulating the ($\nu_e, \nu_\mu, \nu_s$) oscillations already introduced in Eq.~(\ref{eq:H_3_ems_v1}),
\begin{eqnarray} \label{eq:H_3_ems_v2}
\arraycolsep=3pt
\medmuskip = 1mu
     H_{3\nu}^{(e \mu s)}  =
      \begin{bmatrix}
	 V_{\rm{CC}} + k s^2_{14}  & k   s_{14} s_{24}  & k  s_{14} 	\\
	\dagger & k  s_{24}^2 &  k  s_{24}  
	\\
	\dagger & \dagger
	&\bar f V_{CC} + k 
    \end{bmatrix} \,.
\end{eqnarray}
We notice that for a very large  values of 
$\bar f$, the $(4,4)$ entry in the Hamiltonian can be much larger than 
the $(1,4)$ and $(2,4)$ entries. In this limit the sterile state $\nu_s$ can be approximately integrated away,
and the effective dynamics of the $(\nu_e, \nu_\mu)$ system is approximately that of a two-level system. 
Second, we observe that, in the limit  $|\bar f| \to \infty $,
if the condition $[H(2,2)-H(1,1)] =0$ is verified, one has a resonant
enhancement regulated by a mixing angle $\theta_{12}^m$.%
\footnote{We underline that, as we have seen in the previous section, in the full 4-flavor
scheme the mixing angle $\theta_{12}^m$ is replaced by  $\theta_{13}^m$
in conjunction with $\theta_{23}\simeq \pi/2$.}
In such a limit, it is possible to provide an exact analytical 
expression for the 1-2 mixing angle in matter diagonalizing the (1,2) block of the Hamiltonian, 
obtaining
\begin{align}
\label{eq:s2_4nu}
\sin2\theta_{12}^m
& = \frac{2 s_{14} s_{24}}{\sqrt{4 s_{14}^2 s_{24}^2 + (v +s^2_{14} - s^2_{24})^2}}\,,
\end{align}
with $v = \frac{V_{CC}}{k}$, and the conversion probability of muon to electron neutrinos can be expressed as
\begin{align}
\label{eq:Pee_4nu}
P_{\mu e}  = \sin^2 2\theta_{12}^m \sin^2 \phi_{12}^m\,, 
\end{align}
where we have introduced the oscillation phase
\begin{align}
\label{eq:phi_4nu}
\phi_{12}^m = \frac{kL}{2} \sqrt{4 s_{14}^2 s_{24}^2 + (v +s^2_{14} - s^2_{24})^2}\,,
\end{align}
These expressions explicitly show that the origin of the mixing angle in matter $\theta_{12}^m$ 
and of the oscillation phase $\phi_{12}^m$ is  purely dynamical and 
completely unrelated to their value in vacuum. The two-level ($\nu_e,\nu_\mu$)
conversion is an effective phenomenon mediated by the far pseudo-sterile state $\nu_s$,
in which case it behaves as a virtual mediator.

A more general formula for the oscillation probability, valid for the large finite values of $\bar f$,
which are of phenomenological relevance, can be obtained by introducing a 2-dimensional effective 
Hamiltonian which retains the second-order corrections in the active-sterile mixing after the integration of the fourth state.%
 \footnote{The effective Hamiltonian can be obtained using the partitioning technique introduced 
 independently in nuclear physics by Feshbach~\cite{Feshbach} and in quantum chemistry  by L\"owdin~\cite{Lowdin} 
 (see~\cite{Many_Body,Book_Brillouin} for textbook treatments).}
 Introducing the projector operators $P = \mathrm{diag} (1,1,0)$ and $Q = \mathrm{diag} (0,0,1)$, the 3-dimensional 
 Hilbert space can be partitioned in the target 2-dimensional subspace (P-space) of interest for the active flavors
  and its orthogonal 1-dimensional complement (Q-space) of the sterile flavor.
 The full 3-flavor Hamiltonian $H \equiv H_{3\nu}^{(e \mu s)}$ in Eq.~(\ref{eq:H_3_ems_v2})  can be expressed as
\begin{align}
H  = PHP + PHQ + QHP + QHQ \,
=
 \begin{bmatrix}
	 H_{PP}  & H_{PQ}  \\
	 H_{QP} &  H_{QQ}  
    \end{bmatrix} \,,
\end{align}
where we denote with $(e,\mu)$  the indexes of $H_{PP}$ and with $(s)$ that of $H_{QQ}$.
This decomposition allows us to derive an effective 2-dimensional Hamiltonian~\cite{Feshbach,Lowdin}
\begin{align}
H_{e\mu}^{\mathrm{eff}}(\lambda)  = H_{PP} + H_{PQ} (\lambda - H_{QQ})^{-1}H_{QP} \,,
\end{align}
which depends on the eigenvalue $\lambda$. For the active neutrinos one has  $|\lambda| \ll |H_{QQ}|$,
since  $H_{QQ} \equiv H_{ss} = f V_{CC} + k $ is very large and one has for the elements of the effective Hamiltonian
\begin{align}
(H_{e\mu}^{\mathrm{eff}})_{\alpha\beta} \simeq H_{\alpha\beta} - \frac{H_{\alpha s}H_{s\beta}}{H_{ss}} \,.
\end{align}
Taking into account the expression of the matrix elements of $H$ in Eq.~(\ref{eq:H_3_ems_v2}),  one finds that 
the two-level effective Hamiltonian is diagonalized by the mixing angle 
\begin{align}
\label{eq:s2_4nu_eff}
\sin2\theta_{12}^m
& = \frac{2 |\bar f| s_{14} s_{24}}{\sqrt{4 \bar f^2 s_{14}^2 s_{24}^2 + [\bar f (v +s^2_{14} - s^2_{24}) +1]^2}}\,,
\end{align}
and the conversion probability of muon to electron neutrinos can be expressed as
\begin{align}
\label{eq:Pee_4nu_eff}
P_{\mu e}  = \sin^2 2\theta_{12}^m \sin^2 \phi_{12}^m\,, 
\end{align}
where the oscillation phase is given by
\begin{align}
\label{eq:phi_4nu_eff}
\phi_{12}^m = \frac{\Delta \lambda L}{2} \,,
\end{align}
with
\begin{align}
\label{eq:Dlabda}
\Delta \lambda =  \frac{|V_{cc}|}{|\bar f  v +1|}{\sqrt{4 \bar f ^2s_{14}^2 s_{24}^2 + [\bar f (v +s^2_{14} - s^2_{24}) +1]^2}}\,,
\end{align}
representing the difference of the two eigenvalues of the effective Hamiltonian.
The expressions in Eqs.~(\ref{eq:s2_4nu_eff}-\ref{eq:Dlabda}), which in the limit $|\bar f| \to \infty $ return the asymptotic results found in 
Eqs.~(\ref{eq:s2_4nu}-\ref{eq:phi_4nu}), make even more explicit the role of the new potential in
 the dynamics of the novel effective 2-flavor resonance.%

\begin{figure*}[t!]
\vspace*{0.05cm}
\hspace*{-0.1cm}
\includegraphics[height=7.70cm,width=7.87cm]{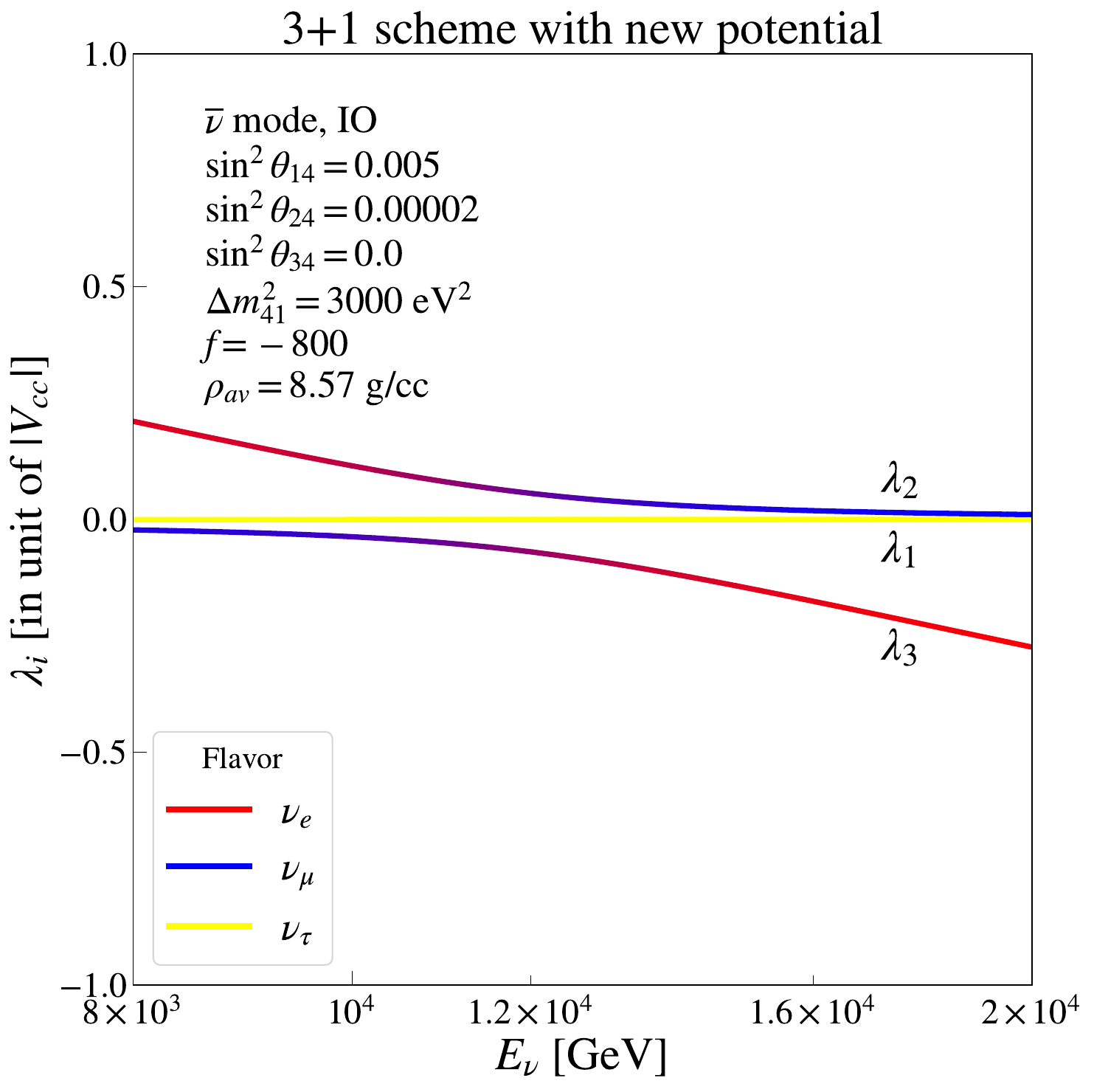}
\includegraphics[height=7.70cm,width=7.87cm]{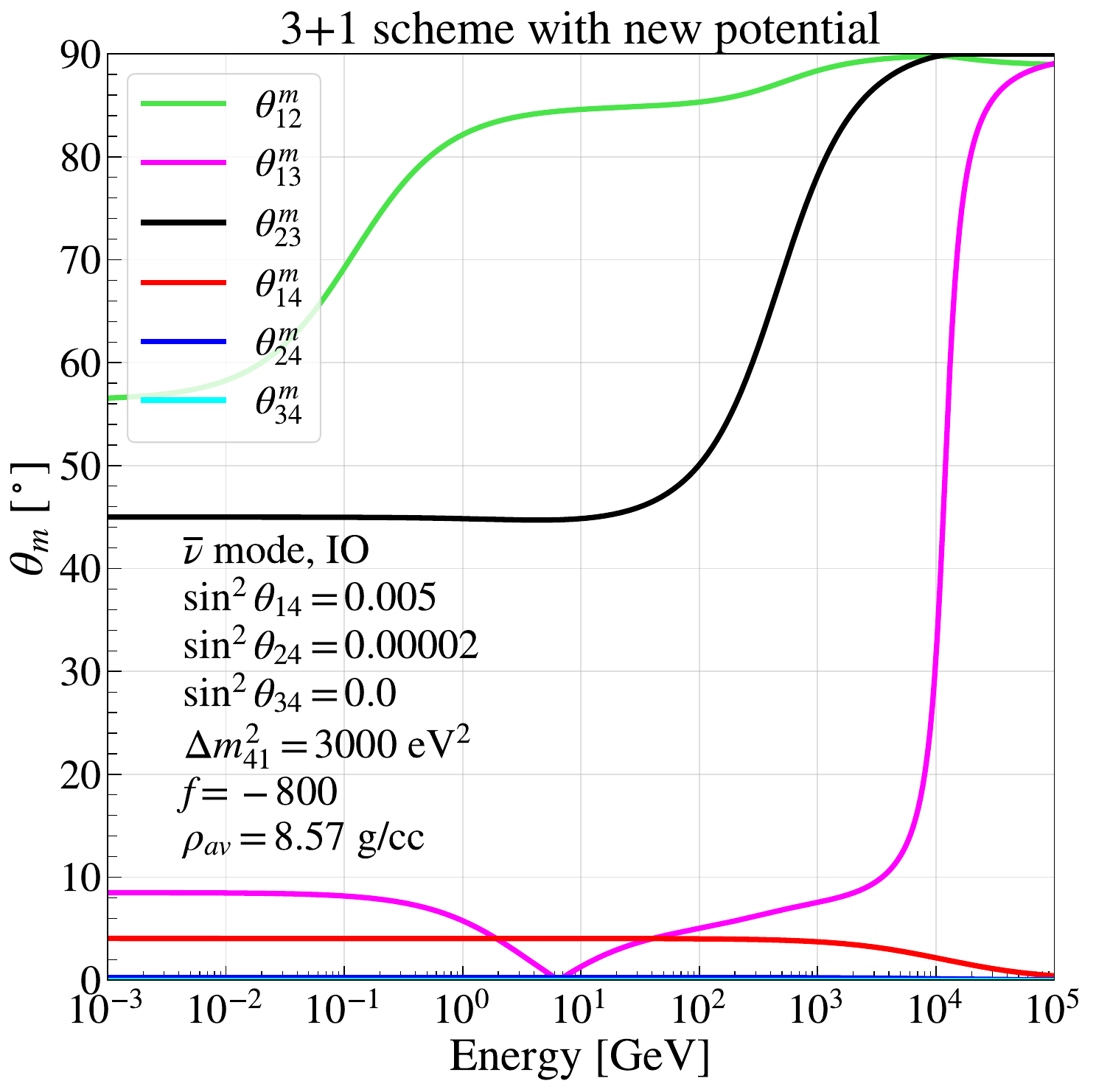}\\
\vspace*{0.5cm}
\includegraphics[height=7.70cm,width=7.87cm]{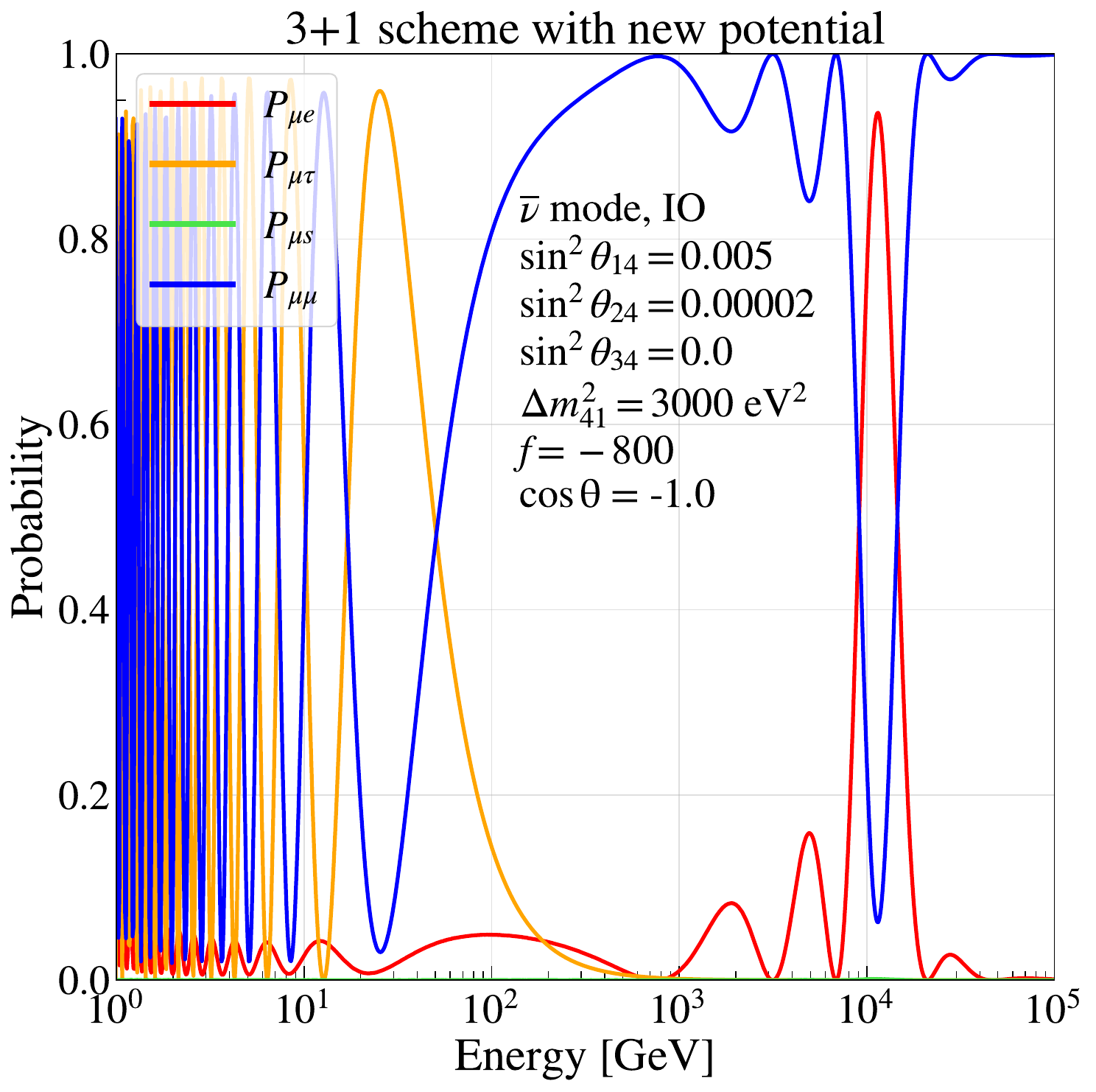}
\includegraphics[height=7.90cm,width=7.87cm]{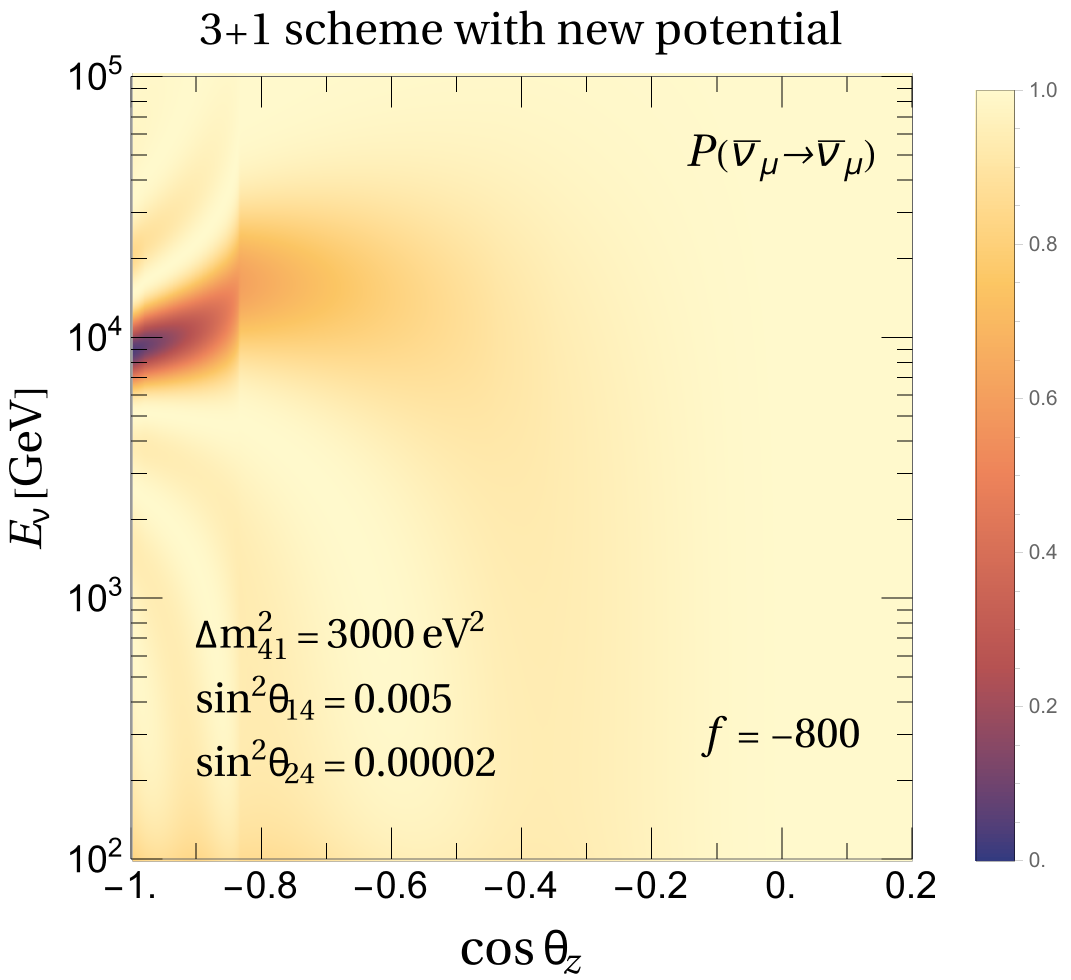}
\vspace*{0.0cm}
\caption{Illustration of the IceCube resonance for parameters in the dark $\theta_{12}$ octant solution as indicated in the plots.}
\label{fig_IceCube_prob_dark}
\end{figure*} 

 The resonance condition, for sufficiently large values of $\bar f $, given the hierarchical pattern $s^2_{24} \ll s^2_{14} $ 
 is approximated by  $V_{CC}= -k s^2_{14}$, which implies that the resonance cannot occur in the neutrino channel because
it requires $V_{CC} < 0$. 
Considering our dark NSI-like solution,
for which $\varepsilon_{ee}^\e  = \bar f s^2_{14}  \simeq -2$ and 
$|\bar f| \simeq 2/s^2_{14}$,
we see that for sufficiently small values of $s^2_{14}$, the parameter $\bar f$ 
can be very large. 
We illustrate this $(\nu_e, \nu_\mu)$ resonant behavior for one selected choice of parameters.
We find that $s^2_{14} = 0.02$ (adopted until now) and the value of $f \simeq - 200$, 
which jointly provide $\varepsilon_{ee}^\e  = \bar f_\e s^2_{14} \simeq -2$, 
are not phenomenologically interesting because such an electron neutrino mixing
at the resulting rescaled value of  $\Delta m^2_{41} \simeq 600$ eV$^2$ (needed to maintain
fixed the resonance energy at $\sim 10$ TeV), is excluded by KATRIN,
which imposes the upper bound  $s^2_{14} \lesssim 0.005$ at 95\% C.L. (see Fig.~3 in~\cite{KATRIN:2025lph}).
This leads us to move towards smaller (larger) values of $s^2_{14}$  ($|f|$), and thus 
to larger values of $\Delta m^2_{41}$, where the bounds from KATRIN get relaxed.
Our final choice is to adopt the values $s^2_{14} = 0.005$,  $f = -800$ (which still provide $\varepsilon_{ee}^\e 
= \bar f_\e s^2_{14} = -2$) and $\Delta m^2_{41} = 3000$ eV$^2$. These parameters are
compatible with KATRIN and also with Mainz~\cite{Kraus:2012he} and Troitsk~\cite{Belesev:2012hx,Belesev:2013cba}, 
which are sensitive in the region of such very high
values of  $\Delta m^2_{41}$ (see Fig.\,1 in~\cite{Giunti:2019fcj} for their combined fit). Note that the large 
value of $f$, leads us to select an extremely small
value of $s^2_{24} \simeq 2\times 10^{-5}$, which serves to respect the bounds on the $\varepsilon_{e\mu}$
and $\varepsilon_{\mu\mu}$ NSI-like couplings imposed by the low-energy ($E\lesssim$ 1 TeV)
phenomenology.

In Fig.~\ref{fig_IceCube_prob_dark} we show the numerical results obtained for  the ``dark resonance'' of IceCube.
The left-upper panel represents the behavior of the eigenvalues. The largest eigenvalue, 
which is positive and very large is not represented in the plot, where we report only the three
smaller eigenvalues. Note that the order of the three eigenvalues is different with 
respect to the case considered in the previous section, because for consistency with the degeneracy theorem,
in the dark solution we have fixed the NMO to be the IO. However, 
this is relevant only at low energies while it is completely irrelevant for the high-energy dynamics
which is independent of the 3-flavor parameters. We can observe that one of the eigenvalues is constant and 
equal to zero because we have set $\theta_{34}=0$ and the $\nu_\tau$ is not
involved in the flavor conversion. We can observe that the other two eigenvalues 
display a clear avoided crossing, which is induced by the off-diagonal 1-2 coupling which
at the resonance takes the value $k_\mathrm{res} s_{14}s_{24}= |V_{CC}| s_{24}/s_{14}  \simeq 0.22\, |V_{CC}| $.
At the resonance, as confirmed by the varying color along the eigenvalues curves, 
there is maximal admixture of $\nu_e$ and $\nu_\mu$ flavors. 
The right-upper panel reports the mixing angles in matter.
For consistency we have fixed $\theta_{12}$ at the octant symmetric ``dark'' best fit value $\sin^2 \theta_{12}=0.69$,
albeit this is irrelevant at high energies. We can see that, differently from the resonance encountered
in the ordinary (non-dark) scenario, only the mixing angle $\theta_{13}^m$ becomes maximal,
confirming that the effective dynamics is a 2-flavor one, albeit it is mediated by the far pseudo-sterile state.
The left-lower panel depicting the muon channel probabilities shows that there is a pure 
$(\nu_e, \nu_\mu$) resonance as expected. All these findings confirm our analytical description. 
For completeness, in the right-lower panel we report the oscillogram of the muon anti-neutrino disappearance
probability, from which we can infer that the dark resonance is more pronounced for core-crossing trajectories.

We must observe that the dark solution would have little impact in the electron disappearance
short-baseline anomalies since the implied $\theta_{14}$ mixing angle is very small. 
However, we think that the illustration of the resonance occurring for the dark octant solution is conceptually
important independently  of its possible phenomenological relevance, because of  three different reasons.
First, as already underlined, this resonance offers the possibility to determine
indirectly the octant of $\theta_{12}$ and consequently distinguishing the NMO
in the framework in which pseudo-sterile neutrinos oscillate with a potential proportional
to electron or proton number density. 
In fact, confronting
the oscillation probabilities and the oscillogram with those obtained for the benchmark parameters
used for the lower-octant solution 
discussed in Sec.~\ref{Sec:IceCube} (see the right panels in Figs.~\ref{fig_IceCube_prob_1} and \ref{fig_oscillogram_mm}),
we see that the behavior is quite different. Also, having at disposal a detector able to distinguish neutrinos from antineutrinos
would be very important as this resonance occurs always in the antineutrino channel.
Second, we realize that the unconventional dark $(\nu_e, \nu_\mu)$ resonance occurs in the absence 
of the conventional resonances in the $(\nu_e, \nu_s)$ and  $(\nu_\mu, \nu_s)$ sub-systems. 
Hence, we learn that the new resonant phenomenon mediated by the sterile neutrinos in the presence 
of the new potential can manifest independently of the coexistence of conventional resonances.
Third, it must be noted that the dark resonance cannot be predicted on the basis of the bare level crossing diagram, 
which shows no other 1-3 resonance apart from the standard 3-flavor one occurring at energies of a few GeV.
Finally, we note that for a non-zero value of $\theta_{34}$ a companion dark resonance in the
$(\nu_e, \nu_\tau)$ channel is present, as expected based on our discussion in the previous section.
However, such a resonance is of less phenomenological interest given the involved flavors. 

A final important remark is in order. In this section, we have discussed the novel
 ($\nu_e,\nu_\mu$) resonance motivated by the generalized mass-ordering degeneracy~\cite{Coloma:2016gei}.
 However, it should be underlined that, as evident from the analytical treatment, 
 the properties of this resonance are independent of the value of the solar mixing 
 angle $\theta_{12}$ and of the NMO choice. Hence, 
the resonance occurs identical for both octants of $\theta_{12}$ and for both choices of the NMO.
Also, we point out that the realization of the resonance is not limited
to values of  $\varepsilon_{ee}^\e  = \bar f s^2_{14}  \simeq -2$ and can occur for different 
choices of such coupling. In particular, we stress that far from the condition 
$\varepsilon_{ee}^\e  = \bar f s^2_{14}  \simeq -2$, it is possible to realize the resonance 
also in the neutrino channel. Therefore, such a resonance has potentially a wide
phenomenological interest. A full numerical simulation would be needed to correctly identify all the
regions of the parameters ($\Delta m^2_{41}$, $f$, $s^2_{14}$, $s^2_{24}$) of interest for IceCube.%
\footnote{ 
For completeness, we mention that there is also a $(\nu_e, \nu_s)$ resonance occurring in the neutrino channel, 
which however is of scarse phenomenological interest given the involved flavors.}

 \section{Interaction of pseudo-sterile neutrinos with asymmetric dark matter}
\label{Sec:ADM}

In the treatment provided in the previous sections we have assumed that the sterile neutrino
potential is produced by their interaction with ordinary matter.  The question arises if the such an interaction
can involve the background dark matter (DM) instead of the ordinary matter. This is possible
only under very particular circumstances. First of all, one has to assume that the DM is asymmetric
(see~\cite{Zurek:2013wia} for a review). In this case, as already investigated in various works, 
both active~\cite{Horvat:1998ym,deSalas:2016svi,Choi:2019zxy,Penacchioni:2020xhg,Smirnov:2021zgn,Huang:2021zzz,Salla:2022dxc,Lin:2023xyk,Gherghetta:2023myo,Chauhan:2025hoz}
and sterile neutrinos~\cite{Miranda:2013wla,Capozzi:2017auw,Lopes:2020hem,Chattopadhyay:2025ccy}
can entail MSW effects. In this scenario, the discrepancies recorded in LBL, 
low-energy and high-energy atmospheric neutrinos,
should be interpreted as a hint of the refraction of the pseudo-sterile neutrinos on the DM particles 
encountered  along the path traversed by neutrinos from  source to  detector. 
The explanation of the anomalies provides the strength of the needed potential, which should be fine-tuned to be
approximately ten times larger than the standard potential in the Earth crust/mantle. This potential 
should be generated by an extremely low matter density equal to the density of the DM halo in the solar 
neighborhood $\rho_\chi (r_\s)\simeq 0.4$ GeV/cm$^3$.
This low density should be counterbalanced by an enormous strength of the new interaction $G_X = f_\chi G_F$, with $f_\chi\simeq n_{e}/n_{\chi} \sim10^{24}$,  pointing towards an ultralight vector mediator. 

A potential issue arises concerning  solar neutrinos since the DM can be accreted by the Sun and accumulated at its center. 
In general one may expect an over-density of DM inside the Sun with respect to its average Halo value felt
by terrestrial neutrinos, which would imply big 
MSW effects in solar neutrinos, which however are not observed.
In fact, we have seen that solar neutrinos can tolerate only the small effects considered in Sec.~\ref{Sec:Solar}.
In order to circumvent this difficulty 
one has to avoid that the DM particles are accreted
in the Sun, which can be obtained adopting a mass of the DM well below the evaporation threshold~\cite{Gould:1987ju} ($m_\chi^\mathrm{evp}
\simeq 4$ GeV) and/or supposing that the DM interaction with ordinary matter is negligibly small. In these
conditions the Sun would be transparent to DM and, apart from small effects induced by gravitational 
focusing (see~\cite{Lee:2013wza,Kim:2021yyo}), no over-density with respect to the surrounding Halo would be present in its interior.
Under this assumption, sterile 
neutrinos traveling from the Sun center to the Earth, would propagate 
in a medium of constant density with a potential similar in size (approximately ten times larger) 
to the standard potential in the Earth crust/mantle. This potential is too small to impact solar neutrino transitions, 
as we have explicitly checked with a numerical simulation. 
As a result, in this scenario solar neutrinos will behave as in the ordinary 3+1 scheme just 
providing the upper bound on $\theta_{14}$. 

\begin{figure*}[t!]
\vspace*{0.05cm}
\hspace*{-0.1cm}
\includegraphics[height=7.87cm,width=8.87cm]{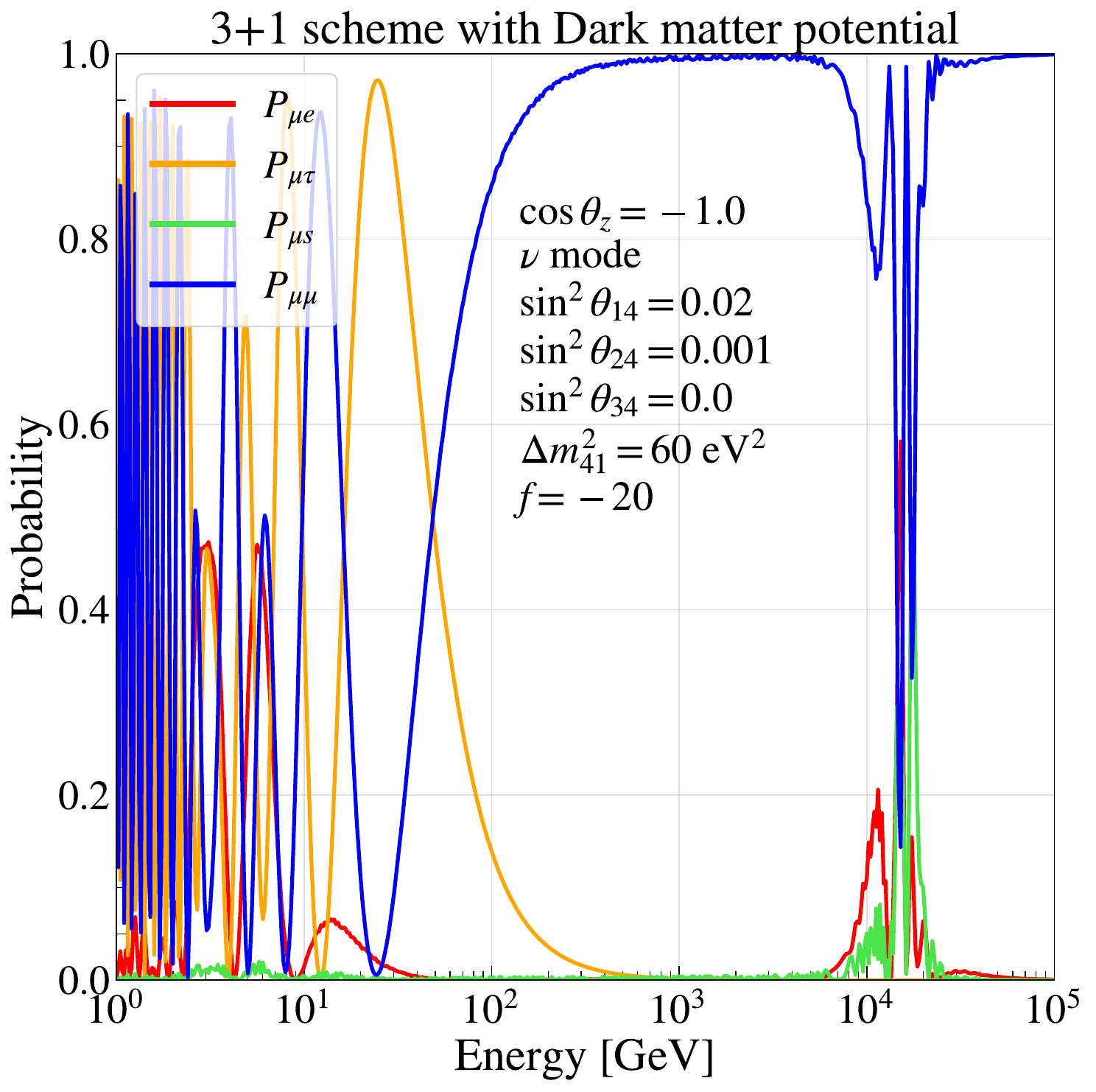}
\includegraphics[height=7.87cm,width=8.87cm]{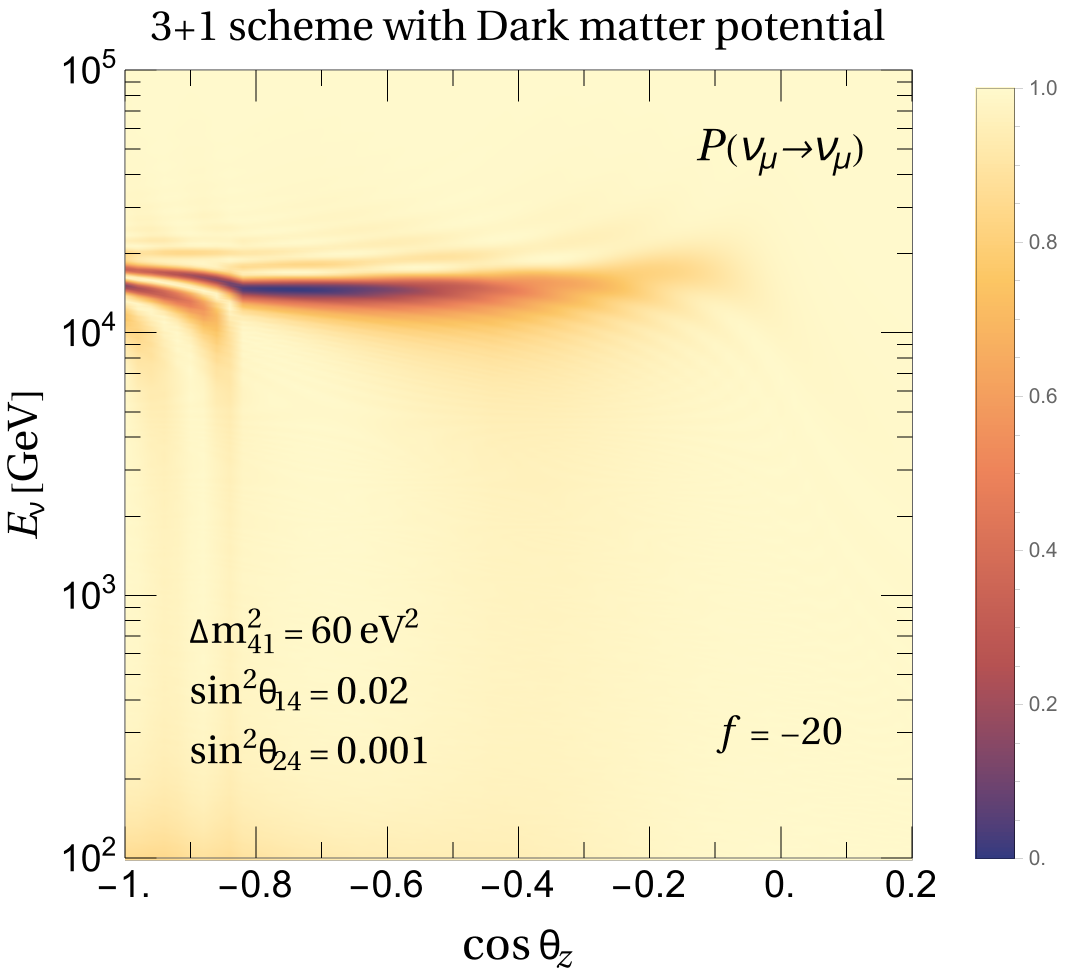}
\vspace*{-0.3cm}
\caption{The figure illustrates the IceCube resonance assuming that the new matter potential is 
proportional to a constant density of dark matter particles. The left panel represents the
oscillation probabilities in the muon neutrino channel for a core-crossing trajectory. The right
panel represents the oscillogram for the muon neutrino survival probability.}
\label{fig_ADM}
\end{figure*}

It is important to note that atmospheric neutrinos are expected
to distinguish between a potential related to ordinary matter from a potential due to dark matter. 
In fact, only in the first case, in which the density varies abruptly along the
path due to the layered structure of the Earth (in particular due to mantle-core-mantle transition),
one has parametric effects. We have seen that  the parametric enhancement has a huge impact on the IceCube resonance for core-crossing trajectories,
providing the resonance with distinctive features (see the right-upper panel of Fig.~\ref{fig_IceCube_prob_1}). 
In the case of MSW on a dark matter background this structure
would be absent. In order to be more quantitative, we show in Fig.~\ref{fig_ADM} the behavior of the IceCube resonance
assuming a new potential having a (negative) constant value in all points of the Earth equal to twenty times the 
value of the standard neutral current potential in the mantle $V_S = 20 V_{NC}^\mathrm{mantle}$.
The left plot shows the probabilities relevant for the muon neutrino channel for a core crossing trajectory.
A comparison with the similar plot obtained for a potential proportional to the (layered) neutron
number density presented in the right-upper panel of Fig.~\ref{fig_IceCube_prob_1}, shows a
drastic reduction of the parametric effects. This is confirmed by the oscillogram displayed in the
right panel. A comparison with the results obtained in the ordinary case (see right panel in Fig.~\ref{fig_oscillogram_mm}), 
shows also a different shape of the oscillograms, in particular the slope in the [$\cos(\theta_z), E]$ plane. 
Therefore, in principle, a spectroscopic study of the high-energy
resonance would allow the distinction between the two cases. Similarly, a precision study of the 
parametric resonances in multi-GeV atmospheric neutrinos could help in distinguishing between 
the two possibilities.%
\footnote{We stress that in the case of a matter potential related to ADM, in addition to the resonance encountered 
in the three-level regime (Sec.~\ref{Sec:IceCube}), also that found in the effective two-level regime (Sec.~\ref{Sec:Dark}) 
can be of relevance in the lower octant of $\theta_{12}$, because there is more freedom on the size of the coupling
$\varepsilon_{ee}^\e = \bar f_\e s^2_{14}$, which in the ADM scenario is basically unconstrained by solar data.}

\section{Discussion and Future Perspectives}
\label{Sec:Discussion}

\subsection{Possible origin of the new matter potential}
\label{SubSec:Origin}

Although we remain agnostic on the nature of the interaction leading to the new matter
potential, here we comment about its possible origin. 
The simplest scenario entails a hidden $U(1)$ gauge group involving
a new vector boson $Z^\prime$. In this model, in the case in which the potential is related to
ordinary baryonic matter, one expects a possible signal in coherent scattering experiments. 
However, these constraints can be evaded by choosing a sufficiently small value of the mediator mass $m_{Z'}$. 
This is possible because the matter potential represents
a process with zero-momentum exchange, while in the scattering experiments the exchanged momentum
is finite. 
Of course, it remains to be understood how the pseudo-sterile neutrinos
impact astrophysics and cosmology, which is a highly non-trivial question.
While the ordinary sterile neutrinos are severely constrained by cosmology,
the situation is less obvious if they interact with ordinary or dark matter.
A common mechanism invoked to render sterile neutrinos compatible with cosmology are 
self-interactions~\cite{Hannestad:2013ana,Dasgupta:2013zpn,Archidiacono:2014nda,Saviano:2014esa,Mirizzi:2014ama,Chu:2015ipa,Cherry:2016jol,Archidiacono:2016kkh,Chu:2018gxk,Farzan:2019yvo,Cline:2019seo,Archidiacono:2020yey},
which seem capable to suppress their mixing. Also unstable pseudo-sterile neutrinos
decaying in visible or invisible products may have a noticeable impact~\cite{Dentler:2019dhz}.
Sterile neutrinos with a new potential have been recently investigated in~\cite{Denton:2018dqq}, where it has been 
shown that they are compatible with a series of bounds from laboratory
and cosmology provided that the mass of the vector mediator is extremely light ($\sim 10$ eV).
According to the same work, self interactions of sterile neutrinos suppress active-sterile mixing
and their production before neutrino decoupling and BBN era, provided that sterile neutrinos have 
contact NSI-like interactions with dark matter particle with an asymmetric dark matter particle.
Considered the highly involved situation of the status of pseudo-sterile neutrinos in cosmology 
we think that the last word on these exotic neutrino states should be left to oscillations 
experiments.

\subsection{Dirac vs Majorana nature of the pseudo-sterile neutrinos}
\label{SubSec:Nature}

We briefly comment on the nature (Dirac vs Majorana) of the sterile neutrino within the 
model under consideration. It is well known that light sterile neutrinos, if Majorana particles,
may be observable in neutrinoless double beta decay~\cite{Goswami:2005ng,Goswami:2007kv,Barry:2011wb,Li:2011ss,Rodejohann:2012xd,Girardi:2013zra,Giunti:2015kza,Liu:2017ago,Deepthi:2019ljo,Jana:2024xmc}.
In fact, sterile neutrino mass contributes to the effective Majorana mass $m_{\beta\beta}$ with a complex term
proportional to $|U_{e4}^2| m_4 e^{i\alpha_4}$, where $\alpha_4$ is a CP-phase related to Dirac and Majorana phases.
The recent study~\cite{Jana:2024xmc} takes into account the most restrictive bounds coming from
KamLAND-Zen, $m_{\beta\beta} < (28 - 122) $\,meV at the $90$\% C.L., where the interval corresponds to
the uncertainty in the nuclear matrix element. According to this work, a Majorana 
sterile neutrino with $|U_{e4}|^2 \simeq 0.01-0.03$ and mass-squared $\Delta m^2_{41} \simeq 60$\,eV$^2$,
as that considered in the present work, is excluded or at least strongly disfavored by data, especially
for the case of normal neutrino mass ordering.  We stress that if $\theta_{12}$ lies in the dark octant 
it is expected to have a substantial impact (in particular for the NO case) in neutrinoless double beta 
decay as shown in~\cite{Vishnudath:2019eiu,Ge:2019ldu}, but this 
is true only in the 3-flavor framework with large ordinary NSI. Within the 3+1 framework, 
where the effective Majorana mass $m_{\beta\beta}$ is dominated by the sterile contribution, the
octant of $\theta_{12}$ is expected to have a minor impact on the limits.
Therefore, the pseudo-sterile neutrinos we are considering, if $s^2_{14}$ is appreciable, are most probably of the Dirac type.
This conclusion can be evaded if $s^2_{14} \ll 0.01$, or within 3+2 and 3+3 schemes where the contribution
to the effective Majorana mass of the additional mass eigenstates  $\nu_5$ and  $\nu_6$ can in principle
lead to cancellations. However, for this scenario to work, one needs fine tuning or an underlying symmetry
forcing the effective Majorana towards values close to zero.

\subsection{Possible manifestations of pseudo-sterile neutrinos in MiniBooNE}
\label{SubSec:Decay}

In the study of the oscillation phenomenology of the pseudo-sterile neutrinos we
have considered basically all kinds of anomalies, with the exception of the
SBL accelerator anomalies such as LSND and MiniBooNE. 
Regarding this last one, it must be noted that MicroBooNE
has recently excluded the electron neutrino appearance~\cite{MicroBooNE:2024tym}
as a possible explanation of the excess. 
We observe that our model
produces a negligible signal since the transition probability is proportional to 
 the effective appearance mixing angle $\sin^2 2 \theta_{\mu e} = 4U_{e4}^2 U_{\mu4}^2 \simeq 10^{-4}$,
and therefore it has no role in the appearance of electron neutrinos.
 However, pseudo-sterile neutrinos may manifest in MiniBooNE in different  ways.
 Here we will mention two (possibly intertwined) options. 

i) {\em Neutral current interactions of pseudo-sterile neutrinos with emission of photons}.
The existence of a new matter potential giving rise to the 
coherent forward scattering, a process involving zero exchanged momentum $q^2 =0$, 
strongly suggests that the pseudo-sterile states may also interact in neutral-current scattering processes 
where a finite $q^2$ exchange is involved. An obvious  candidate to mediate such an 
interaction is a new vector $Z'$ boson related to a symmetry group $U(1)$.
The explanation of MiniBooNE excess
may require a non-trivial coupling of the $Z'$ with baryons and photons. 
In our model,  one has a guaranteed flux of pseudo-sterile neutrinos at the detector proportional to
 $P(\nu_\mu \to \nu_s) \simeq 2 U_{\mu4}^2$, where we have assumed averaged oscillations.
In this condition a minimal dark sector containing a vector $Z'$ and a scalar $\phi$ 
can produce a neutral current scattering with emission of photons.
A similar process has been recently proposed in~\cite{Dutta:2025fgz}, 
where the active neutrinos have been supposed to be coupled to the new mediators.  
We believe that the same model could work even more naturally by replacing 
the active neutrinos with the pseudo-sterile ones.

ii) {\em Visible decay of pseudo-sterile neutrinos}.
 We have already underlined that the small values of $|U_{\mu4}|^2\simeq 0.001$ preferred by the data
would eliminate the tension between IceCube and the most stringent negative $\nu_\mu$ disappearance searches
of MINOS/MINOS+~\cite{MINOS:2017cae} and NOvA~\cite{NOvA:2024imi}.
Such low values of $|U_{\mu4}|^2$, in principle, present another advantage, having 
been invoked to explain the LSND and MiniBooNE
excesses in models in which the sterile neutrino is endowed with visible decay~\cite{Palomares-Ruiz:2005zbh,Bai:2015ztj,Dentler:2019dhz,deGouvea:2019qre,Hostert:2024etd}.
However, to this regard, it must be noted that a recent study~\cite{Hostert:2024etd}
has shown that MicroBooNE results
are in tension with the visible decay explanation of MiniBooNE.
We note that the visible decay scenario may coexist with the neutral current scattering with  emission 
of photons since the same vector and/or scalar particles can mediate both processes.

\subsection{Implications for flavor conversion in a medium with varying density}
\label{SubSec:VarDens}

In our present study of the resonant behavior focused on matter effects in the Earth, 
along the neutrino trajectory the density is constant or presents
layers of constant density. In this case, the dynamics of the resonance is determined by the mixing angles
in matter and by the oscillating phases. The situation would be different in the case of neutrinos
traversing a matter profile with varying density. In this case, one may interpret the eigenvalues 
as the energies of the instantaneous mass eigenstates in matter as a function of the density. Looking
from this perspective at the eigenvalues plots, one expects that for large values of $f$ 
a strong adiabatic conversion between the flavors $\nu_\mu$ and $\nu_e$ will occur traversing
the resonance region. 
As a consequence, a muon neutrino produced at high density (above the resonance 
condition) would be transformed in an electron-neutrino at lower density (below the 
resonance condition).  In addition, a non-zero value of $\theta_{34}$ would imply 
an important role of $\nu_\tau$ in the conversion phenomenon. Therefore, one expects that
the model will have implications for neutrinos propagating in dense environments.
As a matter of fact, the recent work~\cite{Dev:2023znd}  has shown that even in the standard 3-flavor scheme there is an 
observable impact of matter effects in high-energy neutrinos produced in dense environments, leading to
observable deviations of the flavor composition detected on the Earth, with respect to the case of pure vacuum oscillations.
In the presence of pseudo-sterile neutrinos the situation is expected to change appreciably.
We think that this can be a potential signature of our model, which can be investigated 
with future upgraded versions of the Neutrino telescopes such as IceCube-Gen2~\cite{IceCube-Gen2:2021tmd}.

\subsection{Signatures of the model}
\label{SubSec:Smoking}

The proposed scenario has several signatures in planned neutrino experiments. 
In case of  appreciable values of $|U_{e4}|^2 \simeq \sin^2\theta_{14}\sim 0.01-0.03$, 
the pseudo-sterile neutrinos should show up with a distinctive kink in the $\beta$-decay 
energy spectrum measured by KATRIN~\cite{KATRIN:2025lph}. Hence, one of 
the most effective smoking guns of the model is non-oscillatory. 
We notice that KATRIN in~\cite{KATRIN:2025lph} has presented the analysis of five measurement
campaigns. The collaboration has already
collected data in other eleven campaigns~\cite{KATRIN_talk_2025},
which cumulatively contain a statistics four times larger than the first five ones. 
Hence, we can expect a noticeable 
improvement of the sensitivity  in the near future. 

Considering the most obvious oscillation feature, which is the $L/E$ dependency of oscillations in vacuum,
we must notice that it is difficult to be observed. In fact,
the high indicated value of $\Delta m^2_{41} \sim$ 60 eV$^2$ would give rise 
to a mere energy-independent deficit in all existing short-baseline oscillation experiments.
The observation of the oscillation pattern at energies of a few MeV would require a baseline 
of few centimeters. In principle one could circumvent this obstacle by considering a 
movable finely segmented detector with an exceptional spatial resolution. Indeed,
some of the existing experiments are already using this kind of approach. However, from the contour plots 
of the exclusion regions~\cite{Acero:2022wqg}, it seems that all the current experiments
 loose sensitivity for $\Delta m^2_{41} \gtrsim$ 10 eV$^2$.
Perhaps an upgraded version of the existing detectors may be able to improve the sensitivity. 
Considering larger GeV energies obtained in accelerators, the new SBL project
nuSCOPE~\cite{Acerbi:2025wzo} has an impressive sensitivity in the range
of high $\Delta m^2_{41} $, and would provide the 
possibility to probe directly the tiny values of the sterile mixing angles
implied by the new scenario by measuring the electron and muon 
disappearance as well as the muon appearance channels (see~\cite{nuSCOPE_talk_2025}).

Considering signatures involving the presence of matter effects, we highlight what we have called  the matter-vacuum synergy.
In fact, this concept applied to T2K (quasi-vacuum) and NOvA (matter-dominated)
has been an important guide for us to build the model and will have an even
more important role in the future. In fact, having at our disposal two new experiments, Hyper-Kamiokande~\cite{Hyper-Kamiokande:2018ofw},
and DUNE~\cite{DUNE:2015lol},
which will substantially upgrade the existing ones would allow us to maximally exploit the matter-vacuum interplay.
The first experiment would work as as a solid (quasi-vacuum) anchor for the second  (matter-dominated) one,
making it possible  to better identify the new matter effects. 
In addition, it is worthwhile to mention that DUNE, with a broad band spectrum, would 
 be particularly sensitive to the spectral energy distortions induced by the NSI-like couplings and
 together with Hyper-Kamiokande and ESS$\nu$SB~\cite{ESSnuSB:2021azq} will constitute the ideal setup for exploring
 the NSI-like Hamiltonian shown in Eq.~(\ref{eq:Hdyn_2}) and Eq.~(\ref{eq:Hdyn_2_v2}),
 which is another very distinctive signature of the model.
We have shown that the model predicts an excess of electron-like events due to the modification
of the standard 3-flavor resonance occurring at a few GeV in atmospheric neutrinos. Dedicated analyses
performed by Super-Kamiokande and IceCube DeepCore may provide interesting
indications about this feature. The future detectors Hyper-Kamiokande, DUNE and
KM3NeT-ORCA~\cite{KM3Net:2016zxf} are expected to be extremely sensitive to the behavior of multi-GeV atmospheric
 neutrinos, and will be able to perform a precision spectroscopy of the resonance.
 However, one should keep in mind that in  LBL and low-energy atmospheric neutrinos,
 the effects of the 3+1 model with the new matter potential are indistinguishable from those induced
 by NSI couplings, albeit such couplings have a well definite pattern in our model, being arranged
 in a non-generic rank-1 matrix.
 Ultimately, the 4-flavor origin of the observed effects can be proved only by looking
 at high-energy multi-TeV atmospheric neutrinos or by exploiting kinematical methods outlined above
 (SBL oscillations and $\beta$-decay). On the other hand, we underline that purely kinematical experiments 
 cannot reveal the new matter potential, whose identification makes necessary the explorations performed 
 with LBL, atmospheric neutrino experiments and with neutrinos of astrophysical origin.
 
Concerning the resonance at around 10 TeV, we have already underlined the limits of our work, 
in which we could not perform a full data analysis. An accurate official analysis would
be indispensable to establish the quality of the fit,
which is of the foremost importance to confirm the existence of the new matter
potential. Strictly speaking, a good quality of the fit would not provide per se 
a distinctive signature of the model but only a sort of degenerate solution with the ordinary 3+1 scheme.
In principle, one should be able to distinguish sufficiently well the new resonance from the usual one,
which is probably a much more difficult task. The new resonance predicts the existence of cascade 
events generated by electron neutrinos, whose identification is a realistic target for IceCube. 
We note that a precision study of the resonance  
would be possible with an experiment of new conception such as the Neutrino Kaleidoscope recently 
proposed in~\cite{Kamp:2025yzq}, where a collimated beam of TeV neutrinos~\cite{Bojorquez-Lopez:2024bsr}
produced in a muon collider~\cite{InternationalMuonCollider:2025sys} would be detected in one of the existing or planned Neutrino
Telescopes such as IceCube, KM3NeT and P-ONE. The realization of such an ambitious experiment 
would enable us to explore the amplification phenomenon for mantle and core-crossing trajectories, 
using our planet with its layered structure as the ultimate neutrino oscilloscope.
Moving to the sky, the scenario is expected to impact the
flavor composition of astrophysical neutrinos produced in dense environments.
In particular, a substantial  conversion of muon into electron neutrinos is expected.
 All in all, we think that our model is highly predictive and can be efficiently tested by the 
 current and future experiments.

\section{Conclusions}
\label{Sec:Conclusions}

We have considered a 3+1 scenario in which sterile neutrinos feel a 
new matter potential proportional to background density.
We have called such states as {\em pseudo-sterile neutrinos}.
We have shown that the model behaves as an effective NSI-like scenario
at low energies ($E\lesssim $ 1 TeV)  and presents a very distinctive Hamiltonian,
represented by  a non-generic rank-1 matrix. Remarkably, unlike the ordinary NSI, 
the induced NSI-like couplings, being generated via mixing with a 
gauge-singlet state, are not subject to the stringent constraints from $SU(2)$
 gauge invariance involving charged leptons, thus overcoming a well-known 
 structural  limitation of the ordinary NSI. At the phenomenological level, 
 the novel NSI-like couplings
 are able to provide an explanation of the
discrepancy observed between NOvA and T2K  and explain the excess
of $\nu_e$-like events in multi-GeV atmospheric neutrinos as hinted at by
Super-Kamiokande. 

At high energies ($E \simeq 10$ TeV) the new scenario 
reveals its full 4-flavor nature producing a resonance as 
suggested by IceCube. We have shown that the structure of the amplification 
phenomenon is much more complex with respect to the  ordinary 3+1 framework.
Specifically, we have demonstrated that a genuine three-level dynamics can emerge.
In this regime, the pseudo-sterile neutrino participates to new resonant phenomena 
with the active flavors and it cannot be integrated out from the dynamics.
In addition, we have pointed out  that, in the case in which the new matter potential is proportional 
to the electron or proton number density, an interesting resonance exists for the 
solar mixing angle in the dark octant. In this case, the fourth pseudo-sterile state
can be integrated out, giving rise to a different regime, which
can be described by an effective two-level Hamiltonian, of which we have provided a full
analytical derivation. This resonance seems phenomenologically
less interesting as it involves very small values of $\theta_{14}$, which are
not relevant for the short-baseline anomalies. However, a full numerical study is warranted 
to better evaluate its potential interest.
Finally, we have discussed the possibility
that the new matter potential is generated by the interaction of the pseudo-sterile states 
with a background of asymmetric dark matter particles pointing out
the role of parametric enhancement in the Earth in discriminating such a scenario.

The three-level regime seems to be the most interesting from the phenomenological point
of view. In order to be of maximal relevance for the anomalies, the new scenario requires 
that the new potential is approximately twenty times larger than the 
standard neutral-current potential and has negative sign $f = V_S/|V_{NC}| \simeq -20$.
 The preferred value of the mass-squared splitting lies around 
$\Delta m^2_{41} = 60~$eV$^2$. Both $|f|$ and  $\Delta m^2_{41}$ are 
roughly uncertain by a factor of a two and are positively correlated since their
ratio is required to be approximately constant in order to fix the IceCube resonance energy
at $\sim 10$\,TeV.
The data also indicate that the admixture of the muon neutrinos
with the fourth sterile neutrino mass eigenstate is one order of magnitude
smaller than the mixing of the electron neutrinos ($|U_{e4}|^2 \simeq 0.01-0.03$,
$|U_{\mu 4}|^2 \simeq 0.001$).  The small values of
$|U_{\mu4}|^2$  eliminate the tension between IceCube and the other negative $\nu_\mu$ disappearance searches.
Also the values of $|U_{\tau4}|^2$ are limited to be at least one order of magnitude smaller than  $|U_{e4}|^2$.
 The predictions for $\Delta m^2_{41}$  and $|U_{e4}|^2$ will be directly tested in the near future
 by KATRIN, in which one expects  a characteristic kink in the $\beta$-decay energy spectrum. 
Concerning the nature (Dirac vs Majorana) of the pseudo-sterile neutrino, the null results  from $0\nu2\beta$ decay 
strongly favor the Dirac case. However, this conclusion can be evaded in 3+2 and 3+3 frameworks if
an underlying symmetry forces the effective Majorana mass close to zero.

The pseudo-sterile neutrino scenario presents an extremely rich phenomenology,
involving baselines from the centimeter scale to the Earth-Sun distance, 
passing through the hundreds/thousands kilometers traveled by
LBL and atmospheric neutrinos.  Also the interested energies span from 
a fraction of MeV of $pp$ solar neutrinos, to hundreds of TeV of atmospheric
neutrinos. It is plausible  to expect further signatures at even larger baselines and
higher energies in neutrinos of astrophysical origin. Also, additional
pseudo-sterile species are naturally expected to exist, making the picture even more complex and intriguing.
Given such a rich landscape, if the proposed scenario
were to be confirmed, a strenuous effort would be required to the neutrino community
to shape the model and identify the nature of the new fundamental interaction
involving the pseudo-sterile states.
We would like to conclude by recalling that
 the neutrino index of refraction in matter has led to the resolution of the longstanding solar neutrino problem,
and likewise we hope that it can shed light on the modern puzzle of the terrestrial neutrino oscillation anomalies 
in an even more spectacular way.

\begin{acknowledgments}

\noindent  The work of S.S.C. is funded by the Deutsche Forschungsgemeinschaft (DFG, German Research Foundation) – project number 510963981. A.P. acknowledges partial support by the research grant number 2022E2J4RK ``PANTHEON: Perspectives in Astroparticle and Neutrino THEory with Old and New messengers" under the program PRIN 2022 funded by the Italian Ministero dell’Universit\`a e della Ricerca (MUR) and by the European Union – Next Generation EU as well as partial support by the research project {\em TAsP} funded by the Instituto Nazionale di Fisica Nucleare (INFN).

\end{acknowledgments}

\bibliographystyle{utphys}
\bibliography{STERILE_POTENTIAL-References_v1}

\end{document}